\documentclass[aps,nofootinbib,floatfix,showpacs,preprintnumbers,prd,twocolumn,floatfix,nofootinbib,superscriptaddress]{revtex4}
\RequirePackage{lineno}
\usepackage{soul, ulem}
\usepackage{graphicx, epsfig, bm, amsmath}
\usepackage{amssymb, color, float,hyperref}
\usepackage{wasysym}
\usepackage{wrapfig}
\usepackage{lipsum}
\usepackage{pdflscape}
\usepackage{rotating}
\usepackage{scalerel,stackengine}
\stackMath
\newcommand\reallywidehat[1]{
\savestack{\tmpbox}{\stretchto{
  \scaleto{\scalerel*[\widthof{\ensuremath{#1}}]{\kern-.6pt\bigwedge\kern-.6pt}%
    {\rule[-\textheight/2]{1ex}{\textheight}}
  }{\textheight}%
}{0.5ex}}%
\stackon[1pt]{#1}{\tmpbox}%
}

\begin{document}

\title{Search for dark matter signals towards a selection of recently-detected DES dwarf galaxy satellites of the Milky Way 
with H.E.S.S.}

\author{H.E.S.S. Collaboration}
\noaffiliation

\author{H.~Abdallah} 
\affiliation{Centre for Space Research, North-West University, Potchefstroom 2520, South Africa}

\author{R.~Adam}
\affiliation{Laboratoire Leprince-Ringuet, École Polytechnique, CNRS, Institut Polytechnique de Paris, F-91128 Palaiseau, France}

\author{F.~Aharonian}
\affiliation{Max-Planck-Institut f\"ur Kernphysik, P.O. Box 103980, D 69029 Heidelberg, Germany}
\affiliation{Dublin Institute for Advanced Studies, 31 Fitzwilliam Place, Dublin 2, Ireland}
\affiliation{High Energy Astrophysics Laboratory, RAU, 123 Hovsep Emin St Yerevan 0051, Armenia}

\author{F.~Ait Benkhali}
\affiliation{Max-Planck-Institut f\"ur Kernphysik, P.O. Box 103980, D 69029 Heidelberg, Germany}

\author{E.O.~Ang{\"u}ner}
\affiliation{Aix Marseille Universit\'e, CNRS/IN2P3, CPPM, 13288 Marseille, France}

\author{M.~Arakawa}
\affiliation{Department of Physics, Rikkyo University, 3-34-1 Nishi-Ikebukuro, Toshima-ku, Tokyo 171-8501, Japan}

\author{C.~Arcaro}
\affiliation{Centre for Space Research, North-West University, Potchefstroom 2520, South Africa}

\author{C.~Armand}
\email[]{Corresponding authors. \\  contact.hess@hess-experiment.eu}
\affiliation{Laboratoire d'Annecy de Physique des Particules, Univ. Grenoble Alpes, Univ. Savoie Mont Blanc, CNRS, LAPP, 74000 Annecy, France}

\author{T.~Armstrong}
\affiliation{University of Oxford, Department of Physics, Denys Wilkinson Building, Keble Road, Oxford OX1 3RH, UK}

% ...

\begin{abstract}
Dwarf spheroidal galaxy satellites of the Milky Way are prime targets for indirect detection of dark matter with gamma rays due to their proximity, high dark matter content and absence of non-thermal emission processes.
Recently, the Dark Energy Survey (DES) revealed the existence of new ultra-faint dwarf spheroidal galaxies in the southern-hemisphere sky, therefore ideally located for ground-based observations with the imaging atmospheric Cherenkov telescope array H.E.S.S. We present a search for very-high-energy ($E\gtrsim100$ GeV) gamma-ray emission 
using H.E.S.S. observations carried out recently towards Reticulum II, Tucana II, Tucana III, Tucana IV and Grus II satellites. No significant very-high-energy gamma-ray excess is found from the observations  on any individual object nor in the combined analysis of all the datasets. Using the most recent modeling of the dark matter distribution in the dwarf galaxy halo, we compute for the first time on DES satellites individual and combined constraints from Cherenkov telescope observations on the annihilation cross section of dark matter particles in the form of Weakly Interacting Massive Particles. The combined 95\% C.L. observed upper limits reach $\langle \sigma v \rangle \simeq 1 \times 10^{-23}$ cm$^3$s$^{-1}$ in the $W^+W^-$ channel and $4 \times 10^{-26}$ cm$^3$s$^{-1}$ in the $\gamma\gamma$ channels for a dark matter mass of 1.5~TeV. The H.E.S.S. constraints well complement the  results from Fermi-LAT, HAWC, MAGIC and VERITAS and are currently the most stringent in the $\gamma\gamma$ channels in the multi-GeV/multi-TeV mass range.
\end{abstract}


\maketitle

\section{Introduction}
\label{sec:intro}
Precise cosmological measurements~\cite{Adam:2015rua} support the theory that most of the matter in the Universe is composed of non-baryonic cold dark matter (DM).  The search for non-gravitational interactions of DM is one of the major efforts in contemporary fundamental astrophysics. Despite the worldwide multi-faceted efforts that have been deployed over the last decades to detect DM, its nature is presently unknown. 
Many theoretical models~\cite{Bertone:2004pz} have been devised to propose DM particle candidates. Among them is a Weakly Interacting Massive Particle (WIMP) with mass and coupling at the electroweak scale that provides the cold DM density measured in the 
Universe today~\cite{Steigman:2012nb}, which is popularly acknowledged as the {\it WIMP miracle}. Among the experimental strategies devised to detect DM, the indirect searches look for the Standard Model particles produced during the DM annihilation or decay. WIMPs could still annihilate today in dense regions of the Universe producing very-high-energy (VHE, E$\gtrsim$100 GeV) gamma rays in the final states that can be eventually detected by ground-based imaging atmospheric Cherenkov telescopes (IACT) such as the High Energy Stereoscopic System (H.E.S.S.). 

Among the most favorable environments to look for DM annihilation in VHE gamma rays are dwarf spheroidal galaxies  (dSphs) satellites of the Milky Way, with many of them nearby and at high Galactic latitudes. The measured stellar kinematics in dSphs make them the most DM-dominated objects in the Universe. They are composed of old stellar populations and contain little gas which could act as target materials for VHE cosmic rays.  No hint is found for non-thermal processes that could give rise to emission from non-DM scenarios which would serve as background for a DM search in VHE gamma rays~\cite{Viana:2012zz,Winter:2016wmy}. Despite the lower DM signals expected for dSphs compared to the central region of the Milky Way, they have the advantage of negligible background emission to hide a DM signal. 

Numerous dSphs have been discovered via the Sloan Digital Sky Survey~\cite{Abazajian:2008wr} covering the Northern celestial hemisphere. More recently ultra-faint dSphs are being unveiled by the ongoing surveys like PanSTARRS~\cite{PanSTARRS}, and the DES~\cite{DES}, with the prospect of more discoveries with the Large Synoptic Survey Telescope~\cite{LSST}. DES is a southern-hemisphere optical survey providing photometric measurements to detect stellar overdensities with unprecedented sensitivity in the southern sky. The ultra-faint Milky Way satellites newly discovered by DES are consistent with being dSphs while a fraction of them are referred to as dSph candidates in absence of confirmation from spectroscopic measurements. They represent new promising targets for VHE gamma-ray searches for DM annihilations.   

We present here the observations carried out by H.E.S.S. on a selection of DES satellites to search for DM annihilation signals.  The targeted systems are Reticulum II (Ret~II), Tucana II (Tuc~II), Tucana III (Tuc~III), and Grus II (Gru~II), with Tucana IV (Tuc~IV) in the field of view (FoV) of Tuc~III observations. The results of the search for DM annihilation signals 
are presented for individual and combined searches towards these targets. The paper is organized as follows. Sec.~\ref{sec:signals} presents the DM signals expected from the targets. In Sec.~\ref{sec:dataset} and \ref{sec:analysis}, we present the observational datasets, and the data analysis, respectively. Sec.~\ref{sec:results} is devoted to the results. We conclude in Sec.~\ref{sec:summary}.

\section{Dark matter signals}
\label{sec:signals}
\subsection{Dark matter distribution and gamma-ray flux}
The energy-differential gamma-ray flux expected from the self-annihilation of DM particles of mass $m_{\rm DM}$ in the region of solid angle $\Delta\Omega$ can be written as~\cite{Bertone:2004pz}:
\begin{widetext}
\begin{equation}
\label{eq:flux}
\frac{{\rm d} \Phi_\gamma}{{\rm d} E_\gamma} (E_\gamma,\Delta\Omega)=
\frac {\langle \sigma v \rangle}{8\pi m_{\rm DM}^2}\sum_{\rm f}  \text{BR}_{\rm f} \frac{{\rm d} N^{\rm f}}{{\rm d}E_\gamma} \, J(\Delta\Omega) \ ,
\quad {\rm with} \quad  J(\Delta\Omega) =  \int_{\Delta\Omega} \int_{\text{LoS}}\rho^2(s(r,\theta)) ds\, d\Omega \, .
\end{equation}
\end{widetext}
The first term groups the total velocity-weighted annihilation cross section $\langle \sigma v \rangle$ and the sum of the annihilation spectra $dN^f/dE_\gamma$ in the final states $f$ with associated branching ratios $\text{BR}_f$. The second term, often referred to as the $J$-factor, corresponds to the square of the DM mass density $\rho$ integrated over the line-of-sight (LoS) $s$ and $\Delta\Omega$. The distance from the observer to the annihilation location $s$ is given by $r = (s^2+s_0^2 - 2\,s_0\, s\,\cos \theta)^{1/2}$, where $s_0$ is the distance from the target to the Earth and $\theta$ the angle between the direction of observation and the dSph center. The DM mass density is inferred from the measurements of the position and LoS velocity of the stars gravitationally bound in the dwarf galaxy potential well through the Jeans equation~\cite{2008gady.book.....B}. 
The finite number of kinematic measurements of the member stars leads to an uncertainty on the $J$-factor,
see, {\it e.g.}, Ref.~\cite{Geringer-Sameth:2014yza}. 

The expected DM flux is composed of a continuum spectrum extending up to the DM mass, and a line emission feature. The former contribution arises from the hadronization and/or decay of quarks, heavy leptons, and gauge bosons involved in the annihilation process. The latter comes from the prompt annihilation into $\gamma X$ with X= $\gamma$ , h, Z or a non-standard model neutral particle, providing a  spectral line at an energy $E_{\gamma} =  m_{\rm DM}[1-(m_X/2m_{\rm DM})^2]$. Additional gamma rays are produced when the DM particles self-annihilate into charged particles via processes involving virtual
internal bremsstrahlung and final state radiation. Such processes provide a wider line-like feature that peaks at an energy close to $m_{\rm DM}$.

\subsection{Target selection}
Five targets were selected among the 16 newly discovered DES dSphs~\cite{2015ApJ80750B,Li:2015kag}. The selection is based mainly on the DM content and visibility from the H.E.S.S. site. Targets with measured or predicted $J$-factor close to $\log_{10}(J_{<0.5^\circ}/{\rm GeV^2 cm^{-2}})\sim19$ are chosen, with visibility at zenith angles lower than $60^\circ$ spread all over the year. The priority has been given to targets that are confirmed as dwarf galaxies, followed by the galaxy candidates with the largest $J$-factor. The chosen targets are outlined in Tab.~\ref{tab:table3}. The low-luminosity Milky Way satellite Ret II has been discovered using photometric data from the DES~\cite{Koposov:2015cua,2015ApJ80750B}. Located at a distance of 32~kpc from the Sun, it is one of the nearest dSphs after Segue~1 (23~kpc)~\cite{2007ApJ654897B} and Sagittarius (24~kpc)~\cite{2016AcA66197H}. Ret~II is about three times more luminous than Segue 1, which suggests that its DM halo may be more massive than that of Segue~1. This makes it a privileged dSph target for DM searches. Assuming dynamical equilibrium and spherical symmetry, a Jeans analysis of the available kinematic data suggest that the $J$-factor of Ret~II is among the highest of the known dSphs. Its $J$-factor  integrated within a radius of 0.5$^{\circ}$ reaches $\rm \log_{10} (\mathit{J}/GeV^2cm^{-5})$ = 19.6~\cite{Bonnivard:2015tta} 
based on a kinematic sample of 38 member stars. Alternative studies derived mean $\rm \log_{10} (\mathit{J}/GeV^2cm^{-5})$ values as large as 20.5~\cite{Achterbeg:2015dca} and as low as 18.2~\cite{Evans:2016xwx} within a 0.5$^{\circ}$ radius. A systematic study presented in Refs.~\cite{Bonnivard:2015tta,2016MNRAS.462..223B} shows that its $J$-factor determination is stable against assumptions in the Jeans analysis. No hints of tidal disruption~\cite{2015ApJ808108W} or a significant binary star population~\cite{2019MNRAS.487.2961M} in Ret~II have been detected so far. Based on the velocities and metallicities of its stars, Ret~II is confirmed as an ultra-faint dwarf galaxy~\cite{2018ApJ...863...25M}.
Present photometric and spectroscopic data cannot constrain the fraction of binary stars in the kinematic stellar sample.  In the observed absence of tidal disruption and binary stars that would artificially inflate the velocity dispersion, Ret~II is a prime DM target for H.E.S.S.   

Tuc~II is an ultra-faint dSph galaxy satellite of the Milky Way discovered from DES photometric data~\cite{2015ApJ80750B} located at 57~kpc from the Sun. Spectroscopic measurements of member stars~\cite{2016ApJ...819...53W} reveal a low internal velocity dispersion. Assuming dynamical equilibrium, spherical symmetry, and a negligible contamination of binary stars in the stellar sample, the $J$-factor of Tuc~II is calculated as   
$\rm \log_{10} (\mathit{J}/GeV^2cm^{-5})$ = 18.7 within 0.5$^{\circ}$~\cite{2016ApJ...819...53W}, which makes it an interesting DM target among known dSphs.  Spectroscopic observations of member stars of Tuc~II classify it as a dwarf galaxy~\cite{2016ApJ...819...53W}. No tidal disruption or significant binary star population have been measured so far in this system. In Ref.~\cite{Pace:2018tin} the $J$-factor of Tuc II is predicted to be $\rm \log_{10} (\mathit{J}/GeV^2cm^{-5})$ = 19.0 within a 0.5$^{\circ}$ radius~\cite{Pace:2018tin}. The conservative estimate is used in what follows.

Among the dSph candidates discovered by DES~\cite{2015ApJ80750B}, Tuc~III is the nearest low-luminosity Milky Way satellite located at a heliocentric distance of 25~kpc~\cite{Simon:2016mkr}. 
Spectroscopic measurements show a very low velocity dispersion in the member stars, and only upper limits can be safely derived. 
Despite its larger size and lower surface brightness compared to any known globular
cluster, Tuc~III cannot be confirmed as a DM-dominated system, and, therefore, cannot be definitely classified as a dSph. If Tuc~III has a DM halo similar to the one of other satellites with similar stellar mass, the $J$-factor of Tuc~III can be as high as $\rm \log_{10} (\mathit{J}/GeV^2cm^{-5})$ = 19.4 within 0.5$^{\circ}$~\cite{Simon:2016mkr} making it a very promising DM target that can be conveniently observed by H.E.S.S. However, the derivation of the $J$-factor from the modeling of the velocity distribution suffers from systematic uncertainties and $J$-factors as low as $\rm \log_{10} (\mathit{J}/GeV^2cm^{-5}) \simeq$ 17.8 are possible~\cite{Simon:2016mkr}. The predicted values from Ref.~\cite{Fermi-LAT:2016uux} and Ref.~\cite{Pace:2018tin} are $\rm \log_{10} (\mathit{J}/GeV^2cm^{-5})$ = 19.0 and 17.7, respectively, within 0.5$^{\circ}$.

No accurate spectroscopic measurements are available for Gru~II and Tuc~IV. In absence of measurements of the velocity dispersion of member stars, they are classified as likely dSphs~\cite{Fermi-LAT:2016uux}. No $J$-factor can be measured and the 
empirical law from Ref.~\cite{Fermi-LAT:2016uux} is used to provide an estimate of the $J$-factor of Gru~II and Tuc~IV.

The determination of the DM density distribution in dSphs is subject to uncertainties that can significantly affect the $J$-factor estimation.  Due to the finite sample of stellar tracers in dSphs, the statistical uncertainty on the $J$-factor is higher for ultra-faint candidates than that of the classical dSphs such as Sculptor or Draco.
The Jeans equation framework assumes dynamical equilibrium of stellar tracers ({\it e.g.} no tidal disruption), spherical symmetry of the system, light profile parametrization, and radial dependence of the velocity anisotropy. This set of hypotheses may not be valid in the physical systems, which would lead to systematic uncertainties in the $J$-factor determination. In what follows, the statistical uncertainties for the computation of the limits are considered when the spectroscopic measurements are available. Sources of systematic uncertainties in the $J$-factor determination in dSphs are discussed, for instance, in Refs.~\cite{Bonnivard:2014kza,Lefranc:2016dgx}. 

\begin{sidewaystable}
\centering
\begin{tabular}{ l | c | c | c | c | c | c | c | c  }
\hline
\hline
Source & Heliocentric distance & Confirmed dSph & ON region radius &$\overline{\log_{10} J_{<0.5^{\circ}}}$ & $\sigma_{\rm J_{<0.5^{\circ}}}$ & $\overline{\log_{10} J_{\rm ON}}$  & $\sigma_{\rm J_{\rm ON}}$ & Ref. for the \\
name & [kpc] & & [degrees] &[$\log_{10} {\rm (GeV^{5}cm^{-2}})$] &  [$\log_{10} {\rm (GeV^{5}cm^{-2}})$] & [$\log_{10} {\rm (GeV^{5}cm^{-2})}$]   &  [$\log_{10} {\rm (GeV^{5}cm^{-2}})$] & $J$-factor\\
\hline
Reticulum II & 32 & Yes & 0.200 &19.6 & 0.9 & 19.2 & 0.6 & \cite{Bonnivard:2015tta}\\
Tucana II & 58 & Yes & 0.200  &18.7 & 0.8 & 18.4 & 0.7 & \cite{2016ApJ...819...53W}\\
Tucana III$^\dag$ & 25 & No & 0.125 &19.4 & - & 18.8 & \hspace{0.1cm} 0.7$^\ddag$ & \cite{Simon:2016mkr}\\
Tucana IV$^\dag$ & 48 & No & 0.125 &18.7  & - & 18.1 & \hspace{0.1cm} 0.7$^\ddag$ & \cite{Fermi-LAT:2016uux}\\
Grus II$^\dag$ & 53 & No & 0.125 &18.7 & - & 18.1 & \hspace{0.1cm} 0.7$^\ddag$ & \cite{Fermi-LAT:2016uux}\\
\hline
\hline
\end{tabular}
\caption{\label{tab:table3}
Selected Milky Way satellites discovered by DES for H.E.S.S. observations. The second column gives the heliocentric distance of these objects. The third column provides their identification status: systems whose photometry makes them compatible with being a dSph but without spectroscopic measurements are classified as not confirmed. The size of the ON region is given in the fourth column. The fifth and sixth columns provide the mean value of the $J$-factor for an integration angle of 0.5$^{\circ}$ and its associated 1$\sigma$ uncertainty. The uncertainty quoted for Ret~II takes also into account possible sources of systematic uncertainties~\cite{Bonnivard:2015tta} while only the statistical uncertainty is considered for Tuc~II.
The references from which these values are extracted are given in the ninth column. The same quantities are provided in the seventh and eight columns for an integration angle corresponding to the size of the region of interest (ON region). The symbol $^\dag$ marks the systems for which no spectroscopic measurement is available. In such a case, the integration angle for the ON region corresponds to 0.125$^{\circ}$ as relevant for point-like emission searches. $^\ddag$In the absence of measured uncertainty, $\sigma_{\rm J_{\rm ON}}=0.7$ is assumed in order to have an estimate of the effect of this uncertainty on the combined results (see Sec.~\ref{sec:results}).}
\end{sidewaystable}
\begin{sidewaystable}
\centering
\begin{tabular}{ l | c | c | c | c | c | c }
\hline
\hline
Source & RA & Dec & Longitude & Latitude & Live time  & Mean zenith angle  \\
name & [hours] & [degrees] & [degrees] & [degrees] &[hours] & [degrees]\\
\hline
Reticulum II & 03:35:40.8 & -54:03:00 & 266.30 & -49.74 & 18.3 & 42.4\\
Tucana II & 22:52:14.4 & -58:34:12 & 328.04 & -52.35 & 16.4 & 36.1 \\
Tucana III & 23:56:36.0 & -59:36:00 & 315.38 & -56.18 & 23.6 & 39.0\\
Tucana IV & 00:02:55.2 & -60:51:00 & 313.29 & -55.29 & 12.4 & 39.2 \\
Grus II & 22:04:4.8 & -46:26:24 & 351.14 & -51.94 & 11.3 & 29.0 \\
\hline
\hline
\end{tabular}
\caption{\label{tab:table1} Table of Milky Way satellites discovered by DES and consistent with dwarf galaxies observed by H.E.S.S. The  second-to-fifth columns give the distance and the galactic coordinates of each object, respectively.  The last two columns provide the live time on target acquired by H.E.S.S. in the 2017 and 2018 observation campaigns and the mean zenith angle of observations. Uncertainties on the distance of the systems are of about 10\%.}
\end{sidewaystable}

The ultra-faint systems Tuc~III, Tuc~IV and Gru~II lack spectroscopic measurements of their member stars and their $J$-factor can only be predicted. Assuming these objects to be embedded in spherical cuspy DM halo following the relationship between their enclosed mass, velocity dispersion, and half-light radius, an analytic formula to calculate the $J$-factors can be derived~\cite{Evans:2016xwx}. An alternative method based on a distance scaling relationship provides a  compatible estimate of the $J$-factor assuming that the stellar systems lie in DM halos similar to those of known dSphs as shown in Ref.~\cite{Fermi-LAT:2016uux}. For Tuc~III, the $J$-factor value within 0.5$^\circ$ can be as large as $\log_{10}$ J = 19.4 using the lower limit on its mass from the tidal stripping argument~\cite{Simon:2016mkr}. However, the derived value from an upper limit on the LoS velocity dispersion is about two orders of magnitude below. Recent photometric observations classify Tuc~III as an ultra-faint dwarf galaxy~\cite{2018ApJ...863...25M} while spectroscopic measurements failed so far to unambiguously confirm its dynamical status~\cite{Simon:2016mkr}. In absence of lower limits on the velocity dispersion for Tuc~IV and Gru~II, the $J$-factor values in the region of interest (ROI) are derived following the methodology of Ref.~\cite{Fermi-LAT:2016uux} assuming an inner slope of one for a cuspy DM profile~\cite{Evans:2016xwx}. In the case of Tuc~III, Tuc~IV, and Gru~II, no statistical uncertainty can be derived for their $J$-factors. $\sigma_{\rm J_{\rm ON}}=0.7$ is assumed to have an estimate of the degradation of the H.E.S.S. limits towards these objects when considering the $J$-factor uncertainty. Such a value is coherent given the measured uncertainty derived for Ret~II and Tuc~II.

\section{Observations and dataset}
\label{sec:dataset}
H.E.S.S. is an array of five IACTs situated in the Khomas Highland in Namibia, at 1800~m above the sea level. The observatory consists of four 12~m diameter telescopes (CT1-4) at the corner of a square of side length 120~m and a fifth 28~m diameter telescope (CT5) at the middle of the array since 2012.
Given its location in the Southern Hemisphere, H.E.S.S. is best located to observe recently-detected DES dSphs compared to other IACTs. 

The observations presented here were performed with the full five-telescope array (H.E.S.S. Phase 2) towards a selection of recently-discovered dSphs.  Dedicated observations were carried out towards Ret~II, Tuc~II, Tuc~III in 2017 and 2018, and Gru~II in 2018. Previous observations targeted at another source covered the position of Grus II providing a useful exposure at its nominal position.  Given the H.E.S.S. FoV, observations taken towards Tuc~III also include Tuc~IV. 
The dedicated observations were taken in the {\it wobble} mode where the telescope pointing direction is offset from the nominal target position by an angle between 0.5 and 0.7$^\circ$. The observations selected for the data analysis meet the standard run
selection criteria~\cite{Aharonian:2006pe}. Tab.~\ref{tab:table1} summarizes the main observational characteristics of the selected dSphs.
 
After the calibration of raw shower images, the reconstruction of the direction and energy of gamma rays is performed using a template-fitting technique~\cite{2009APh32231D} in which the recorded images are compared to
pre-calculated showers computed from a semi-analytical model. This technique achieves an energy resolution of 10\% and an angular resolution of 0.06$^\circ$ at 68\% containment radius for gamma-ray energies above 200~GeV.
The results presented here have been cross-checked with a different calibration and analysis chain yielding compatible results~\cite{Parsons:2014voa}. 

A  {\it combined} analysis is used to account for the hybrid nature of the observations. Given the configuration of H.E.S.S. Phase II array, a gamma ray can trigger CT5 alone (monoscopic event), or any combination of two of the five telescopes (stereo event). The event reconstruction can be performed in different modes according to the event class. In order to fully benefit from the flexibility of the H.E.S.S. Phase II array, a combined mode exploits both the monoscopic and stereoscopic reconstructions~\cite{Holler:2015tca}. The best reconstruction among the mono (CT5 only), stereo (CT1-5) or H.E.S.S. Phase I-like (CT1-4) reconstruction of each gamma-ray-like event is selected from a $\chi^2$ test. 
Fig.~\ref{fig:significancedistribution} shows for each selected dSph the gamma-ray excess sky map for which the residual background computation is determined using the {\it RingBackground} technique~\cite{Aharonian:2006pe}.  Tab.~\ref{tab:table2} summarizes the available gamma-ray statistics in the source and background regions, the relative size of the background region to the signal region, as well as the significance in the source region. No significant gamma-ray excess is found at the position of the dSph or anywhere in the FoV as shown in Fig.~\ref{fig:significancedistribution}.

\section{Data analysis}
\label{sec:analysis}
\subsection{Region of interest and background measurement}\label{sec:ROIs}
For each selected system, the ON source signal is computed by integrating all the gamma-ray-like events in a disk of angular radius given in the second column of Tab.~\ref{tab:table2}. For Ret~II and Tuc~II, a ON region of $0.2^\circ$ radius is used, divided into two concentric sub-regions of interest (ROIs) of $0.1^\circ$ width each.  
The size of the ON source region is chosen according to the expected DM signal spatial profile versus background as a function of the distance from the object nominal position in order to maximize the sensitivity. In absence of spectroscopic measurements, Tuc~III, Tuc~IV, and Gru~II are considered as point-like sources for H.E.S.S. and therefore only one ROI with a $0.125^\circ$ radius is considered.
\begin{table*}[htbp]
\centering
\begin{tabular}{  l | c | c | c | c |c}
\hline
\hline
Source & ON region size & $N_{\rm ON}$ & $N_{\rm OFF}$ & $\overline{\alpha}$ & Significance\\
name & [degrees] & [counts]  &  [counts]  & & [$\sigma$] \\
\hline
Reticulum II & 0.200 & 949 & 7926 & 8.0 & -0.9 \\
Tucana II & 0.200 & 1170 & 9704 & 8.0 & -1.0  \\
Tucana III* & 0.125 & 689 & 9816 & 15.0 & 0.9 \\
Tucana IV* & 0.125 & 285 & 6550 & 24.1 & 0.6\\
Grus II* & 0.125 & 263 & 4491 & 16.0 & -0.8  \\
\hline
\hline
\end{tabular}
\caption{\label{tab:table2}
Data analysis results for each selected target. The second column gives the size of the ROI. Count numbers measured in the ON and OFF regions are provided in the third and fourth column, respectively. The fifth and sixth column give the ratio of the solid angle size between the OFF and ON regions averaged over all observations, and the measured excess significance between the ON and OFF counts. For the systems marked with the symbol *, the ROI size is chosen as for point-like emission searches. 
}
\end{table*}

\begin{figure*}
\begin{center}
\includegraphics[width=0.4\textwidth]{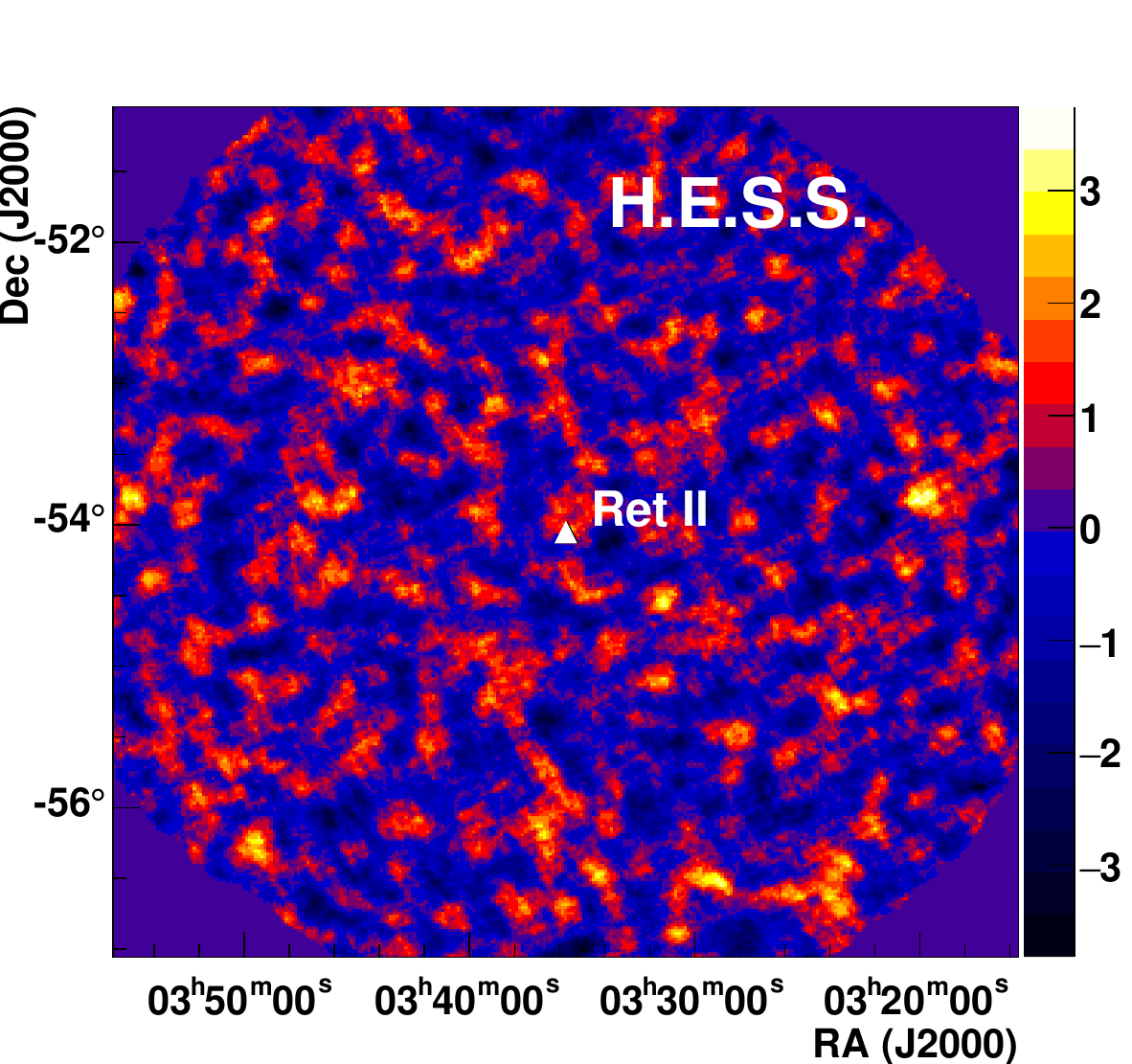}
\includegraphics[width=0.4\textwidth]{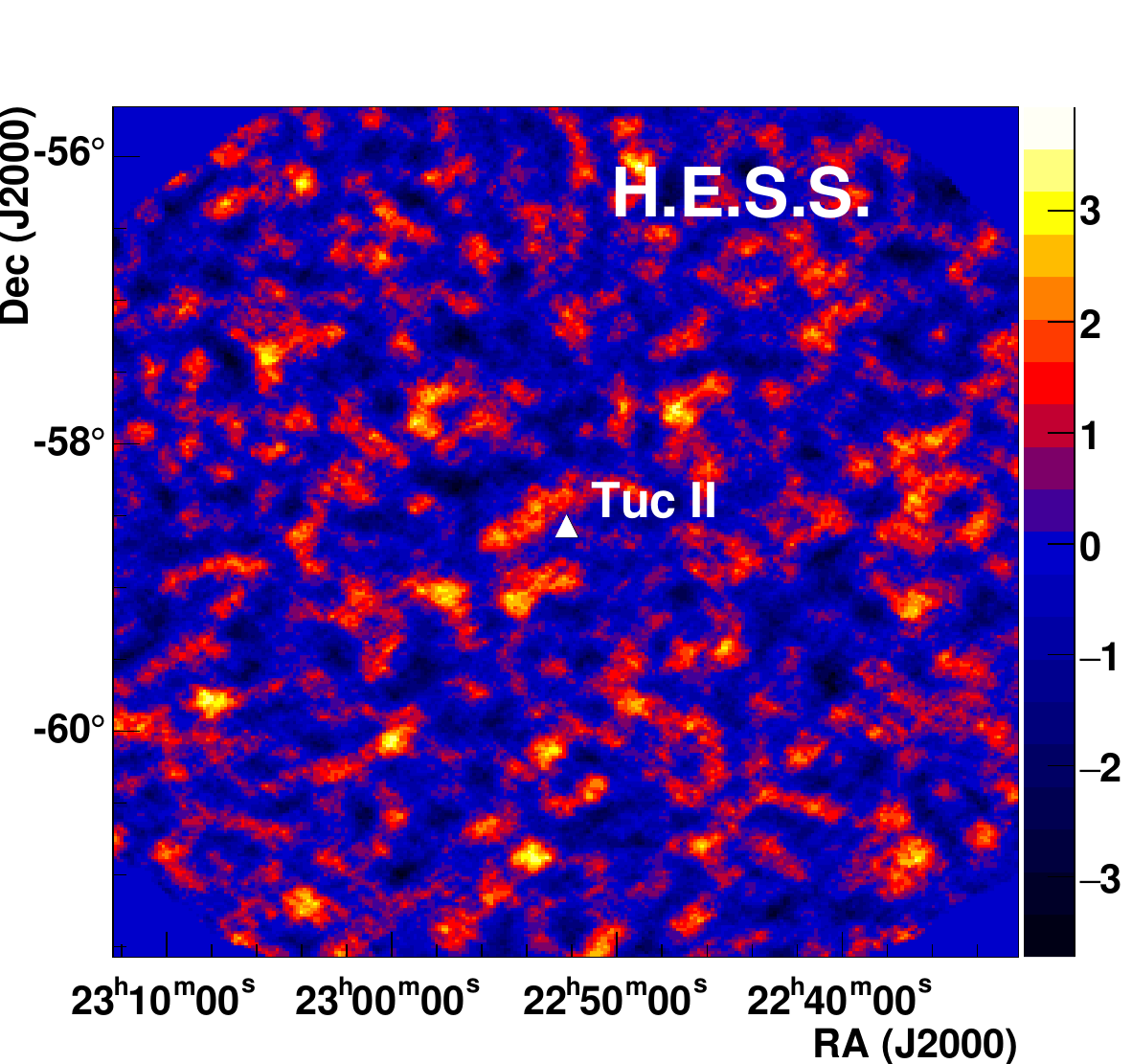}\\
\includegraphics[width=0.4\textwidth]{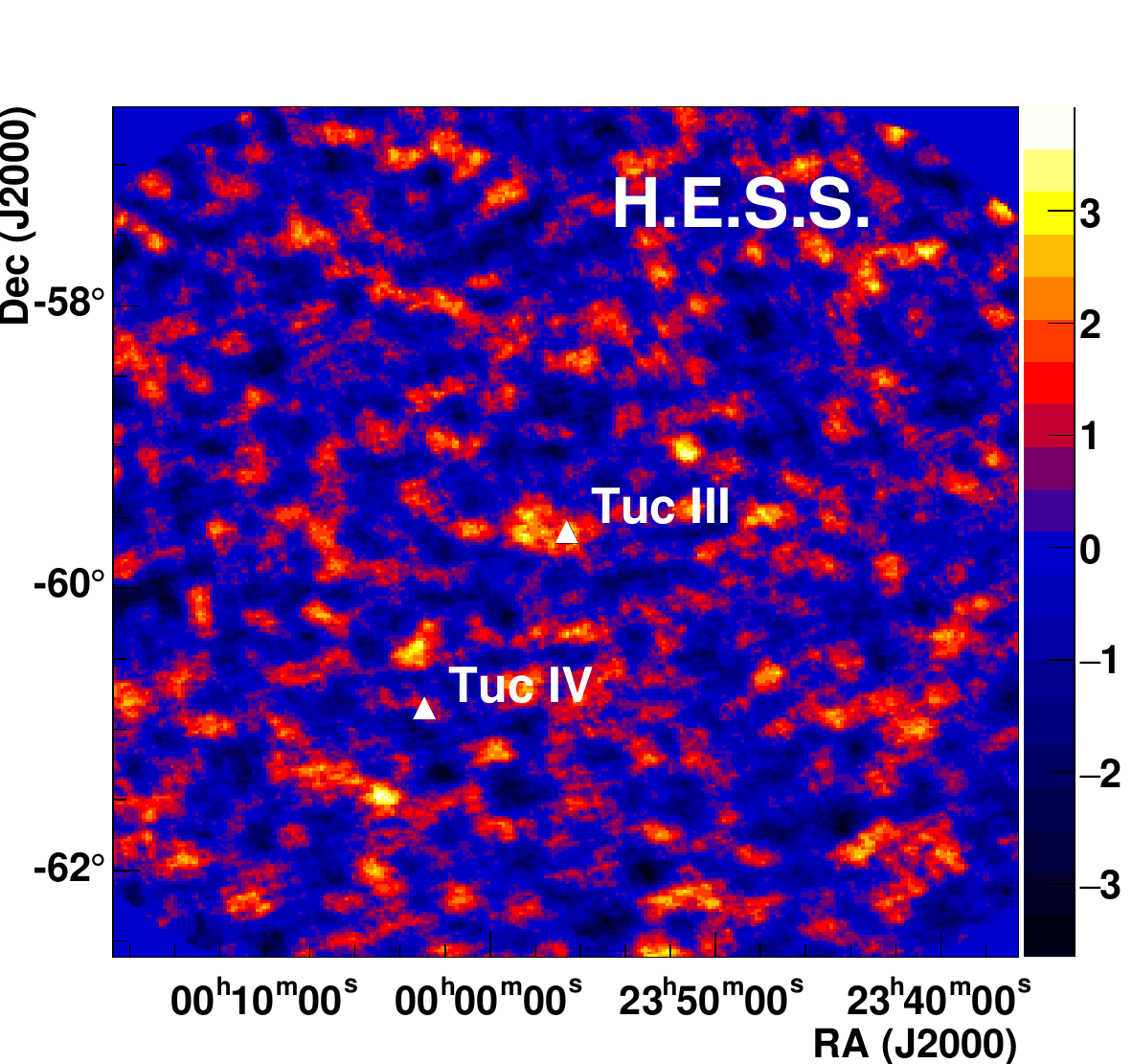}
\includegraphics[width=0.4\textwidth]{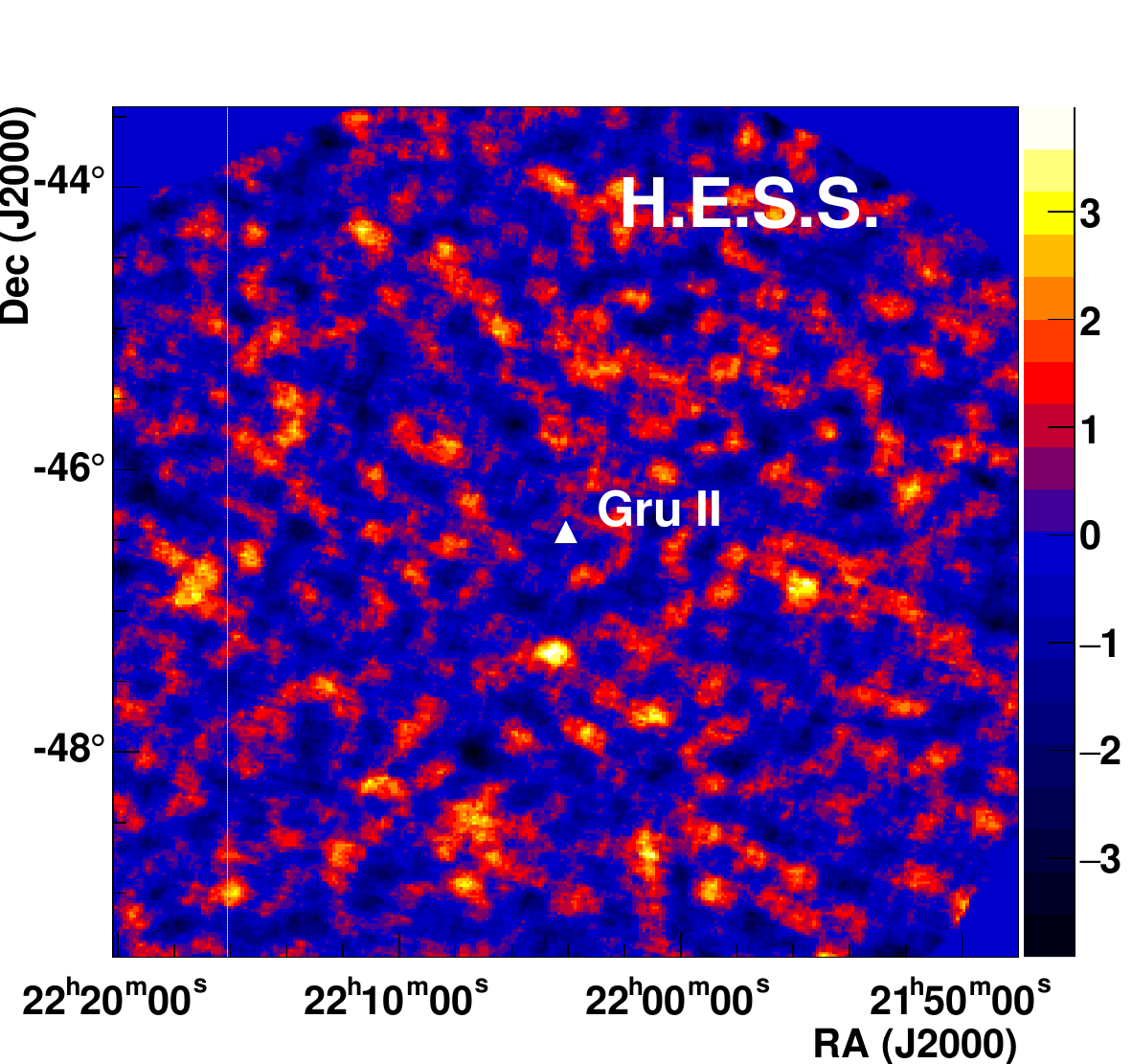}
\caption{Excess significance maps in the FoV of Ret~II, Tuc~II, Tuc~III, and Gru~II, respectively, in Galactic coordinates. Tuc~IV is observed in the FoV of Tuc~III. The nominal position of the systems is marked with a white-filled triangle. The color scale gives the significance of the excess in numbers of standard deviations. No significant excess is observed in any of the FoV.}
\label{fig:significancedistribution}
\end{center}
\end{figure*}

The residual background is measured in the OFF-source regions defined according to the \emph{multiple-OFF} method~\cite{Berge:2006ae}: for each telescope pointing position of the observations, several regions with same solid angle and shape as for the ROIs are defined at the same distance from the pointing position as the ON region. By construction, the centers of the OFF regions lie on a circle of radius equal to the distance between the target position and the pointing position, which leads to identical acceptance in the
ON and OFF regions. This method yields 
smaller systematic errors in the analysis than background determination techniques based on OFF measurements in separated dedicated observations of empty fields of view. 
A disk around the center of the ON region is excluded for the OFF measurements,  with a  radius equal to twice the size of the ON region radius, so that the expected DM signal in the OFF regions is negligible. No additional excluded region is used since no VHE gamma-ray sources are detected in the considered FoV. Given the expected source extension of Tuc~III and Tuc~IV no contamination is expected from one source in the considered signal region of the other. The total number of background events is the sum of all the events that fall in the OFF regions. The parameter $\alpha$ is defined as the ratio between the solid angle size of the OFF and ON regions by $\alpha=\Delta\Omega_{\rm OFF}/\Delta\Omega_{\rm ON}$. The excess sky map is obtained by subtracting the total OFF event count weighted by $1/\alpha$ from the ON event count, and the significance is computed following the statistical approach of Ref.~\cite{1983ApJ...272..317L}. Fig.~\ref{fig:significancedistribution} shows the excess significance sky map for all selected systems. No significant gamma-ray excess is found in the ON source region as well as anywhere else in the sky map. 
Tab.~\ref{tab:table2} provides the size of the ON source region, the number of ON and OFF events, the $\alpha$ values as well as the excess significance for the full ROIs. 

\subsection{Statistical analysis and upper limit computation}
 A two-dimensional (2D)-binned Poisson maximum likelihood analysis is used in order to explore the spatial and spectral characteristics of the expected DM signal with respect to the background. The energy range is divided into 68 logarithmically-spaced energy bins $i$ between 150 GeV and 63 TeV, and the spatial bin $j$ corresponds to the number of ROIs defined for each target. For a given DM mass and annihilation channel, the Poisson likelihood function in the bin $(i,j)$ can be written as
 \begin{widetext}
\begin{equation}
\mathcal{L}_{\rm ij}(N^{\rm S},N^{\rm B}|N_{\rm ON},N_{\rm OFF},\alpha) = \frac{(N_{\rm ij}^{\rm S}+N_{\rm ij}^{\rm B})^{N_{{\rm ON, ij}}}}{N_{{\rm ON, ij}}!}e^{-(N_{\rm ij}^{\rm S}+ N_{\rm ij}^{\rm B})} 
\frac{(N_{\rm ij}^{\rm S'}+\alpha_{\rm j}N_{\rm ij}^{\rm B})^{N_{{\rm OFF, ij}}}}{N_{{\rm OFF, ij}}!}e^{-(N_{\rm ij}^{\rm S'}+\alpha_{\rm j} N_{\rm ij}^{\rm B})} \, .
\label{eq:lik}
\end{equation}
\end{widetext}
$N_{\rm ON, ij}$ and $N_{\rm OFF, ij}$ stand for the number of measured events in the ON and OFF regions, respectively. $N^{\rm B}_{\rm ij}$ is the expected number of background events in the ON region. $\alpha_{\rm j}$ 
is defined as the ratio of the solid angle of the OFF and ON regions for the bin $j$.
$N^{\rm S}_{\rm ij}$ and $N^{\rm S'}_{\rm ij}$ correspond to the number of DM signal events expected in the ON 
and OFF regions, respectively. They are computed by folding the theoretical DM flux 
with the energy-dependent acceptance and energy resolution of H.E.S.S. for the considered data set. The continuum signal spectra are extracted from Ref.~\cite{Cirelli:2010xx}, while the mono-energetic gamma-line signal is a Dirac delta function. The energy resolution of H.E.S.S. is represented by a Gaussian function with a width of $\sigma_{\rm E}/E$ = 10\%. The dependence
of the energy resolution on the observational parameters (mean zenith angle, optical efficiency, off-axis angle) and the analysis selection cuts
have a negligible impact on the results. The same likelihood analysis technique is applied to look both for the continuum and gamma-line signals.

As discussed in Sec.~\ref{sec:ROIs}, $N^{\rm S'}_{\rm ij}$ can be safely taken to $N^{\rm S'}_{\rm ij} \equiv$ 0. The total likelihood
$\mathcal{L}$ is defined as the product of the $\mathcal{L}_{\rm ij}$ over the $ij$ bins.
Since no significant excess between the ON and OFF regions is found in any of the considered systems, upper limits can be computed for any DM mass from a likelihood ratio test statistic (TS)~\cite{2011EPJC711554C} given by: 
\boldmath
\scriptsize
\begin{equation}
TS = 
\begin{cases}
- 2\, \rm \ln \frac{\mathcal{L}(N^{\rm S}(\langle \sigma v \rangle), \widehat{\widehat {N^{\rm B}}}(\langle \sigma v \rangle)
)}{\mathcal{L}(0,\widehat{\widehat{N^{\rm B}(0)}}
)} \,  & N^{\rm S}(\widehat{\langle \sigma v \rangle}) < 0\\
- 2\, \rm \ln \frac{\mathcal{L}(N^{\rm S}(\langle \sigma v \rangle), \widehat{\widehat {N^{\rm B}}}(\langle \sigma v \rangle)
)}{\mathcal{L}(N^{\rm S}(\widehat{\langle \sigma v \rangle}),\widehat{N^{\rm B}}
)} \,  & 0\leq N^{\rm S}(\widehat{\langle \sigma v \rangle}) \leq N^{\rm S}(\langle \sigma v \rangle)\\
0 & N^{\rm S}(\widehat{\langle \sigma v \rangle}) > N^{\rm S}(\langle \sigma v \rangle)
\end{cases}
\label{eq:TS}
\end{equation}
\normalsize
\unboldmath
$\widehat{\widehat {N^{\rm B}_{\rm ij}}}$ is obtained through a conditional maximization by solving $d\mathcal{L}/dN^{\rm B}_{\rm ij} = 0$, 
while $N^{\rm S}_{\rm ij}(\widehat{\langle \sigma v \rangle})$ and $\widehat{N^{\rm B}_{\rm ij}}$
are computed using an unconditional numerical maximization. 
The procedure described in Ref.~\cite{2011EPJC711554C} has been followed in order to compute upper limits 
for positive signals. 
The value of the velocity-weighted annihilation cross section $\langle \sigma v \rangle$ for which $\Delta$TS = 2.71 from the minimum provides one-sided 95\% confidence level (C.L.) upper limits on $\langle \sigma v \rangle$.

Due to the finite number of stellar tracers of the DM-induced gravitational potential in a given DES system $k$, 
the $J$-factor can be treated as a statistical variable. Its uncertainty is modeled as a nuisance parameter which follows a log-normal distribution
\begin{equation}
\begin{aligned}
\mathcal{L}^{\rm J}_{\rm k}(J|\bar{J},\sigma_{ J})=\frac{1}{\sqrt{2\pi}\sigma_{J}\log(10)\times J}\, \\
\times \exp\Big(-\frac{(\log_{10}J-\overline{\log_{10}J})^2}{2\sigma_{J}^2}\Big) \, ,
\label{eq:jfactorlik}
\end{aligned}
\end{equation}
with mean $\bar{J}$ and width $\sigma_{\rm J}$ taken from literature (see Tab.~\ref{tab:table3}). The $J$-factor, $\hat{J}$, that maximizes Eq.~(\ref{eq:jfactorlik}) is derived and then included in the likelihood test as $N_{\rm S} \rightarrow N_{\rm S}\hat{J}/\bar{J}$. 

\section{Results}
\label{sec:results}
\subsection{Upper limits on individual systems}
\begin{figure*}[htbp]
\begin{center}
\includegraphics[width=0.45\textwidth]{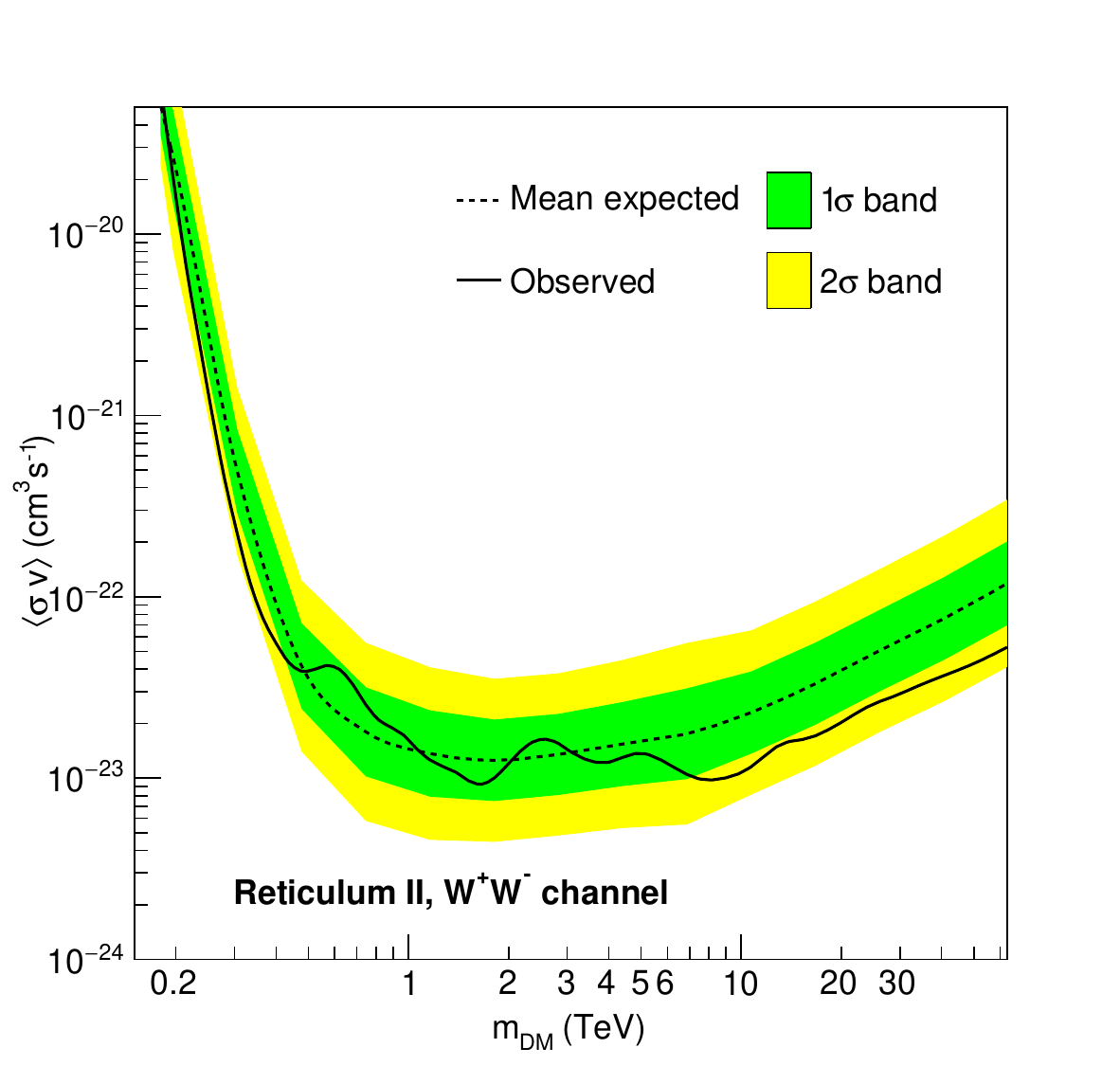}
\includegraphics[width=0.45\textwidth]{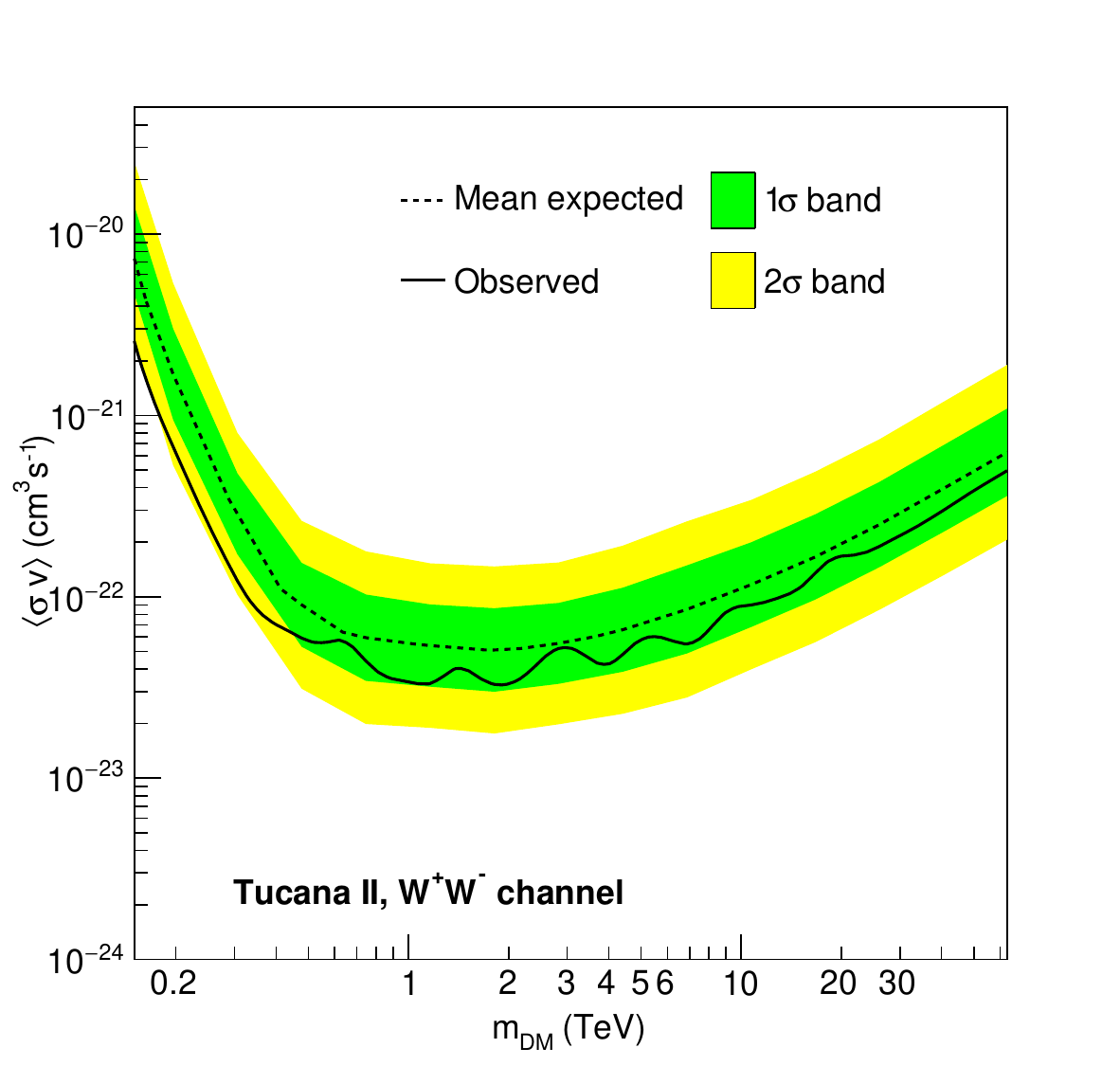}
\includegraphics[width=0.45\textwidth]{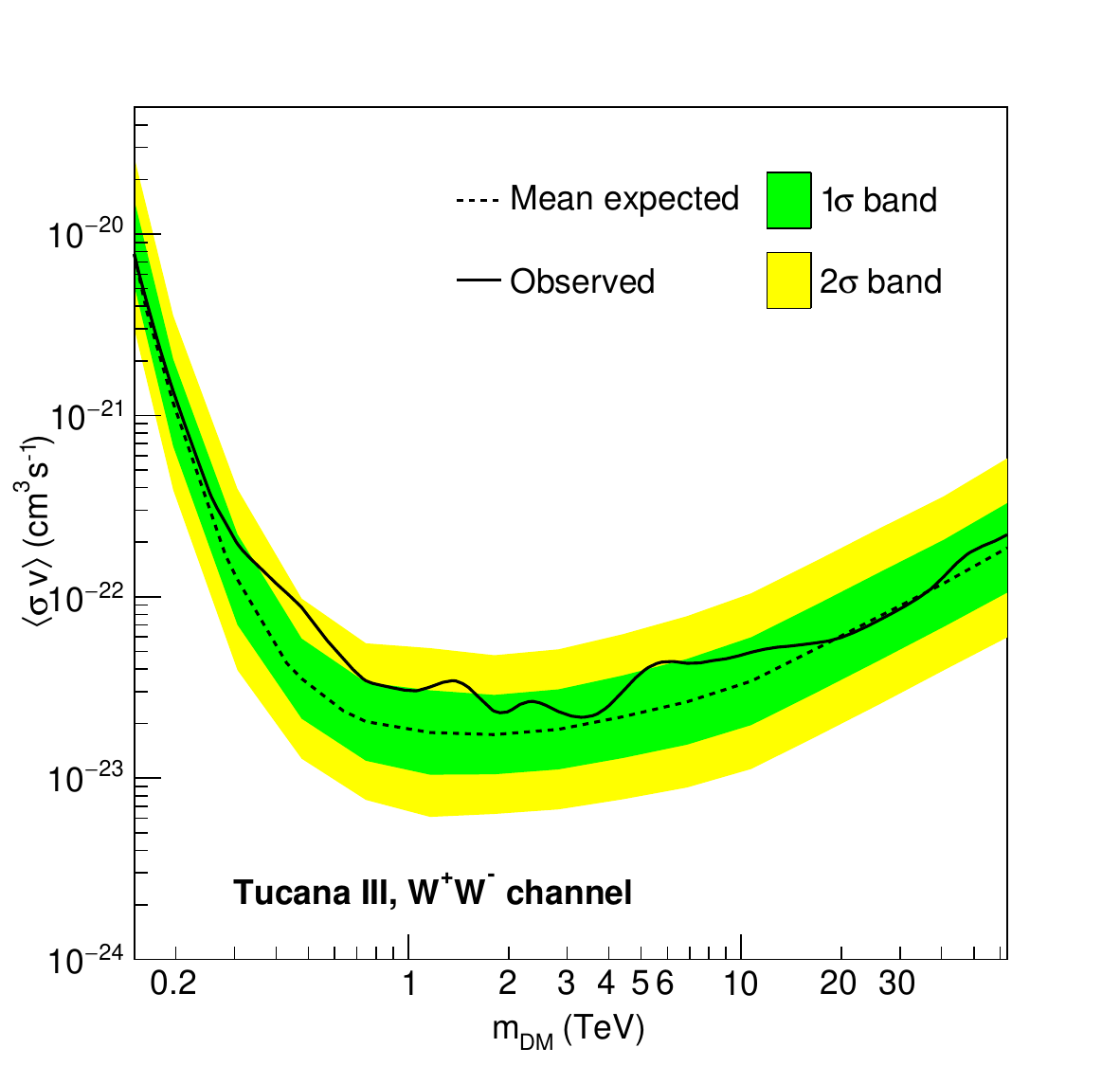}
\includegraphics[width=0.45\textwidth]{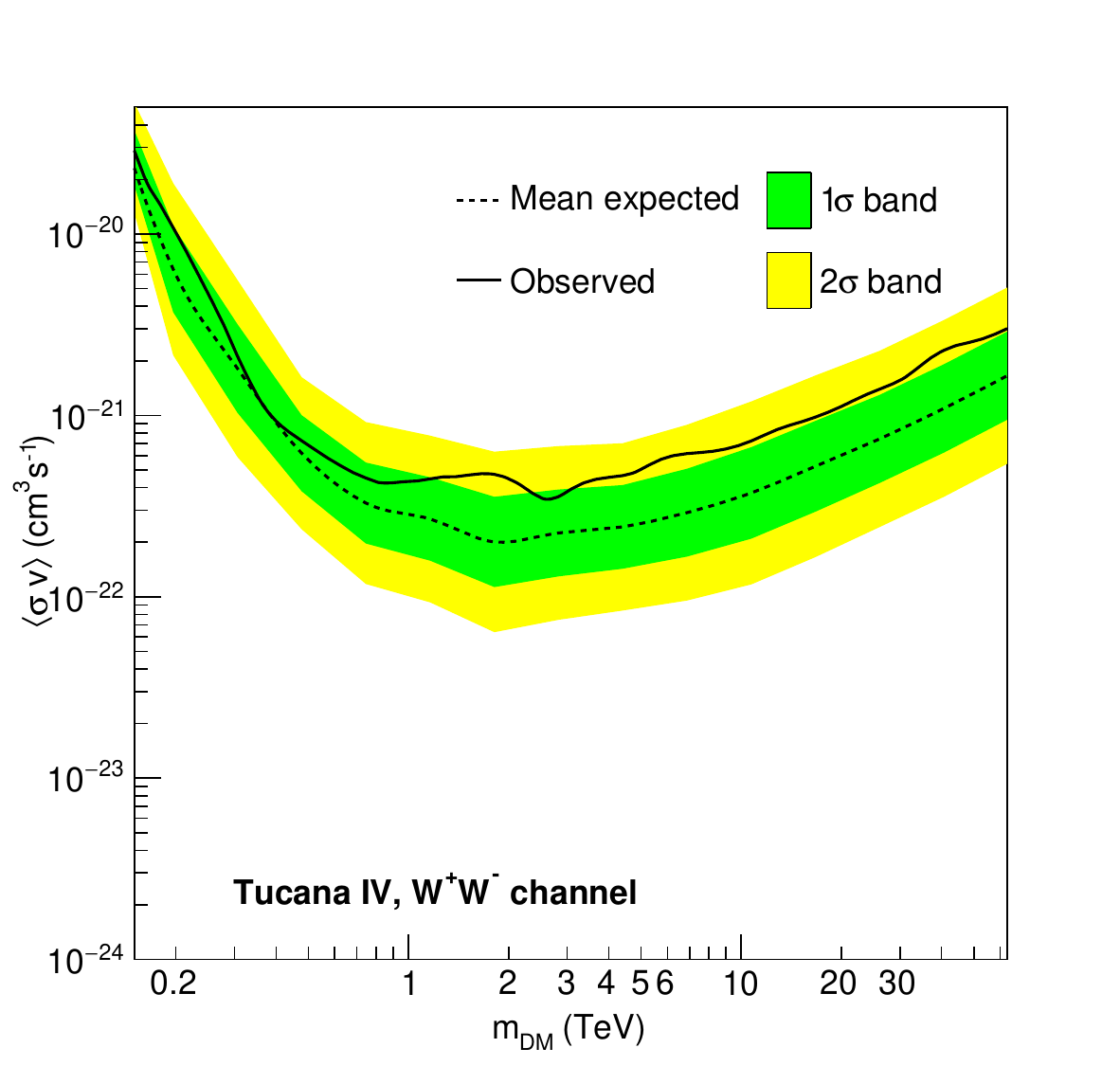}
\includegraphics[width=0.45\textwidth]{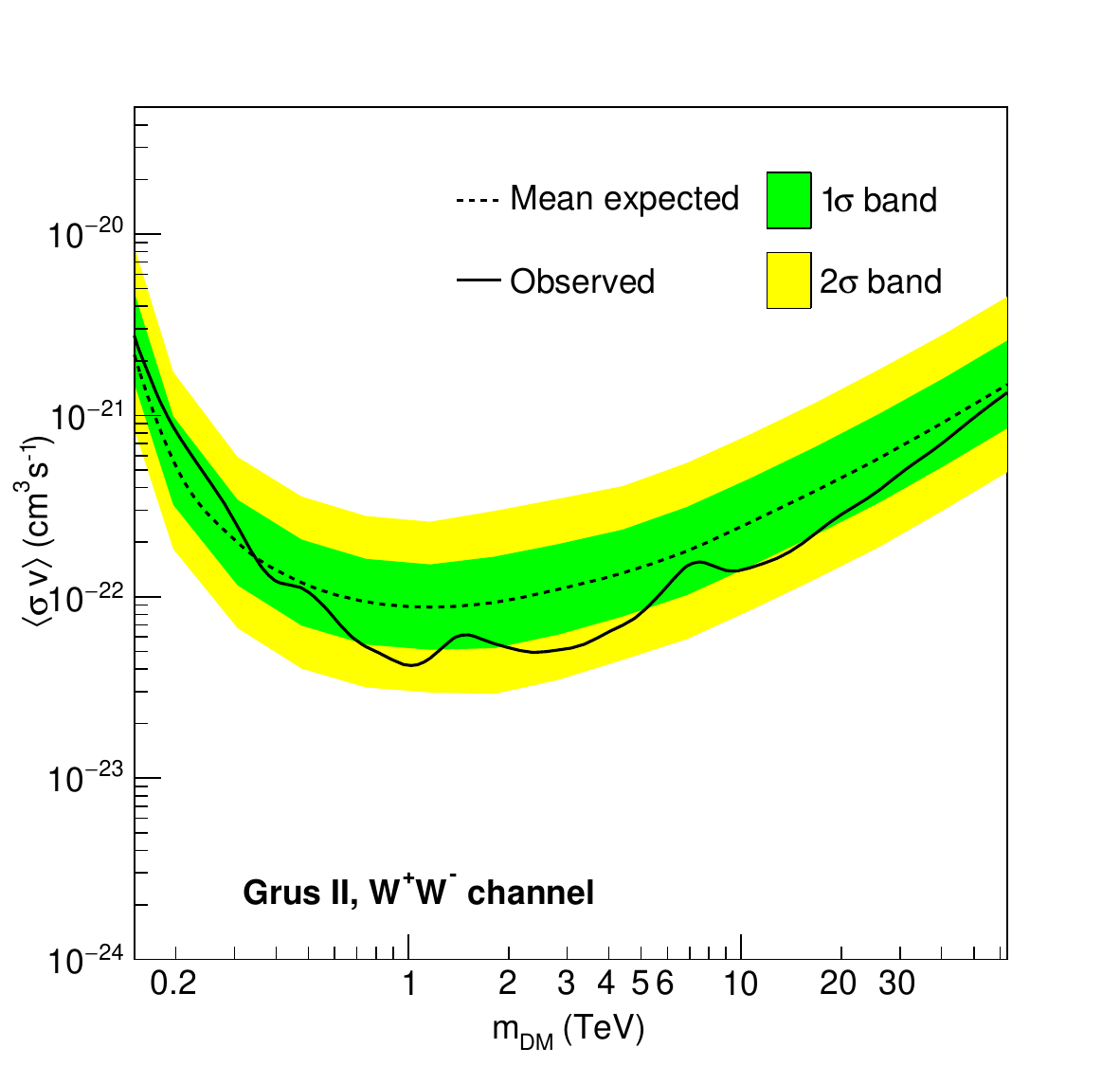}
\caption{95\% C.L. upper limits on the annihilation cross section $\langle\sigma v\rangle$ for Ret~II, Tuc~II, Tuc~III, Tuc~IV, and Gru~II, in the $W^+W^-$ annihilation channel without the uncertainty on the $J$-factor. Observed limits (solid lines) together with mean expected limits (dashed line) and the 1$\sigma$ (green area) and 2$\sigma$ (yellow area) containment bands are shown.}
\label{fig:sigmavi}
\end{center}
\end{figure*}

Since no significant excess is found in the selected DES dSphs in any ROI,  upper limits at 95\% C.L.
on $\langle\sigma v\rangle$ versus the DM mass are derived for each target following Eq.~(\ref{eq:TS}). 
The upper limits as a function of the DM mass are shown in Fig.~\ref{fig:sigmavi} for the $W^+W^-$ annihilation channel. The observed limits  are plotted together with the mean expected limits  and the $1\sigma$ and $2\sigma$  containment bands. Mean expected limits and statistical uncertainty bands are obtained from 100 Poisson realizations of the  background in the ON and OFF regions, respectively. The mean expected limits are given by the mean of the distribution of $log_{\rm10}(\langle\sigma v\rangle)$ obtained in the realizations and the containment bands by its standard deviation.

The best observed limits are obtained for Ret~II. They reach $\langle\sigma v\rangle \simeq 1 \times 10^{-23}$ cm$^3$s$^{-1}$ for a 1.5~TeV DM mass in the $W^+W^-$ annihilation channel. 
In the case of Tuc~III, they reach $\langle\sigma v\rangle \simeq 3 \times 10^{-23}$ cm$^3$s$^{-1}$ for a 1.5~TeV DM mass. Assuming the lower value of the $J$-factor for Tuc~III, the limits degrade by a factor of about 40.
The limits on Tuc~IV and Gru~II are less constraining due to smaller $J$-factors and datasets.

\begin{figure*}[htbp]
\begin{center}
\includegraphics[width=0.45\textwidth]{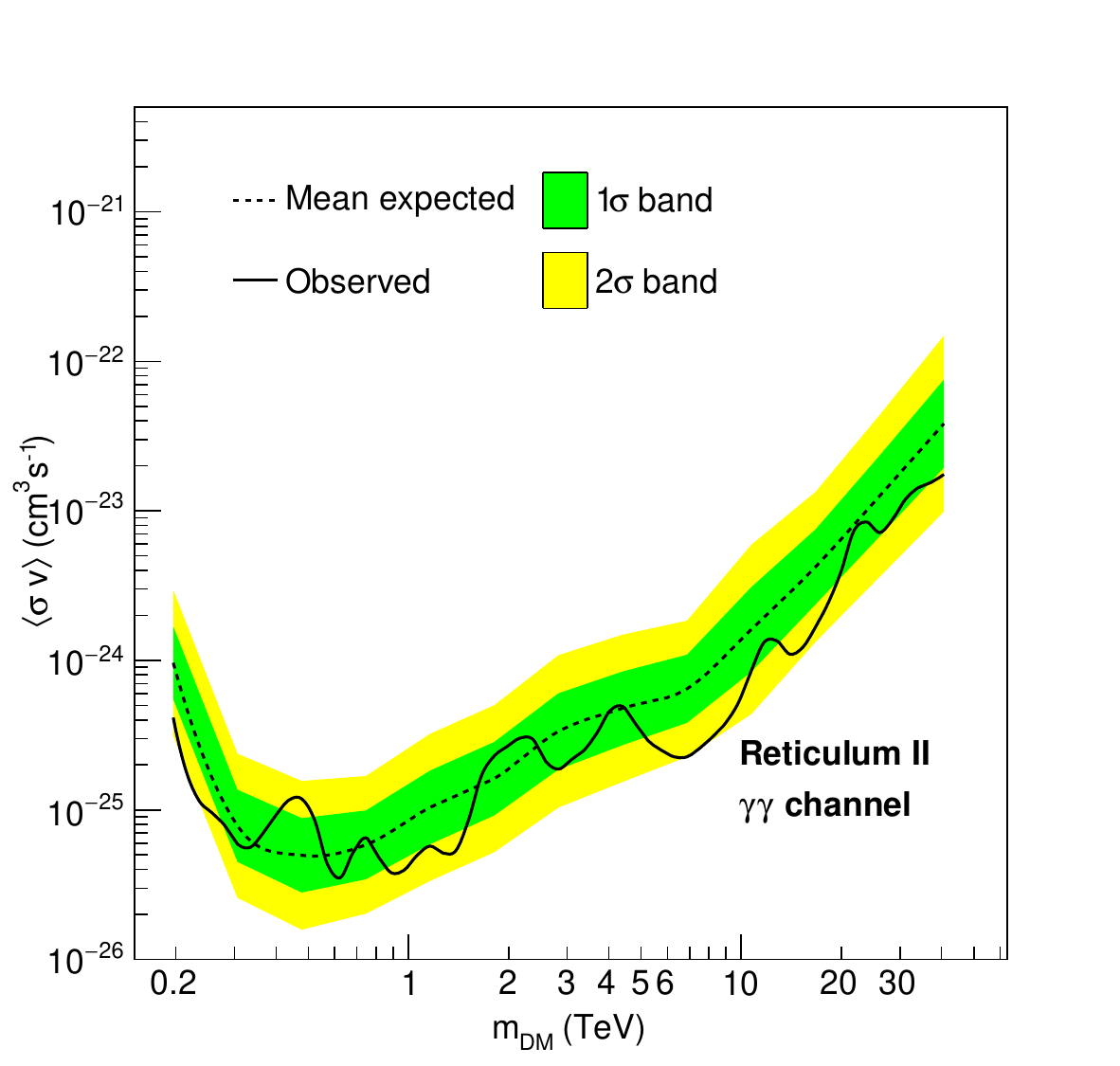}
\includegraphics[width=0.45\textwidth]{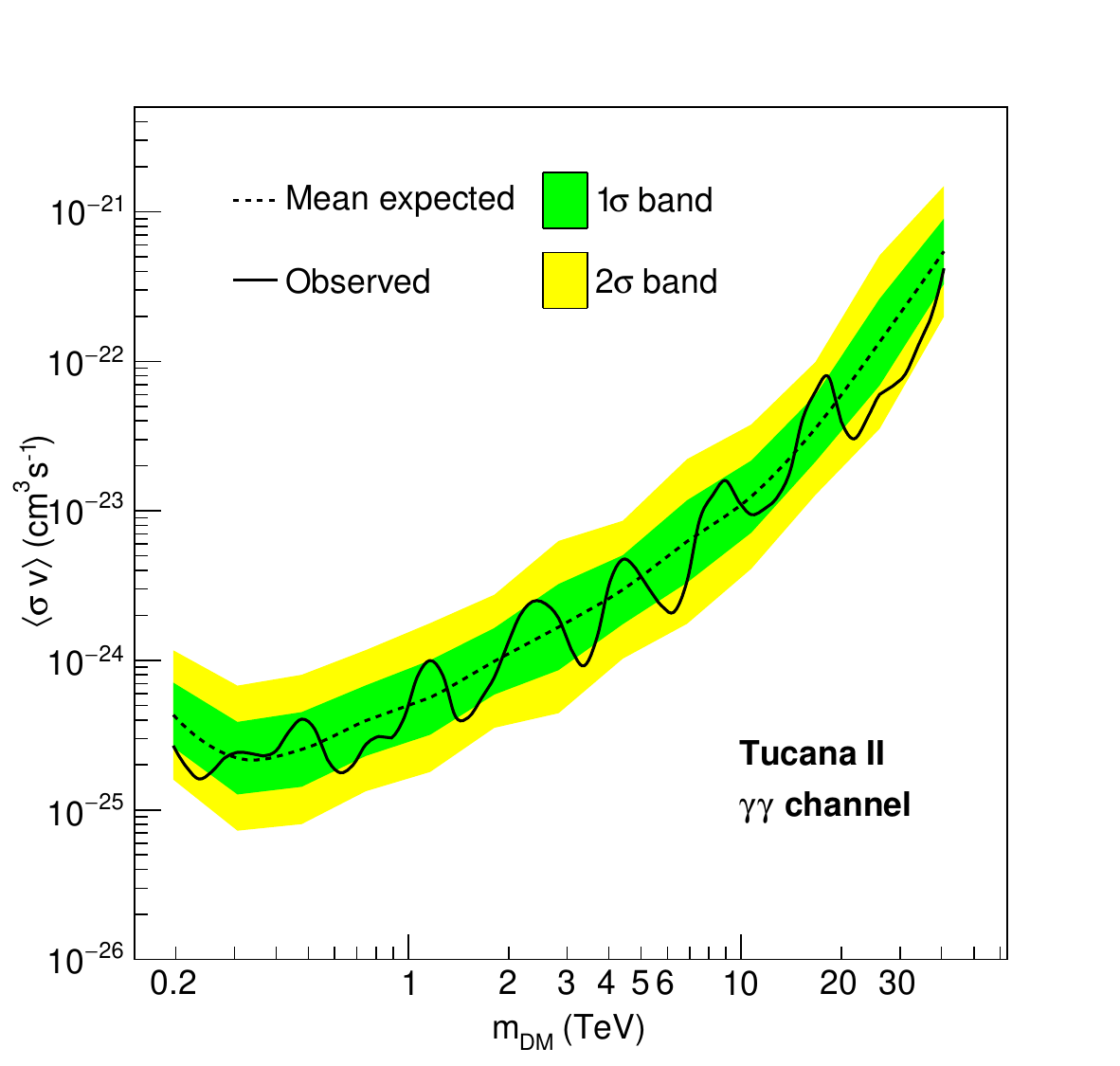}
\includegraphics[width=0.45\textwidth]{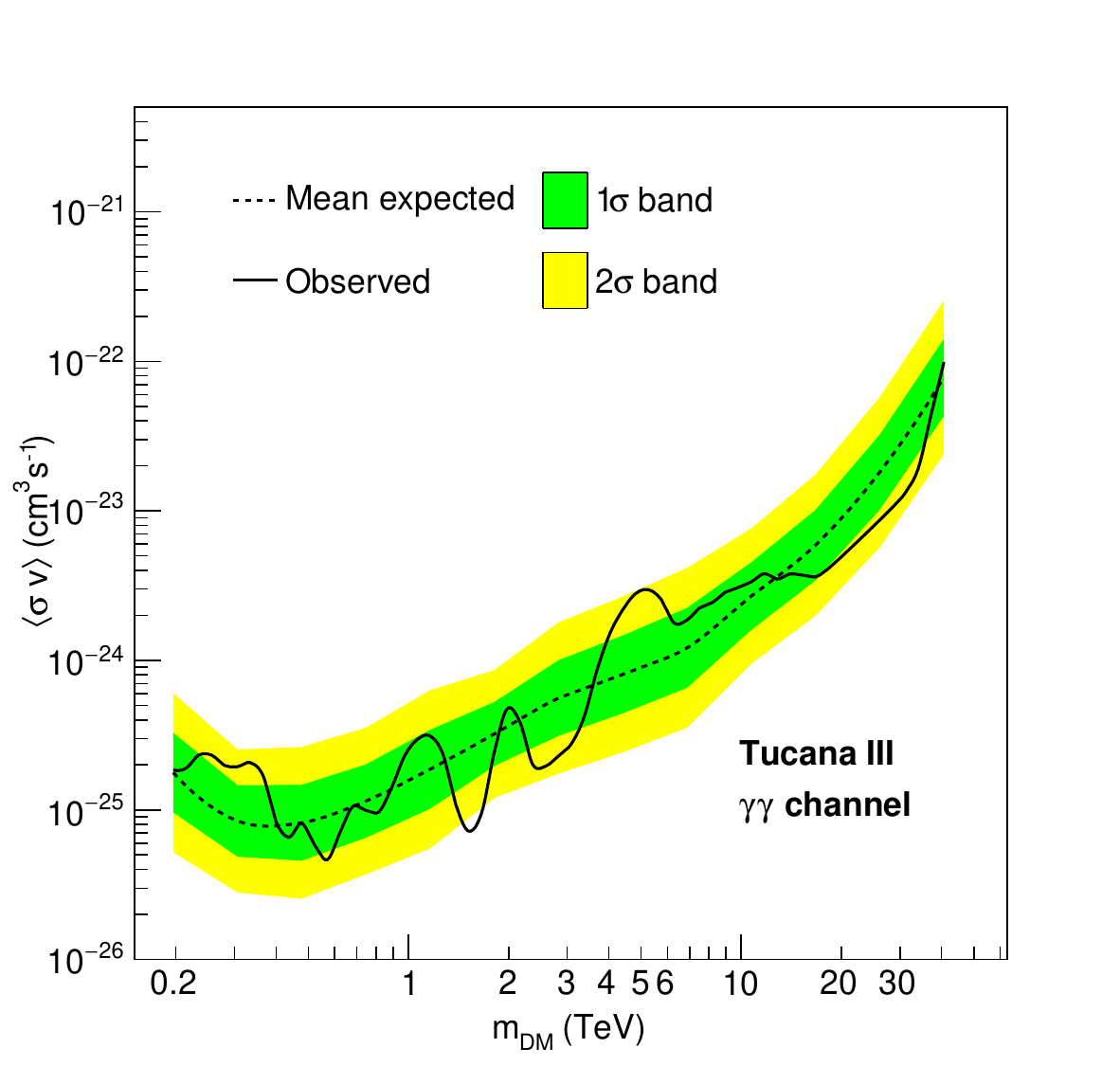}
\includegraphics[width=0.45\textwidth]{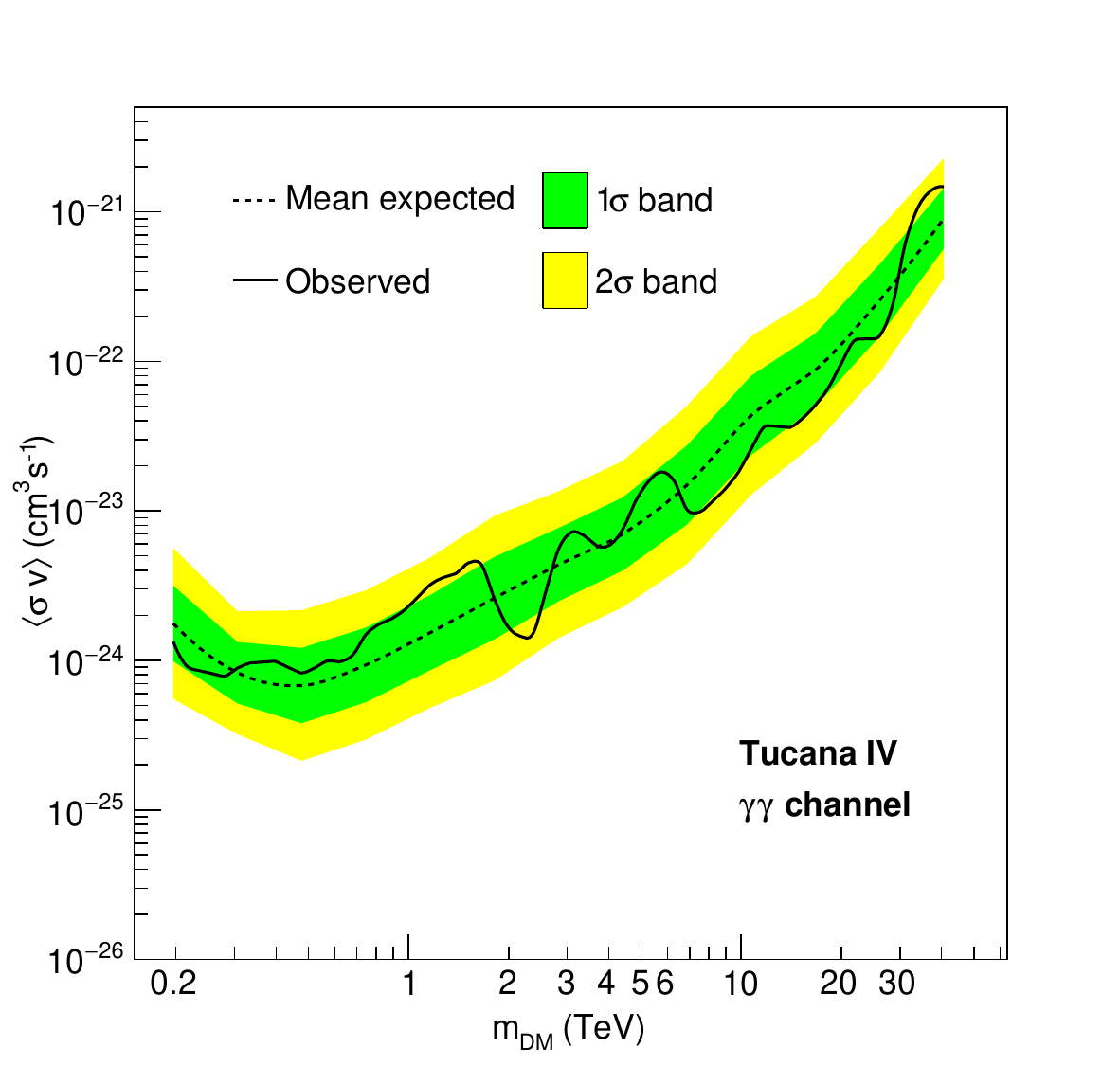}
\includegraphics[width=0.45\textwidth]{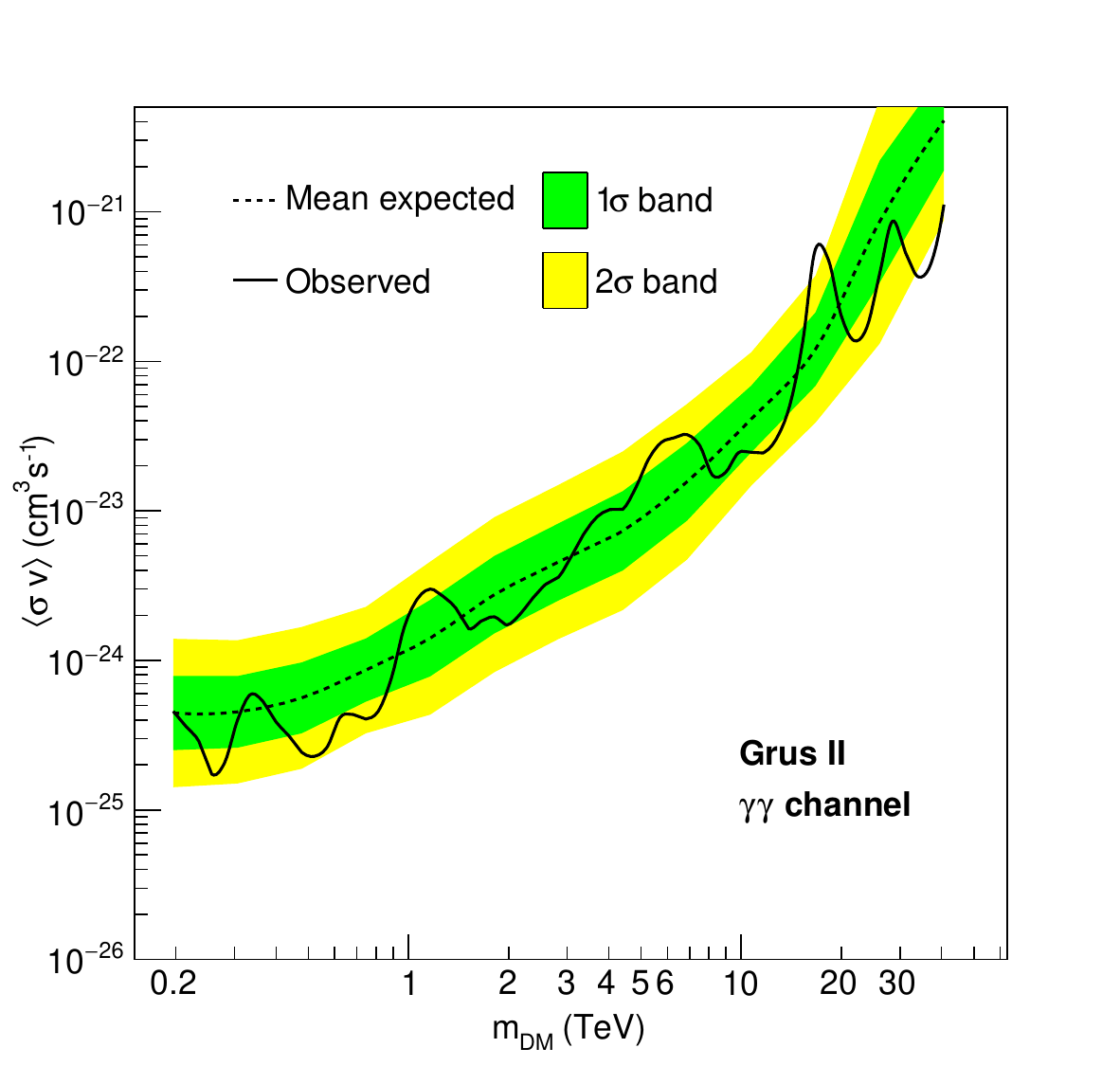}
\caption{95\% C.L. upper limits on the annihilation cross section $\langle\sigma v\rangle$ versus the DM mass $m_{\rm DM}$ for Ret~II, Tuc~II, Tuc~III, Tuc~IV, Gru~II, in the $\gamma\gamma$ annihilation channel, without the uncertainty on the $J$-factor. Observed limits (solid lines) together with mean expected limits (dashed lines), and the 1$\sigma$ (green area) and 2$\sigma$ (yellow area) containment bands are shown. }
\label{fig:sigmavAll_gamma}
\end{center}
\end{figure*}

In addition, a search for monoenergetic gamma-ray lines has been performed on the five targets. The 95\% C.L. observed and mean expected limits together with the containment bands are shown in Fig.~\ref{fig:sigmavAll_gamma} for the five targets. For Ret~II, the observed limit reaches $\langle\sigma v\rangle \simeq 8 \times 10^{-26}$ cm$^3$s$^{-1}$ for a 1.5~TeV DM mass. 

\begin{figure*}[htbp]
\begin{center}
\includegraphics[width=0.45\textwidth]{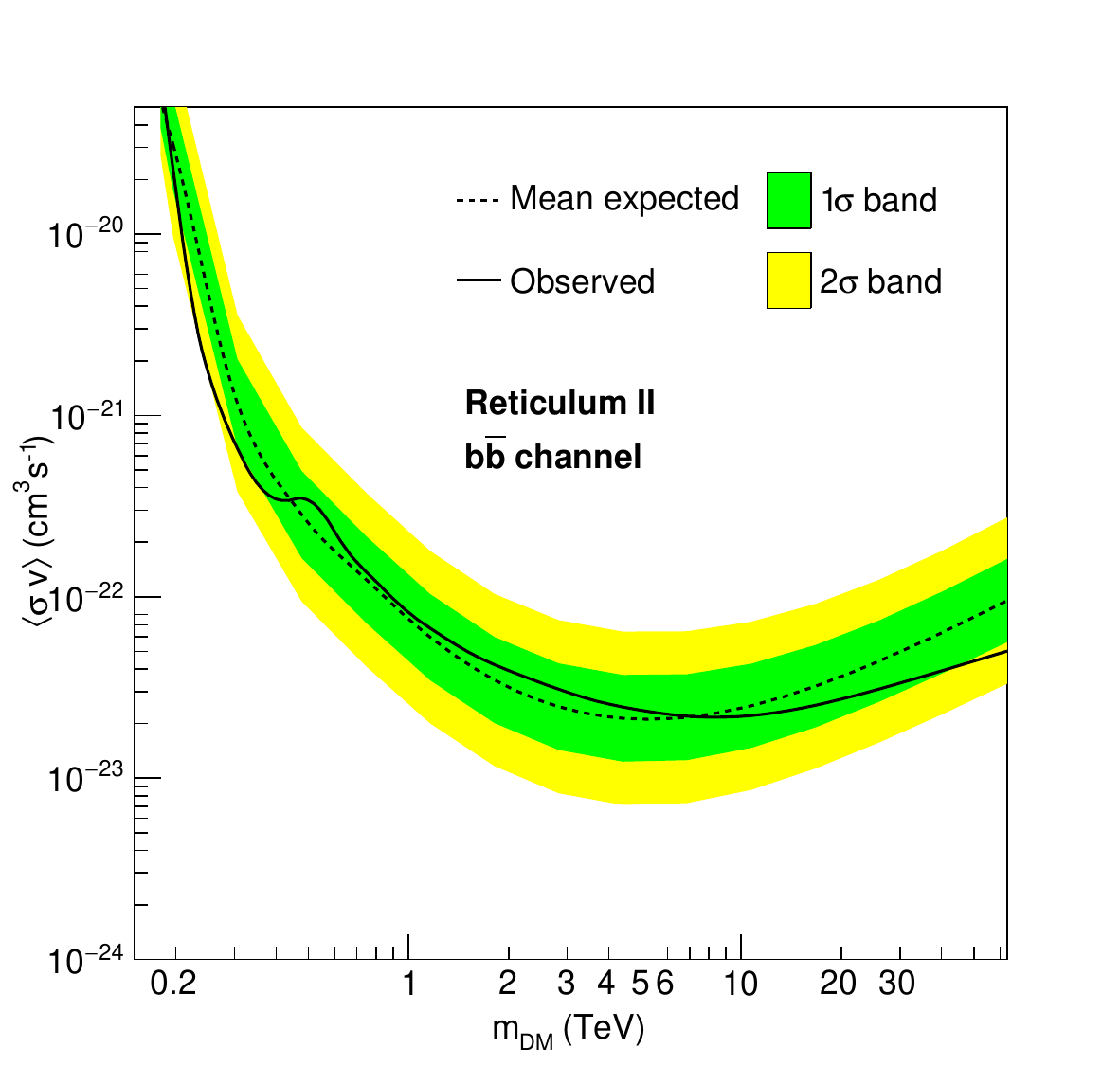}
\includegraphics[width=0.45\textwidth]{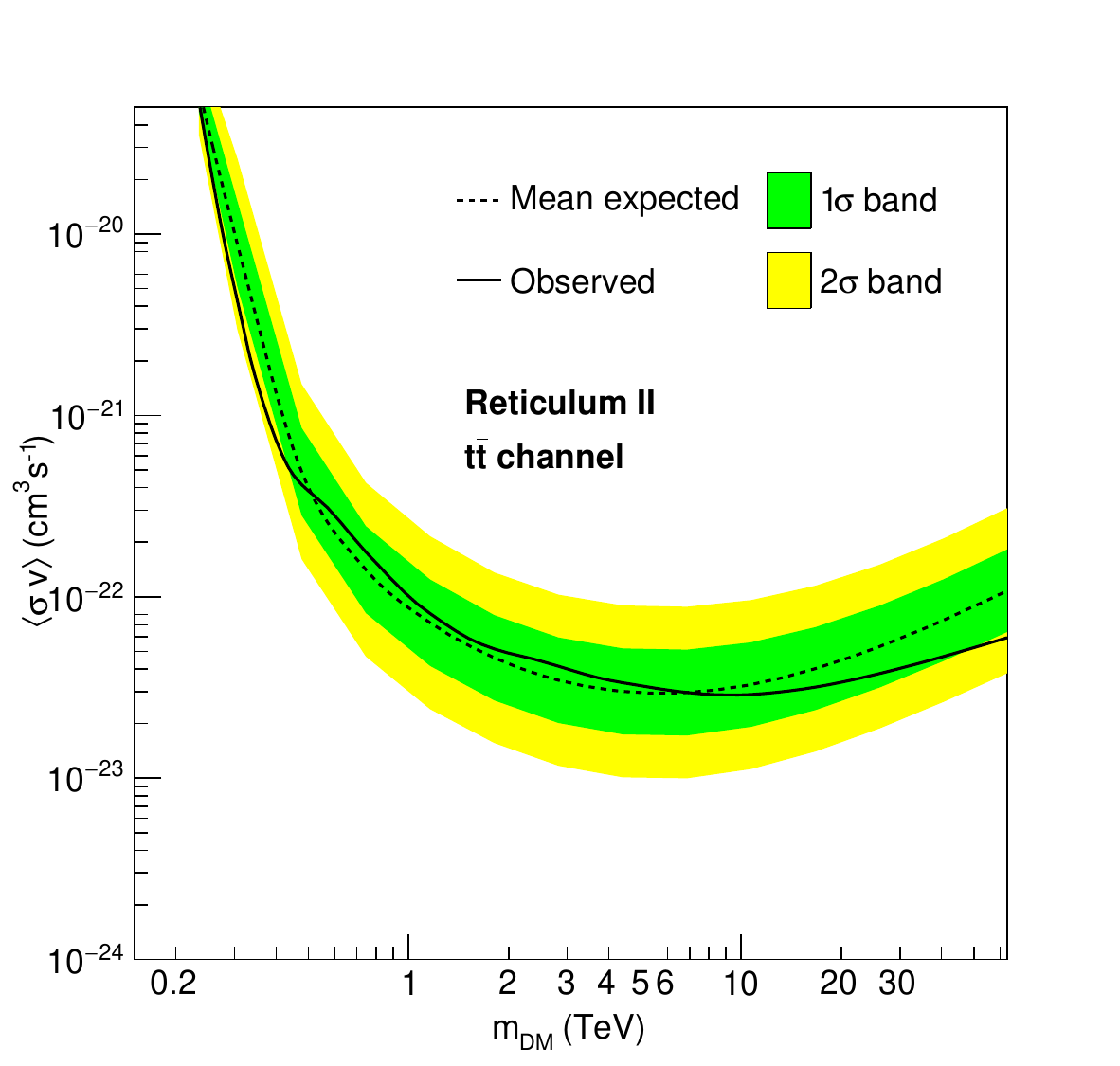}
\includegraphics[width=0.45\textwidth]{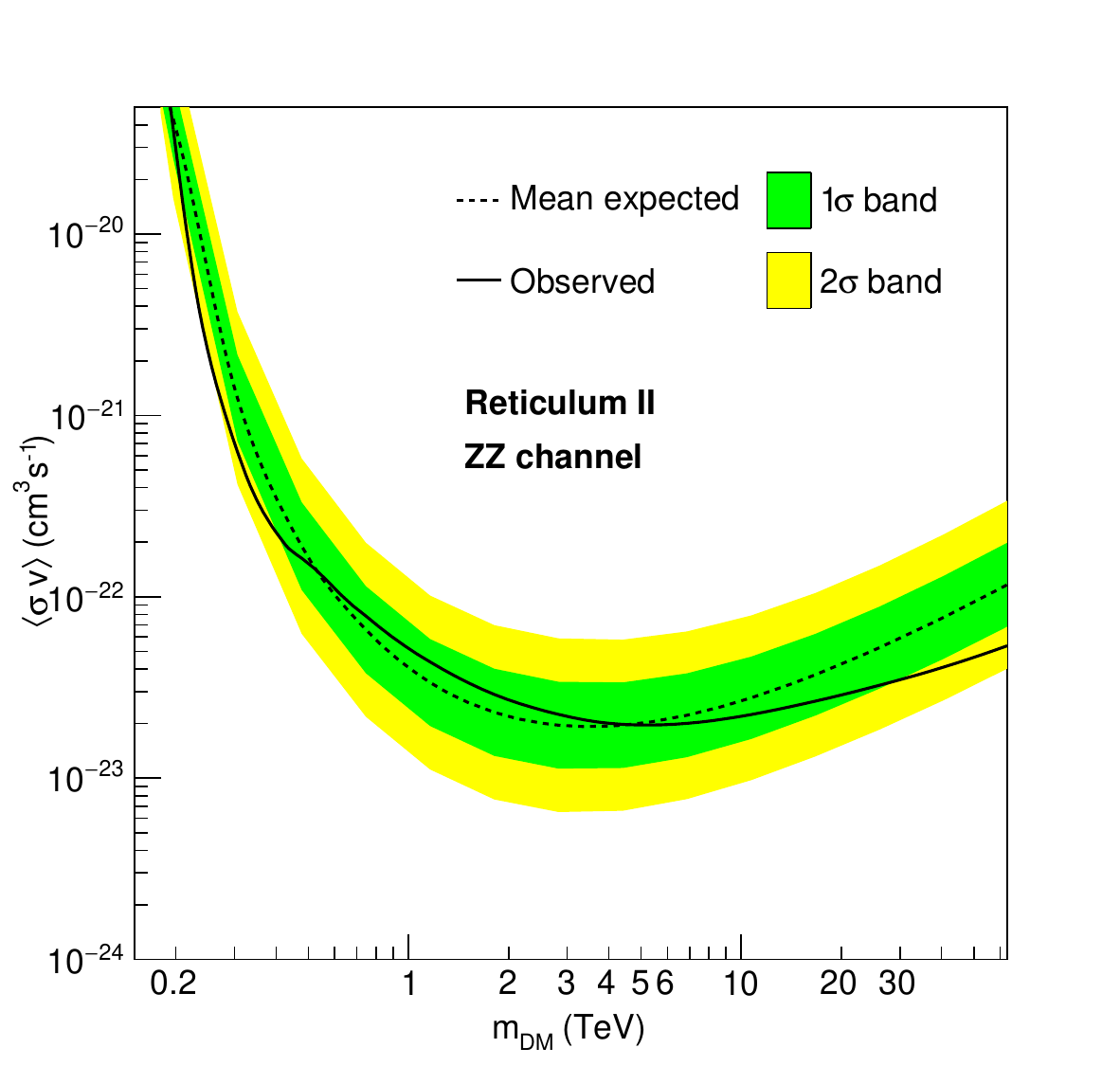}
\includegraphics[width=0.45\textwidth]{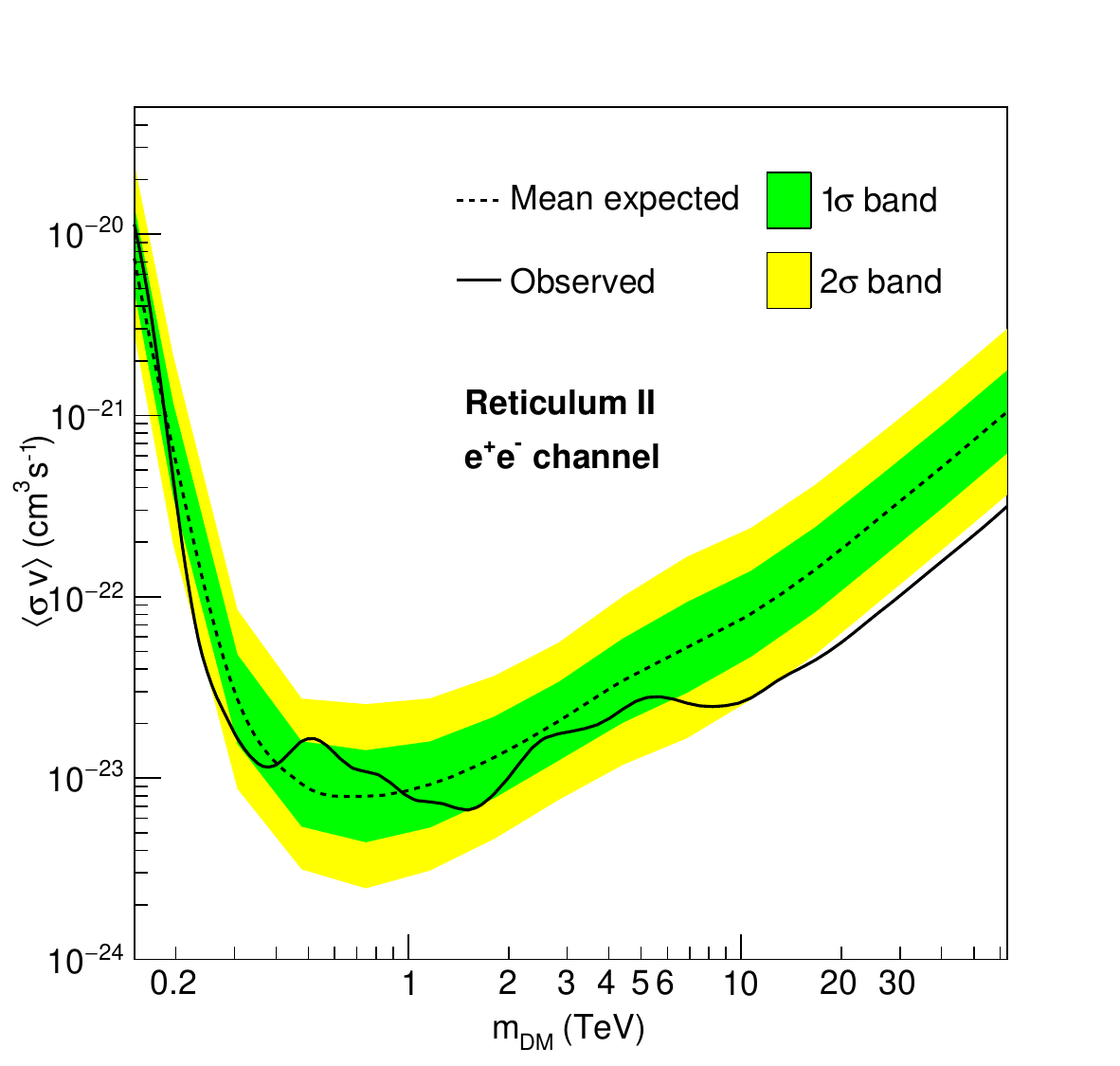}
\includegraphics[width=0.45\textwidth]{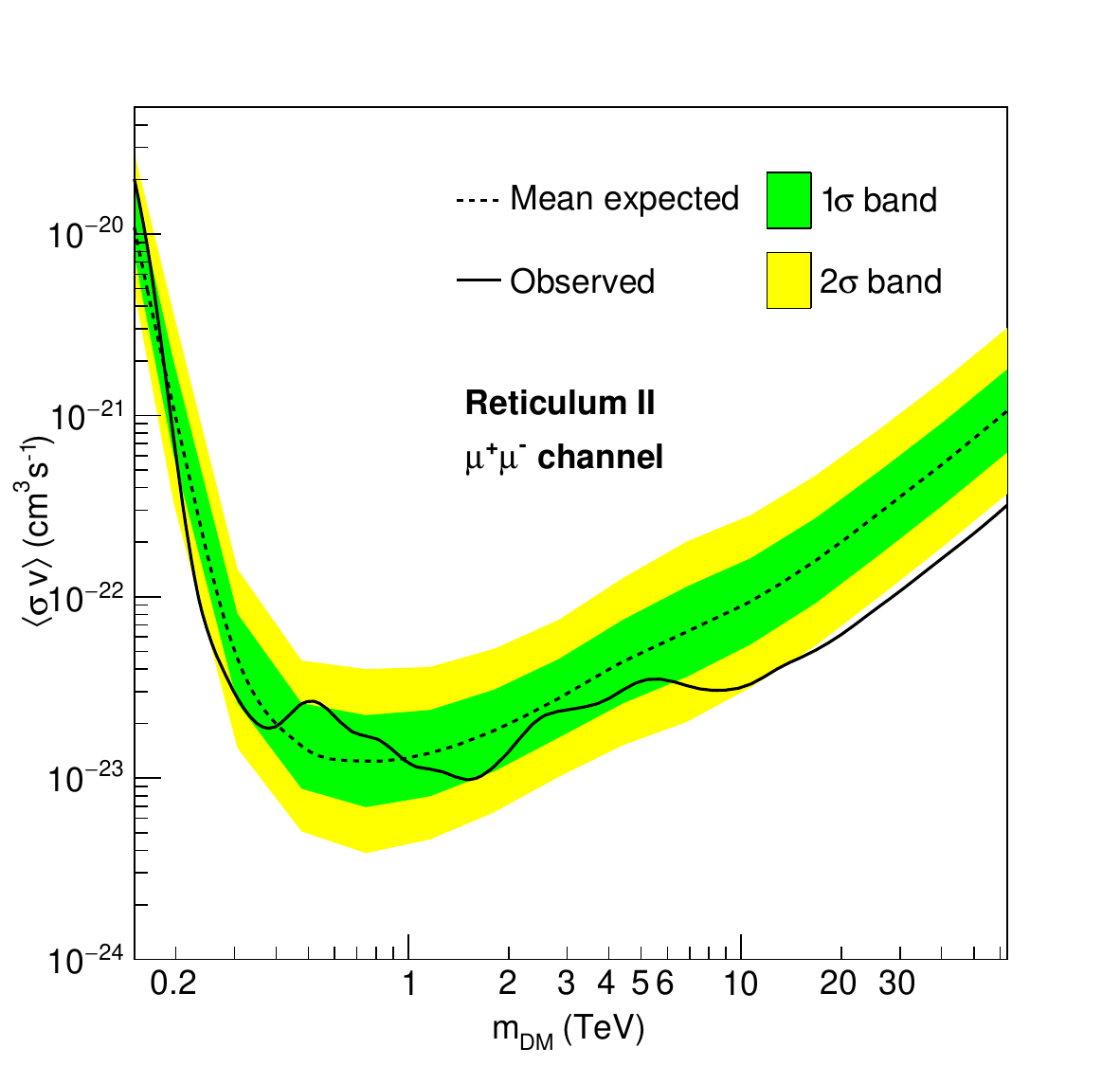}
\includegraphics[width=0.45\textwidth]{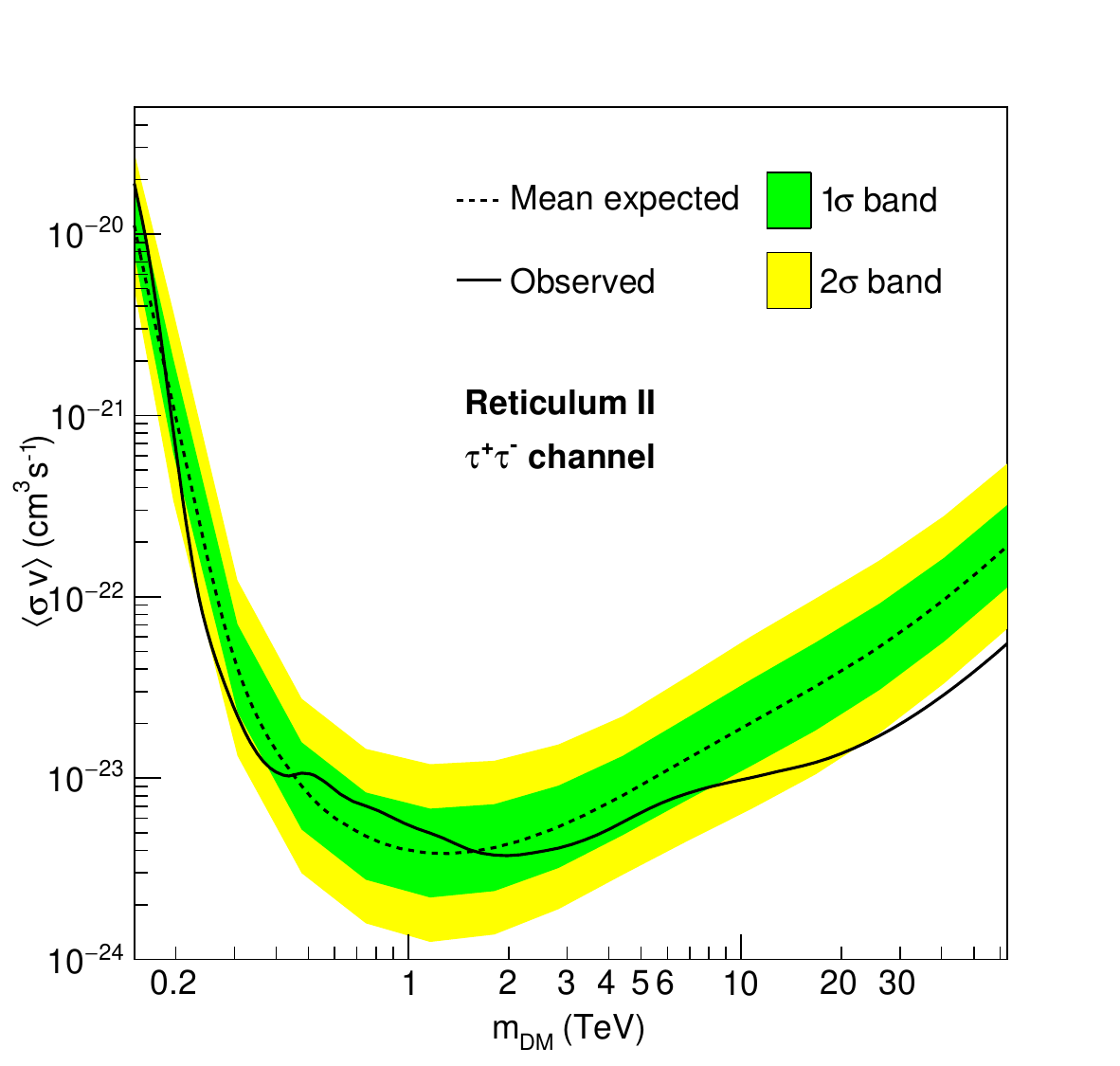}
\caption{95\% C.L. upper limits on the annihilation cross section $\langle\sigma v\rangle$ for Ret~II in the $b\bar{b}$, $t\bar{t}$, $ZZ$, $e^+e^-$, $\mu^+\mu^-$, $\tau^+\tau^-$ annihilation channels, respectively, without the uncertainty on the $J$-factor. Observed limits (solid lines) together with mean expected limits (dashed lines) and the 1$\sigma$ (green area) and 2$\sigma$ (yellow area) containment bands are shown. 
The limits for the other targets can be found in Fig.~\ref{fig:sigmavTucII_all} to Fig.~\ref{fig:sigmavGruII_all}
in the appendix.}
\label{fig:sigmavRetII_all}
\end{center}
\end{figure*}

The search for a DM signal has also been performed in the annihilation channels $ZZ$, $b\bar{b}$, $t\bar{t}$, $e^+e^-$, $\mu^+\mu^-$, and $\tau^+\tau^-$. The 95\% C.L. upper limits on $\langle\sigma v\rangle$ as a function of $m_{\rm DM}$ are shown in Fig.~\ref{fig:sigmavRetII_all} for the most promising target, Ret~II. 

\begin{figure*}
\begin{center}
\includegraphics[width=0.45\textwidth]{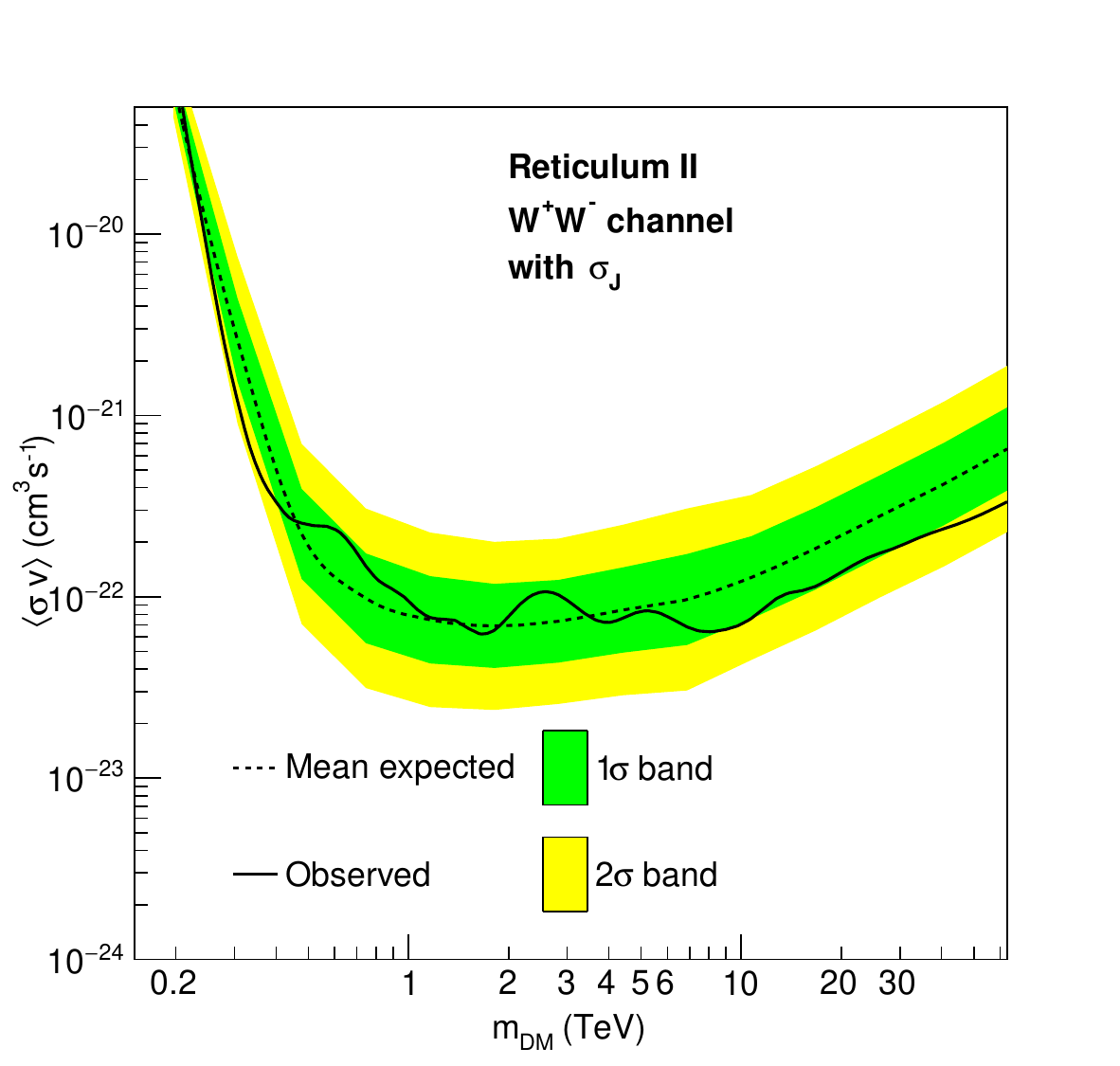}
\includegraphics[width=0.45\textwidth]{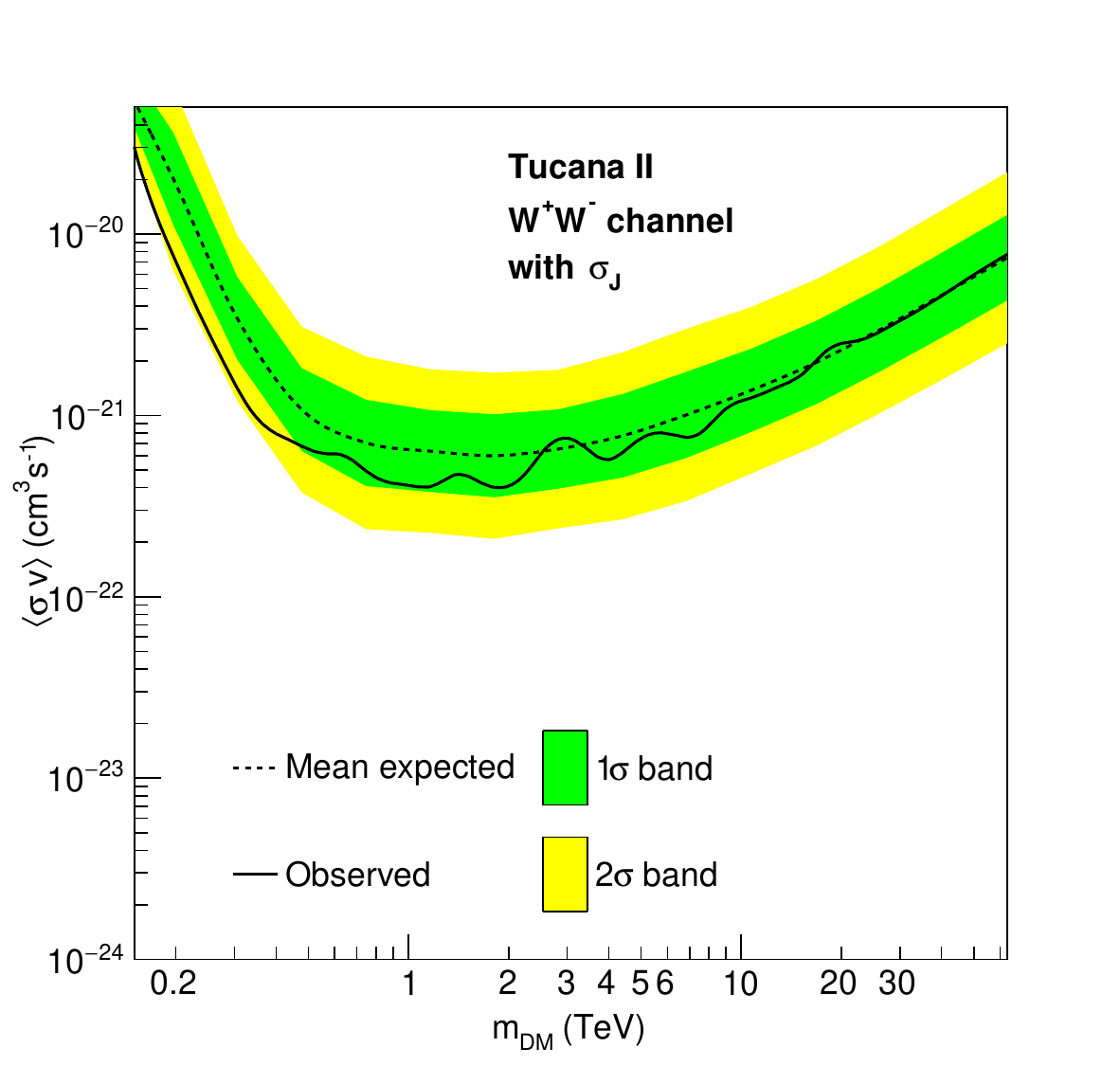}
\caption{95\% C.L. upper limits on the annihilation cross section $\langle\sigma v\rangle$ for Ret~II (left panel) and Tuc~II (right panel) in the $W^+W^-$ annihilation channel including the uncertainties on the $J$-factor. Observed limits (solid lines) together with mean expected limits (dashed lines) with 1$\sigma$ (green area) and 2$\sigma$ (yellow area) containment bands are shown. 
 }
\label{fig:sigmavRetIITucII_W_J}
\end{center}
\end{figure*}

The impact of the uncertainty on the $J$-factor is computed for Ret~II and Tuc~II, as shown in Fig.~\ref{fig:sigmavRetIITucII_W_J} for the $W^+W^-$ channel. The limits degrade by a factor of about six and 12 for Ret~II and Tuc~II, respectively.

\subsection{Combined upper limits}
\begin{figure*}
\begin{center}
\includegraphics[width=0.45\textwidth]{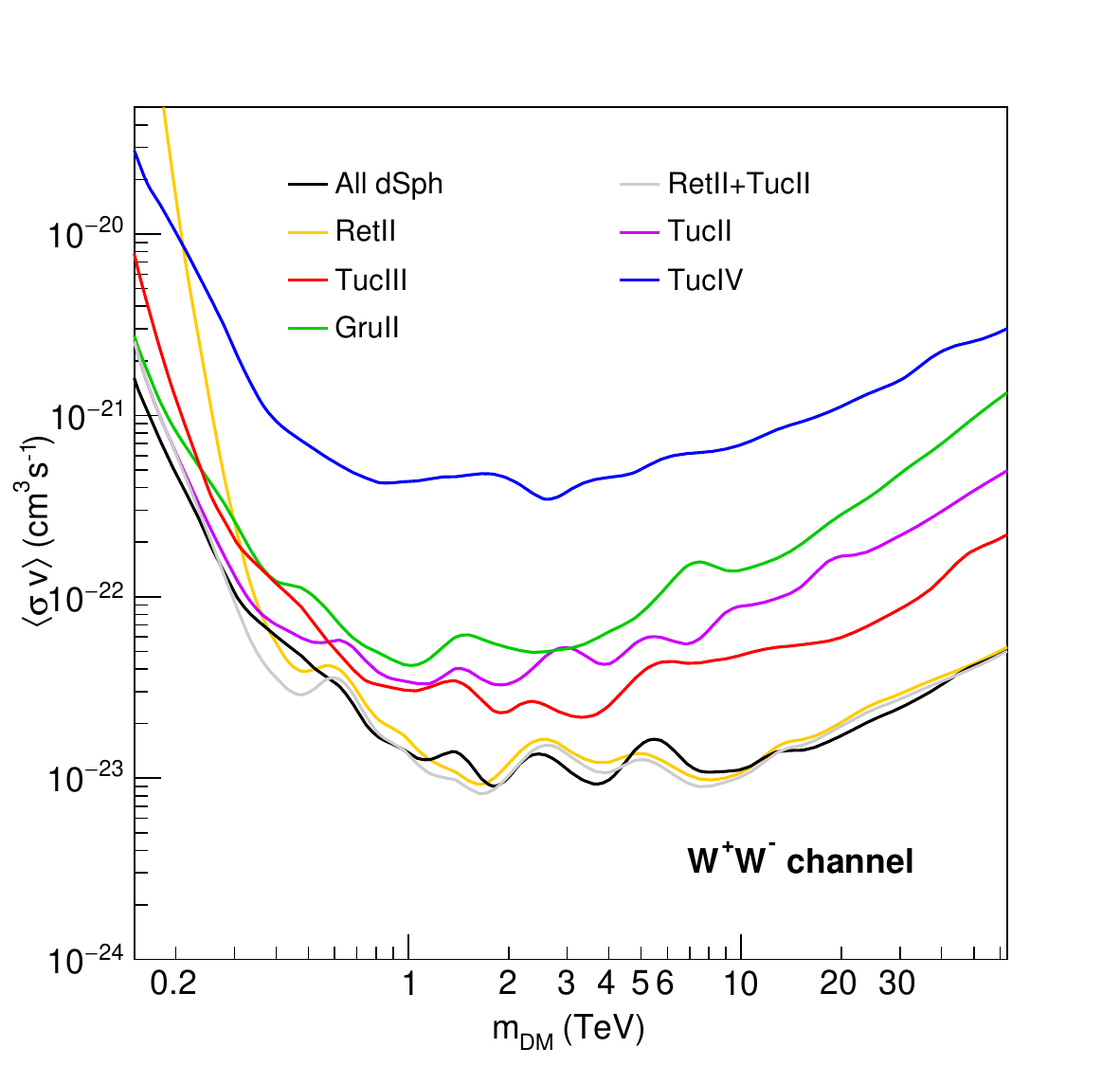}
\includegraphics[width=0.45\textwidth]{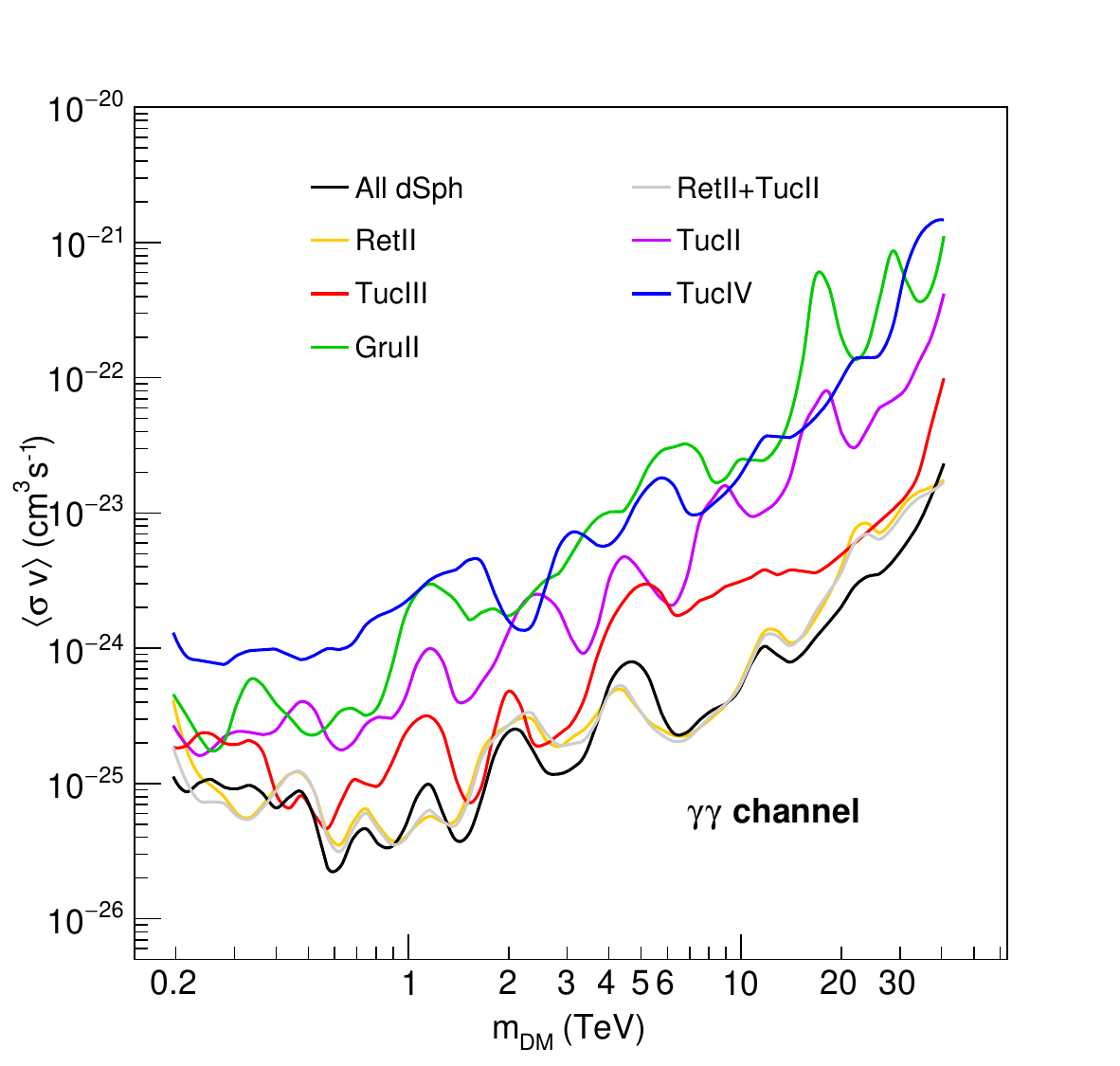}
\caption{95\% C.L. observed upper limits on the annihilation cross section $\langle\sigma v\rangle$ versus the DM mass $m_{\rm DM}$ for the combined analysis in the $W^+W^-$ (left panel) and $\gamma\gamma$ (right panel) annihilation channels, respectively, without the uncertainty on the $J$-factor.}
\label{fig:sigmav_comb}
\end{center}
\end{figure*}

The hypothesis that all targets are in fact gamma-ray emitters, but too faint to be seen with the given exposure, was tested and no overall significant excess was found. 
The combination was performed at the likelihood level, where the total likelihood function writes
\begin{equation}
\mathcal{L}_{\rm joint} = \prod_{\rm k=1}^{N_{\rm targets}}\mathcal{L}_{\rm k}
\end{equation}
where $\mathcal{L}_{\rm k}$ is the likelihood  of each target $k$. A strict joint-likelihood maximization was not performed, but 
the likelihoods were maximized beforehand. The combined observed limits at 95\% C.L. on the $W^+W^-$ and $\gamma\gamma$ channels are shown in the left and right panels of Fig.~\ref{fig:sigmav_comb}, respectively. 
For a 1.5~TeV DM mass, they reach $\langle\sigma v\rangle \simeq 1 \times 10^{-23}$ cm$^3$s$^{-1}$ and $4 \times 10^{-26}$ cm$^3$s$^{-1}$ in the $W^+W^-$  and $\gamma\gamma$ annihilation channels, respectively. These results degrade of about a factor seven when the uncertainty on the $J$-factor is included.

\begin{figure}[htbp]
\begin{center}
\includegraphics[width=0.45\textwidth]{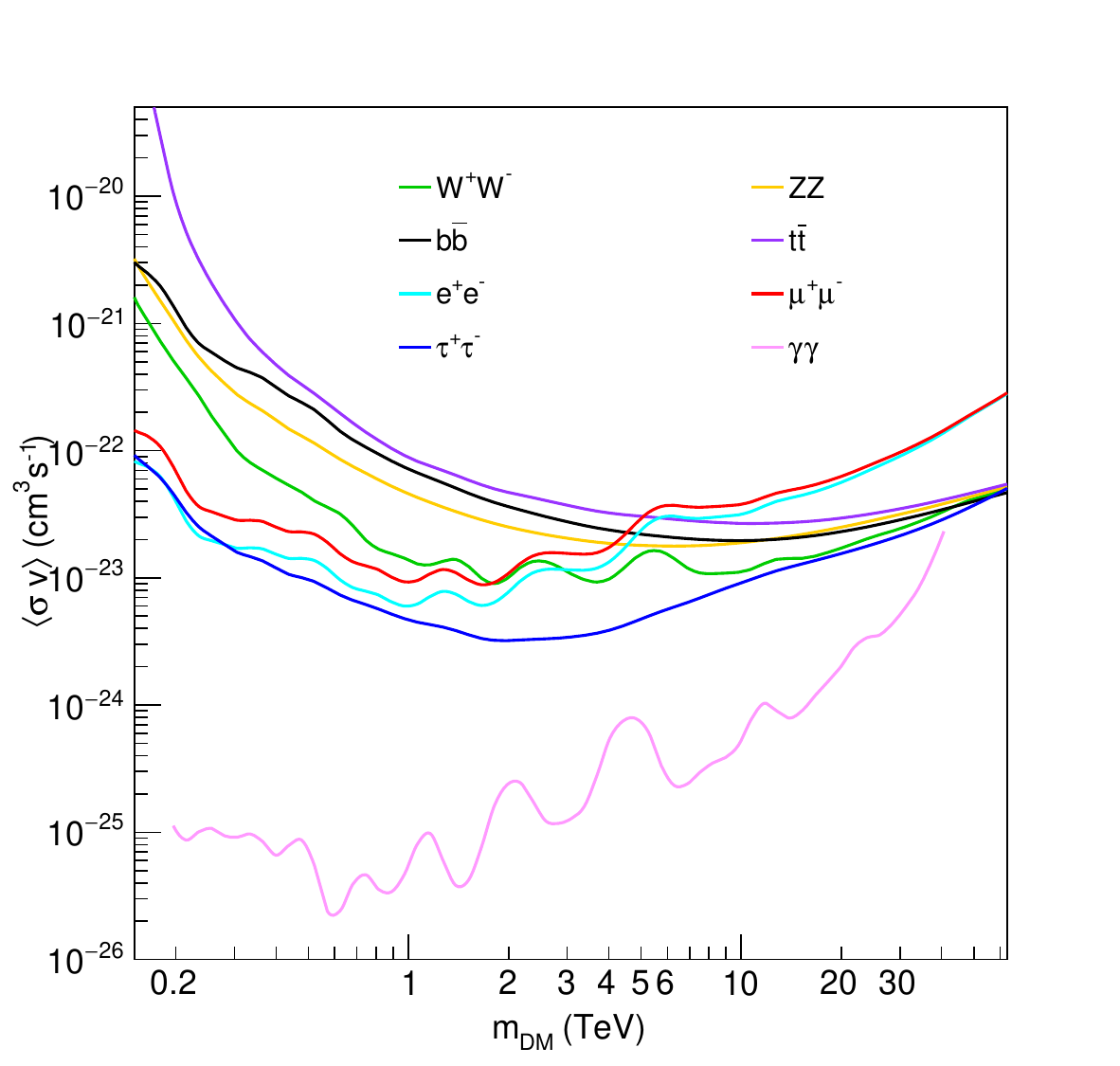}
\caption{Combined 95\% C.L. observed upper limits on the annihilation cross section $\langle\sigma v\rangle$ versus the DM mass $m_{\rm DM}$ for the combined analysis in the $b\bar{b}$, $t\bar{t}$, $W^+W^-$, $ZZ$, $e^+e^-$, $\mu^+\mu^-$, $\tau^+\tau^-$, and $\gamma\gamma$ annihilation channels, respectively, without the uncertainty on the $J$-factor.}
\label{fig:sigmav_comb_allChannels}
\end{center}
\end{figure}

The combination of the two confirmed dwarf galaxies, Ret~II and Tuc~II, is shown as well as the combination of all the five objects. In the former case the limits are driven by Ret~II limits, while in the latter the impact of Tuc~III is also significant. The combined 95\% C.L. observed limits of the five objects are plotted together in Fig.~\ref{fig:sigmav_comb_allChannels} for various annihilation channels.

\begin{figure*}
\begin{center}
\includegraphics[width=0.6\textwidth]{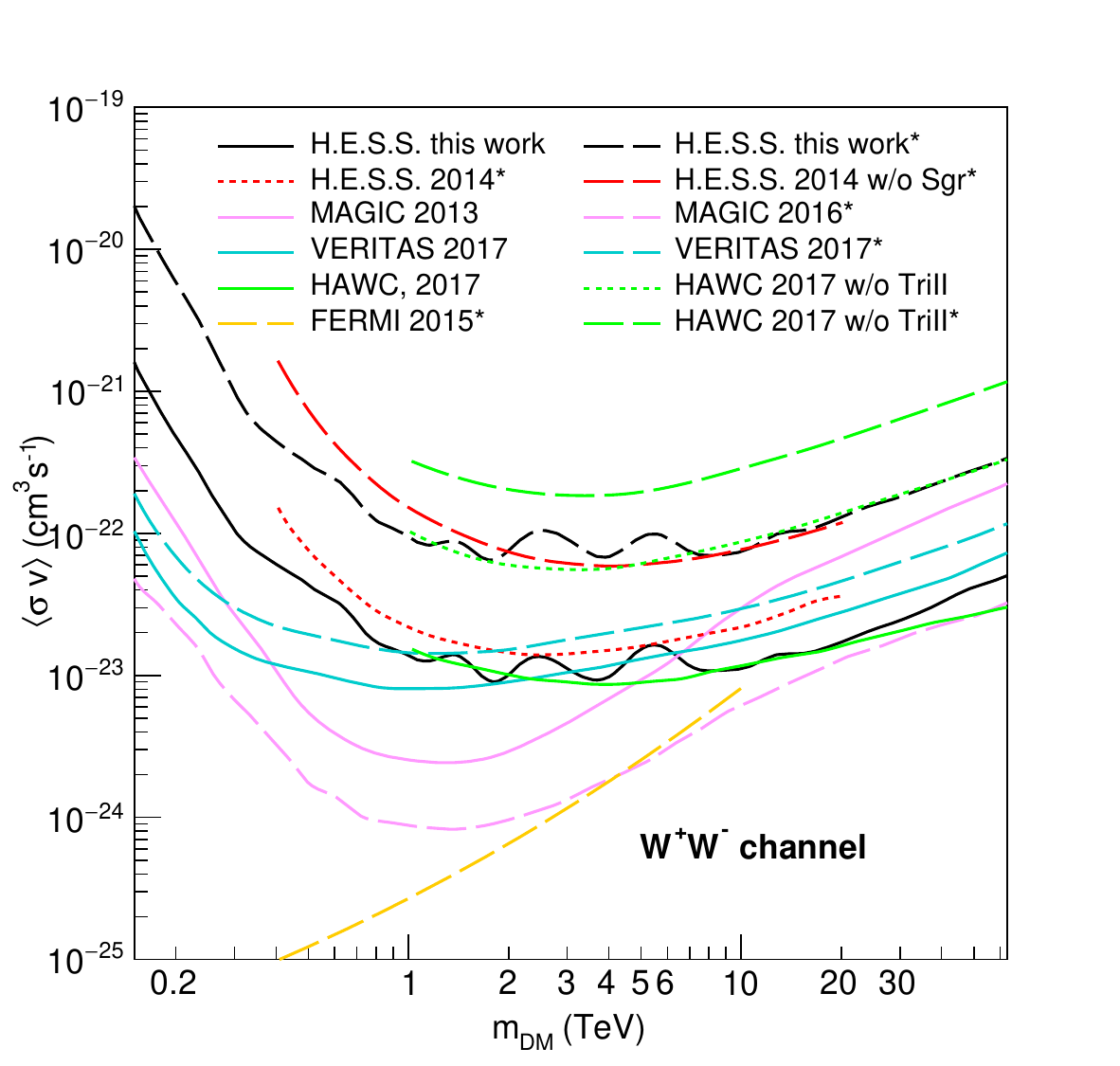}\\
\includegraphics[width=0.6\textwidth]{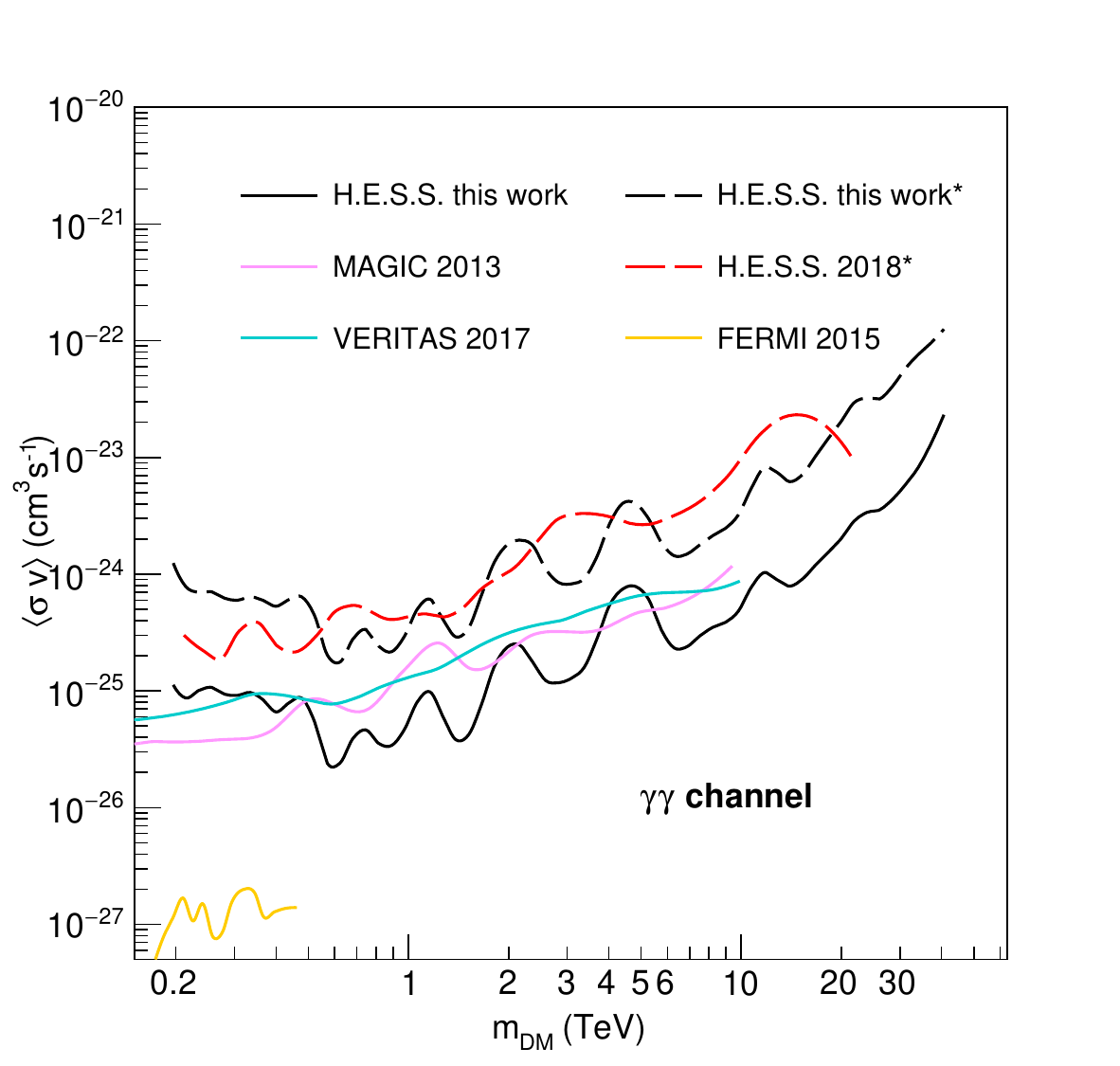}
\caption{Comparison of the observed limits versus the DM mass $m_{\rm DM}$ in the $W^+W^-$ (left panel) and $\gamma\gamma$ (right panel) annihilation channels, respectively. Constraints are shown for HAWC (507 days of data taking, combination of 15 galaxies) with and without Triangulum~II~\cite{Albert:2017vtb}, for Fermi-LAT (6 years of data taking, combination of 15 galaxies for $W^{+}W^{-}$~\cite{Ackermann:2015zua} and 5.8 years of data taken in the Galactic Center region
~\cite{Ackermann:2015lka}), for H.E.S.S. (140~h, combination of five classical galaxies) with and without 90 hours of observations on Sagittarius dSph
~\cite{Abramowski:2014tra,Abdalla:2018mve}, for MAGIC (160~h on Segue I)~\cite{Aleksic:2013xea,Ahnen:2016qkx}, and for VERITAS (128~h, combination of five galaxies)~\cite{Archambault:2017wyh}. The results marked with * include the uncertainty on the $J$-factor. See text for more details.}
\label{fig:sigmavComp}
\end{center}	
\end{figure*}

Constraints on $\langle\sigma v\rangle$ from various experiments are compared in Fig.~\ref{fig:sigmavComp} for the $W^{+}W^{-}$ (left) and $\gamma\gamma$ (right) annihilation channels, respectively. The results obtained by H.E.S.S. in this work combining the five selected DES dSphs, with and without including the uncertainty on the $J$-factor are shown together with previous H.E.S.S. results obtained on a selection of classical dSphs~\cite{Abramowski:2014tra} including the uncertainty on the $J$-factor\footnote{These results are quoted with and without Sagittarius dSph given
that the determination of its dark matter profile is challenging for this tidally-disrupted system. See, for instance, Ref.~\cite{Viana:2012zz}.}. The results from MAGIC on Segue~1\footnote{ The large $J$-factor value used for Segue~1 in the above mentioned results can be overestimated by a factor up to 100~\cite{Bonnivard:2015xpq}.} with (dashed) and without (solid) the uncertainty on the $J$-factor are extracted from Ref.~\cite{Aleksic:2013xea} and Ref.~\cite{Ahnen:2016qkx}, respectively. Results obtained by VERITAS from a combination of five dSphs including Segue~1~\cite{Archambault:2017wyh}, with and without uncertainty on the $J$-factor are plotted together with the stacked limits on 15 dSphs obtained by Fermi-LAT with the uncertainty on the $J$-factor~\cite{Ackermann:2015zua}.
The results  obtained from the HAWC experiment on 15 targets~\cite{Albert:2017vtb}  without the uncertainty on the $J$-factor as well as results that do not include Triangulum~II\footnote{An accurate determination of the $J$-factor of Triangulum II is difficult due to the reduced number of detected member stars (13) and possible tidal stripping~\cite{2017ApJ...838...83K}. The total $J$-factor of Triangulum II used in Ref.~\cite{Albert:2017vtb} is  $\rm \log_{10} (\mathit{J}/GeV^2cm^{-5} )$ = 20.44. Such a large value  is quite speculative and may have been artificially obtained by the presence of a binary star with variable radial velocity~\cite{2017ApJ...838...83K}. The reduced number of member stars makes also the $J$-factor determination more prone to systematic uncertainties.} with and without uncertainty on the $J$-factor are also shown.
In the $\gamma\gamma$ annihilation channel, the previous H.E.S.S. results on a selection of classical dSphs are extracted from Ref.~\cite{Abdalla:2018mve}. The results on Segue~1 from MAGIC ~\cite{Aleksic:2013xea} as well as those by VERITAS on five dSphs including Segue~1~\cite{Archambault:2017wyh}, without uncertainty on the $J$-factor, are plotted. The Fermi-LAT limits on the Galactic Center~\cite{Ackermann:2015lka} are also displayed.
The constraints obtained in the $\gamma\gamma$ annihilation channel from H.E.S.S. are particularly relevant to constraint DM models with enhanced line-like signals in the TeV mass range.

\section{Summary and discussions}
\label{sec:summary}
H.E.S.S. is the first IACT to observe a selection of ultra-faint dwarf satellites of the Milky Way recently discovered by DES to search for a DM annihilation signal with the highest sensitivity among IACTs towards these objects given its position in the Southern hemisphere. The exposure towards the five selected targets, Ret~II, Tuc~II, Tuc~III, Tuc~IV, and Gru~II, amounts to about 80 hours of live time. In absence of a significant excess in any of the object FoV, 95\% C.L. upper limits have been derived on the DM annihilation cross section as a function of the DM mass in various annihilation channels. The strongest limits from an individual object are obtained for Ret~II. Assuming no uncertainty in the $J$-factor, they reach $\langle\sigma v\rangle \simeq 1 \times 10^{-23}$ cm$^3$s$^{-1}$ and $8 \times 10^{-26}$ cm$^3$s$^{-1}$ in the $W^+W^-$ and $\gamma \gamma$ annihilation channels, respectively, for a 1.5 TeV DM mass. Assuming an uncertainty on the $J$-factor, the limits degrade by about a factor seven. 
Using a lower mean value for Ret~II $J$-factor would degrade the limits accordingly. 
The limits from the combined analysis of the five targets are dominated by Ret~II limits assuming the conservative $J$-factor value for Tuc~II. ln the $W^+W^-$ annihilation channel they reach about the same values within the statistical fluctuations.
They go slightly down to $\langle\sigma v\rangle \simeq 9 \times 10^{-24}$ cm$^3$s$^{-1}$ for a 1.5~TeV DM mass when considering only Ret~II and Tuc~II. The combined limits on the five targets in $\gamma \gamma$ reach $\langle\sigma v\rangle \simeq 4 \times 10^{-26}$ cm$^3$s$^{-1}$ for a 1.5~TeV DM. 
Including the $J$-factor uncertainty possibly degrades the combined limits up to a factor of about seven.

The uncertainty on the $J$-factor for ultra-faint systems is challenging to measure or predict. 
The limits derived in this work and similar studies are strongly dependent on the choice of the $J$-factor mean value and its uncertainty for a given system. Most often only the statistical uncertainty on the $J$-factor coming from the finite number of stellar tracers is considered. Only a few studies investigate the impact of the assumptions made to derive the $J$-factor value (see, for instance, Refs.~\cite{Bonnivard:2014kza,Bonnivard:2015vua}). Among the possible sources of systematic uncertainty in the $J$-factor determination~\cite{Lefranc:2016dgx} are the assumptions of a single stellar population or spherical symmetry, a constant velocity anisotropy, and the absence of tidal stripping. The selection of member stars for ultra-faint systems is also complex due to the difficulty to distinguish member stars from interlopers in the foreground. Another caveat when comparing limits including the uncertainty on the $J$-factors from one experiment to another is the treatment of the $J$-factor uncertainty in the derivation of the limits. Here a log-nomal distribution is taken for the $J$-factor likelihood function while some studies use a {\it modified} log-normal distribution~\cite{Fermi-LAT:2016uux,Ahnen:2016qkx}. In addition, the procedure to derive the maximized $J$-factor may differ from one study to another.

The new results obtained by H.E.S.S. are among the most constraining in the $\gamma \gamma$ annihilation channel above 500~GeV. They are comparable to VERITAS and HAWC limits in the $W^+W^-$ annihilation channel in the multi-GeV and multi-TeV DM mass ranges respectively. The constraints obtained in the $\gamma\gamma$ annihilation channel are particularly relevant in the context of DM models with enhanced line signals in the TeV DM mass ranges. Among them are  
models with gamma-ray boxes~\cite{Ibarra:2015tya}, scalar~\cite{Giacchino:2015hvk}, and Dirac~\cite{Duerr:2015wfa} DM models, as well as the canonical Majorana DM triplet fermion known as the {\it Wino}~\cite{Baumgart:2017nsr,Baumgart:2018yed}, and the DM doublet fermion known as the {\it Higgsino}~\cite{Mahbubani:2005pt,Kearney:2016rng,Kowalska:2018toh}. 

The constraints obtained in this work are competitive with other experiments. While the likelihood function definition, the test statistics and the background determination technique may vary from one experiment to another, 
the results complement each other showing the importance of having instrument with different characteristics that observe a different selection of targets. The IACTs are powerful instruments to investigate the multi-TeV DM not accessible to Fermi-LAT. This is particularly true when it comes to search for TeV DM-induced spectral features close to the DM mass. In the $\gamma\gamma$ channel where the expected signal is very sharp the Fermi-LAT sensitivity range cannot extend beyond a few hundred GeV where its detected photon statistics is very low. The excellent energy resolution of H.E.S.S. is crucial to search for monoenergetic signals expected in the $\gamma\gamma$ annihilation channel. In addition, the H.E.S.S. instrument performance enable to cover the broadest DM mass range among the IACTs for line-like signal searches.

Future studies would greatly benefit from high-quality stellar kinematic datasets towards the most promising ultra-faint dSph satellites discovered by DES, such as Tuc~III, Tuc~IV, and Gru~II in order to improve the knowledge of the DM density distribution in these objects. 

\section{Acknowledgments}
The support of the Namibian authorities and of the University of Namibia in facilitating the construction and operation of H.E.S.S. is gratefully acknowledged, as is the support by the German Ministry for Education and Research (BMBF), the Max Planck Society, the German Research Foundation (DFG), the Helmholtz Association, the Alexander von Humboldt Foundation, the French Ministry of Higher Education, Research and Innovation, the Centre National de la Recherche Scientifique (CNRS/IN2P3 and CNRS/INSU), the Commissariat \`a l'\'energie atomique et aux \'energies alternatives (CEA), the U.K. Science and Technology Facilities Council (STFC), the Knut and Alice Wallenberg Foundation, the National Science Centre, Poland grant no. 2016/22/M/ST9/00382, the South African Department of Science and Technology and National Research Foundation, the University of Namibia, the National Commission on Research, Science \& Technology of Namibia (NCRST), the Austrian Federal Ministry of Education, Science and Research and the Austrian Science Fund (FWF), the Australian Research Council (ARC), the Japan Society for the Promotion of Science and by the University of Amsterdam. We appreciate the excellent work of the technical support staff in Berlin, Zeuthen, Heidelberg, Palaiseau, Paris, Saclay, Tubingen and in Namibia in the construction and operation of the equipment. This work benefited from services provided by the H.E.S.S. Virtual Organisation, supported by the national resource providers of the EGI Federation.

\newpage
\bibliography{bibl} 
\appendix
\section*{Appendix: constraints for additional annihilation channels towards Tuc~II, Tuc~III, Tuc~IV and Gru~II}
\begin{figure*}[htbp]
\begin{center}
\includegraphics[width=0.45\textwidth]{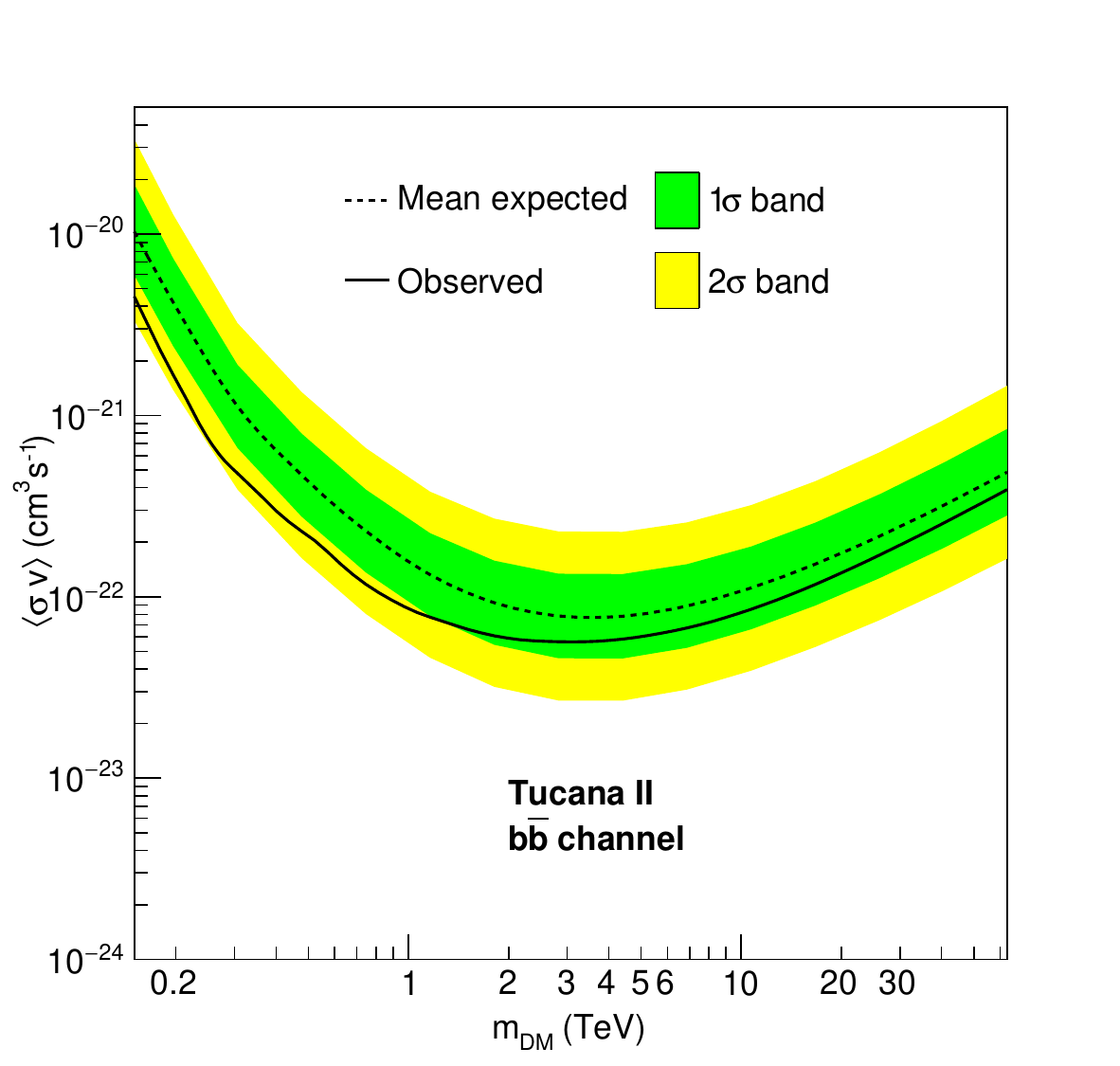}
\includegraphics[width=0.45\textwidth]{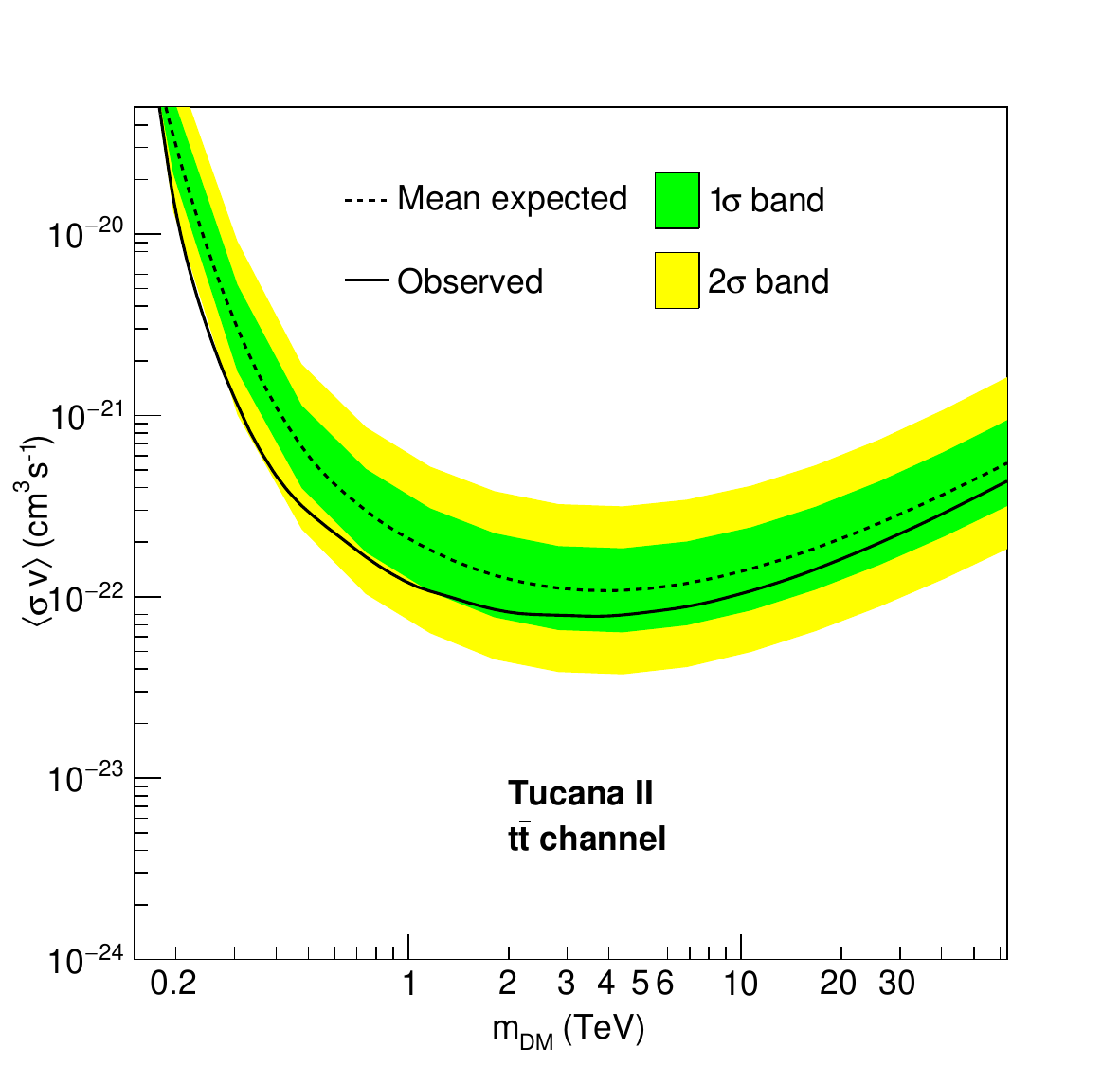}
\includegraphics[width=0.45\textwidth]{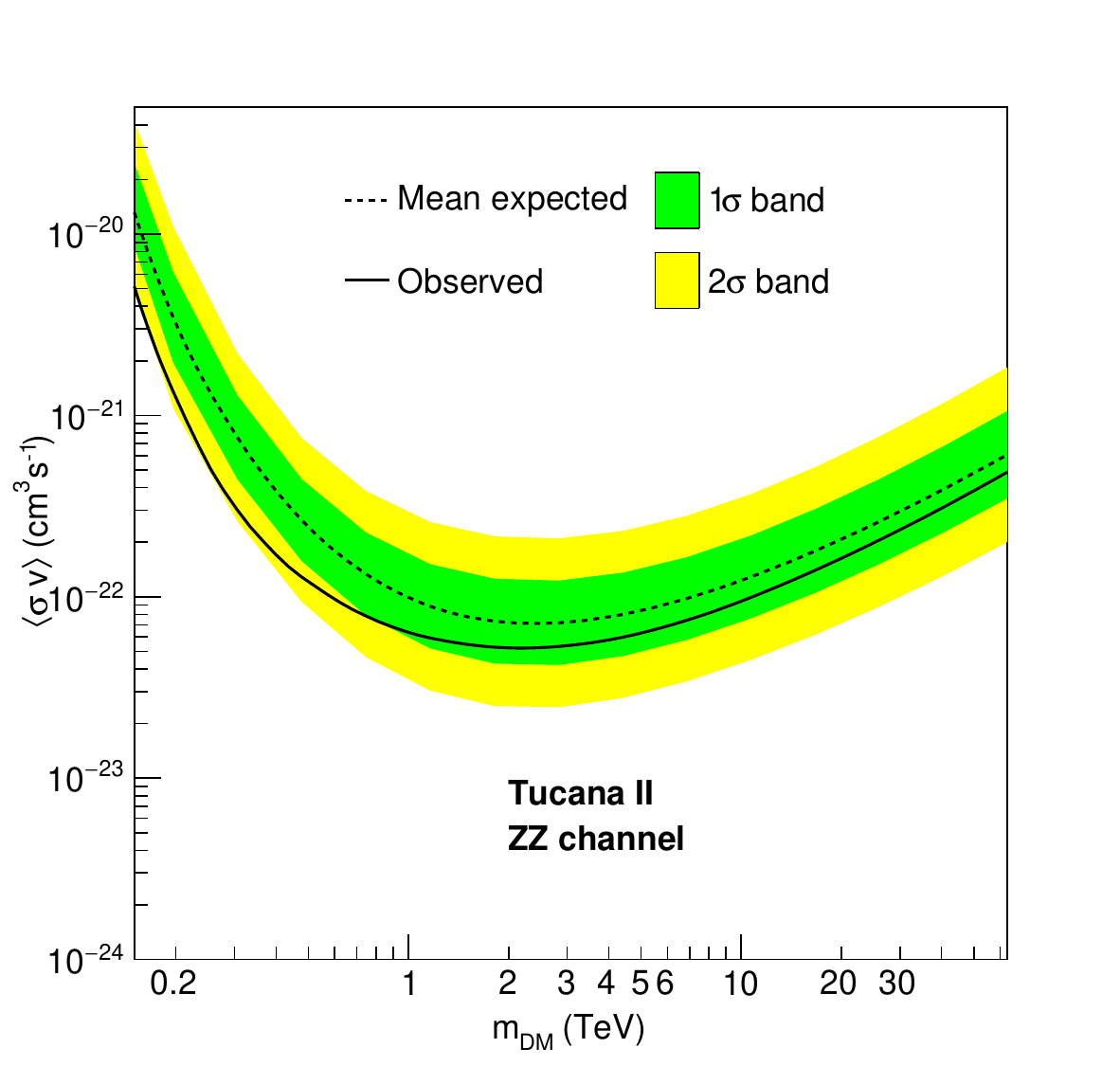}
\includegraphics[width=0.45\textwidth]{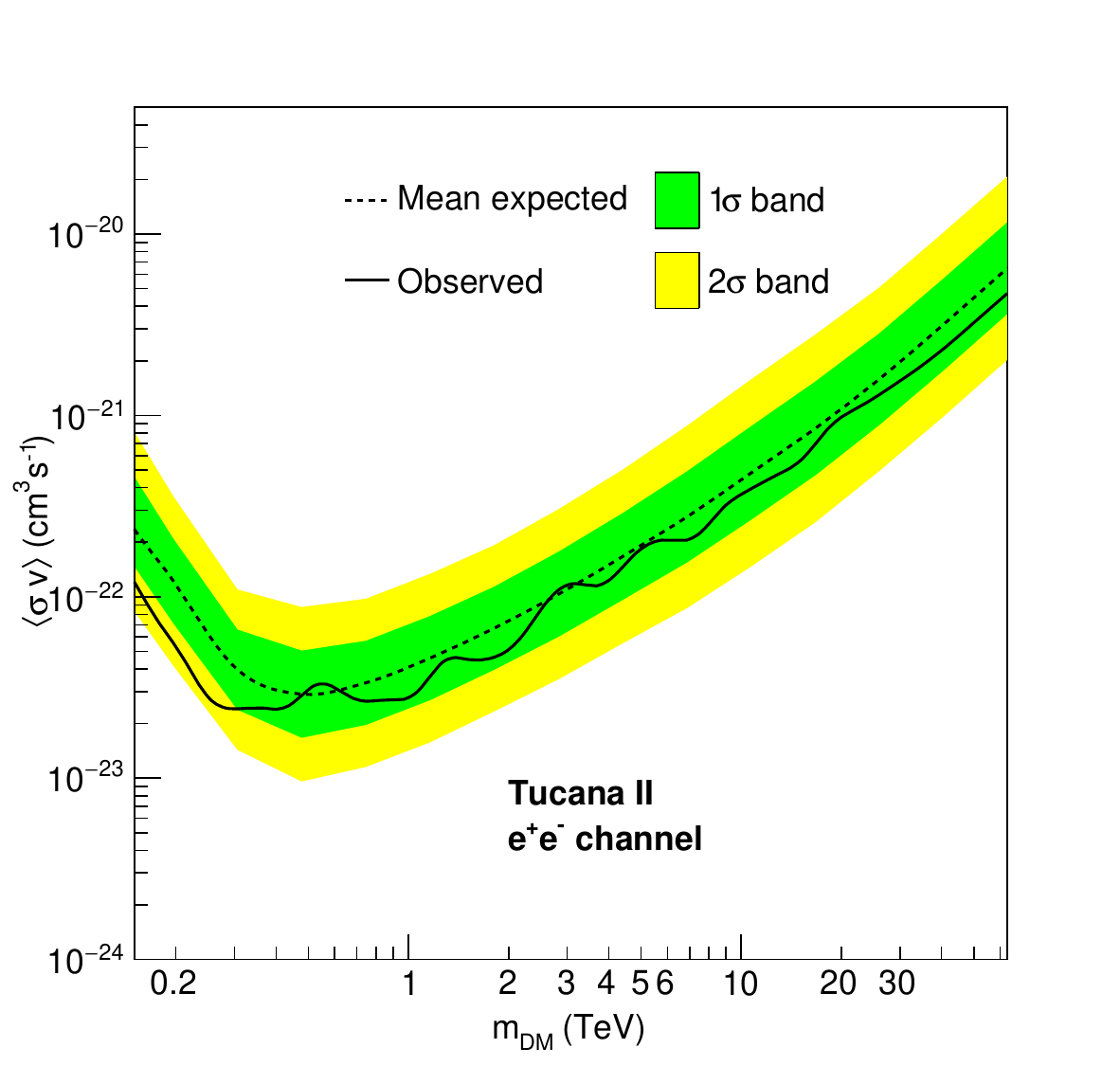}
\includegraphics[width=0.45\textwidth]{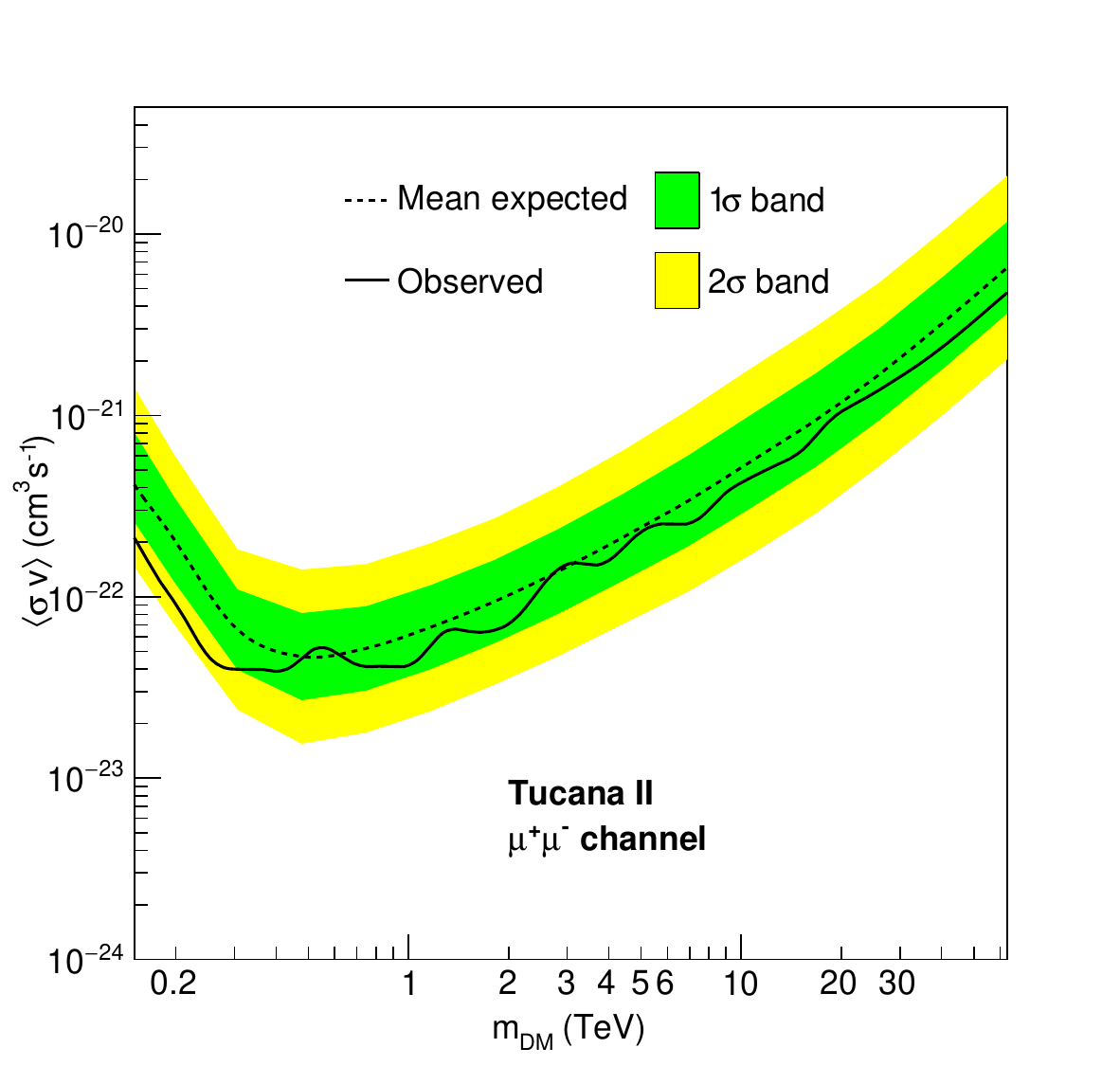}
\includegraphics[width=0.45\textwidth]{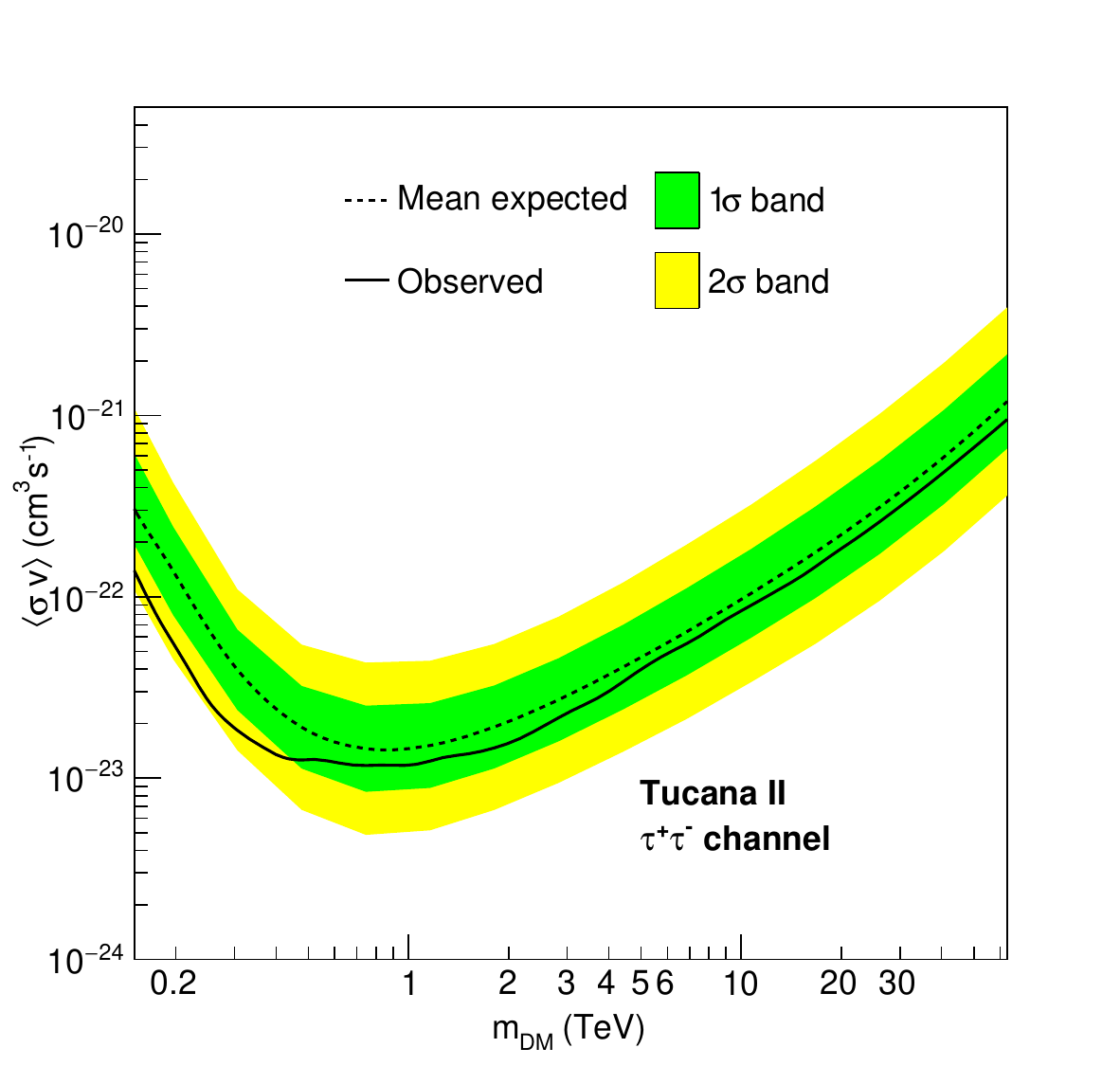}
\caption{95\% C.L. upper limits on the annihilation cross section $\langle\sigma v\rangle$ for Tuc~II in the $b\bar{b}$, $t\bar{t}$, $ZZ$, $e^+e^-$, $\mu^+\mu^-$, $\tau^+\tau^-$ annihilation channels, respectively, without the uncertainty on the $J$-factor. Observed limits (solid lines) together with mean expected limits (dashed line) and the 1$\sigma$ (green area) and 2$\sigma$ (yellow area) containment bands are shown, respectively.}
\label{fig:sigmavTucII_all}
\end{center}
\end{figure*}

\begin{figure*}[htbp]
\begin{center}
\includegraphics[width=0.45\textwidth]{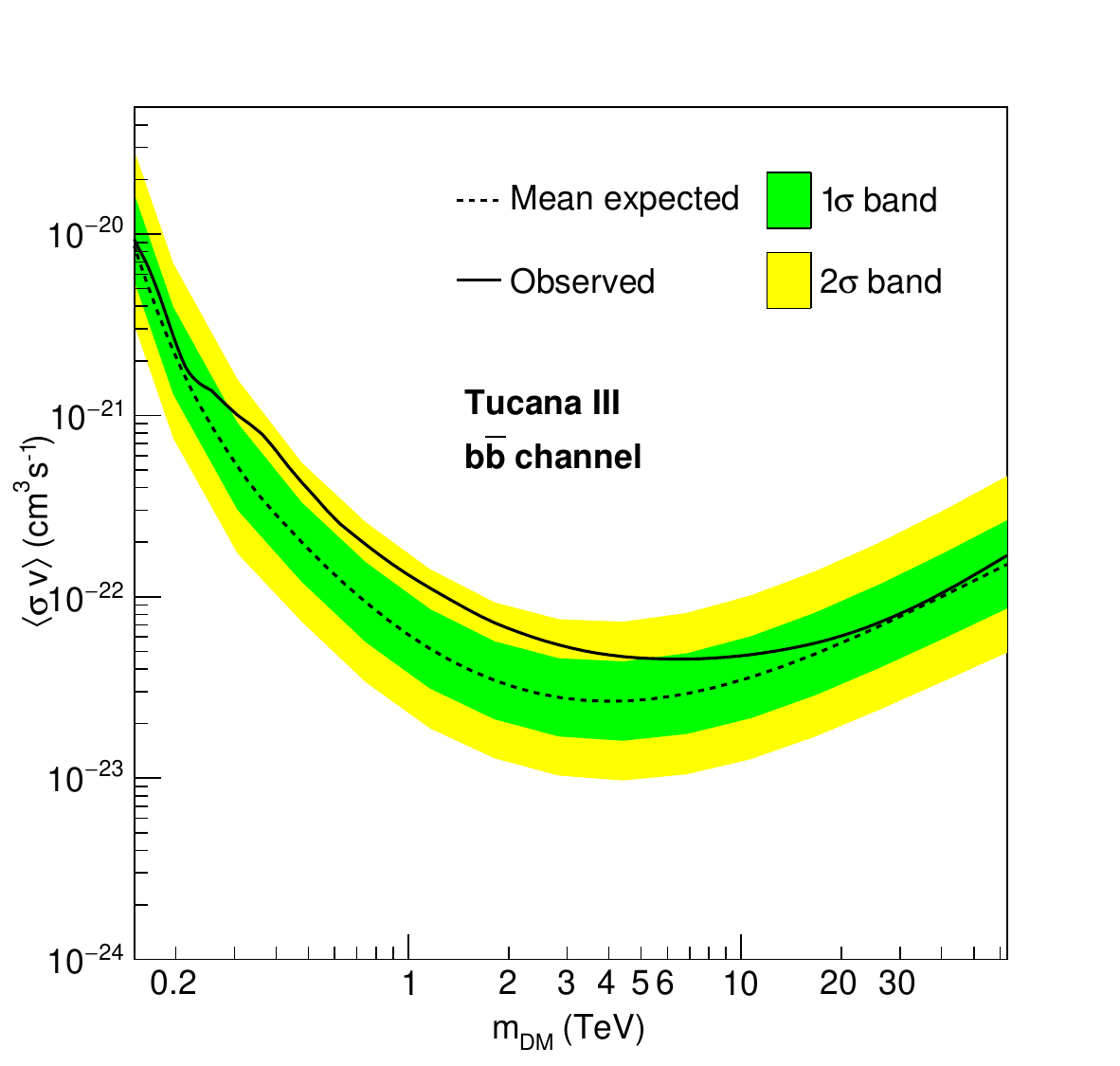}
\includegraphics[width=0.45\textwidth]{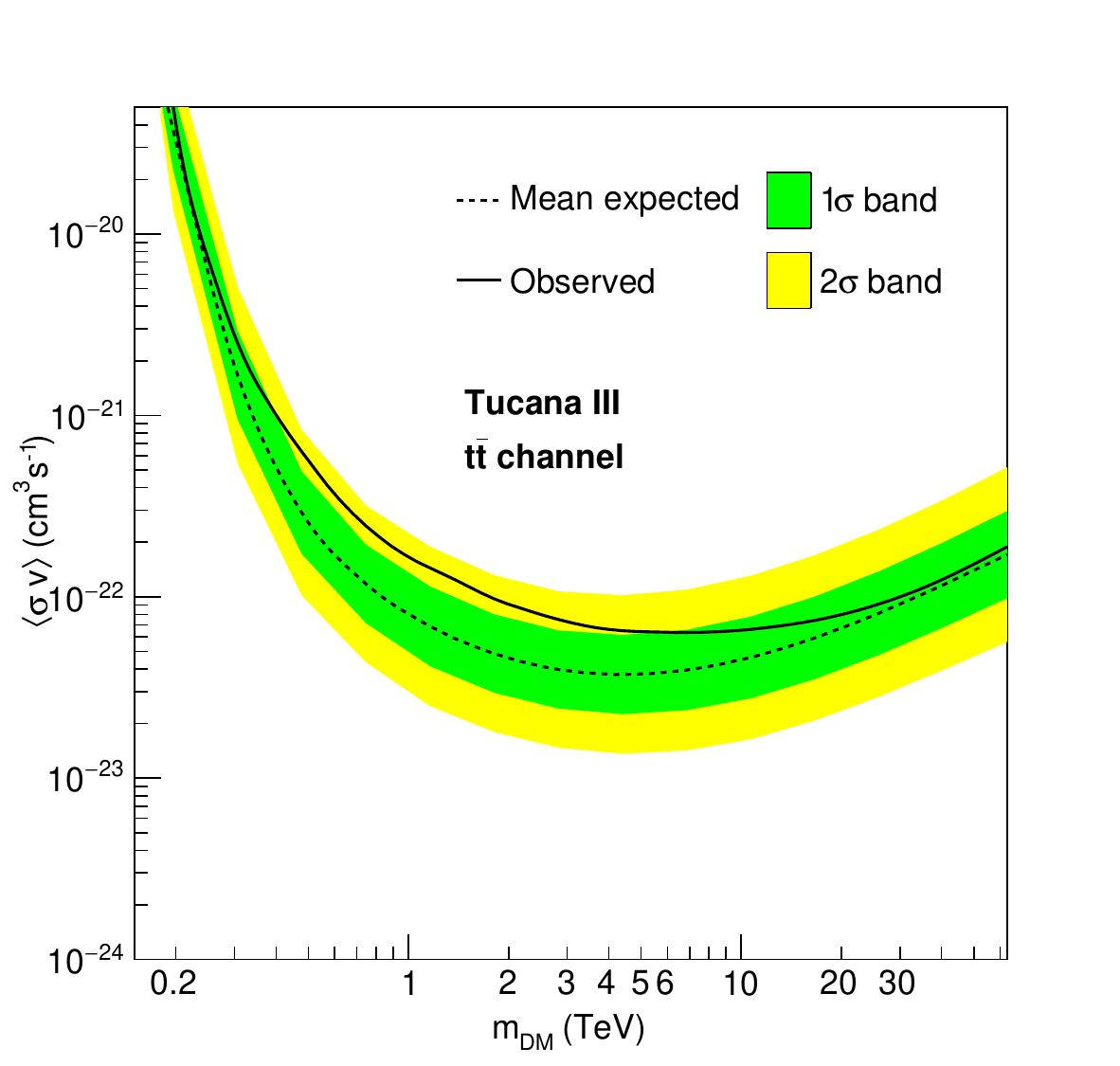}
\includegraphics[width=0.45\textwidth]{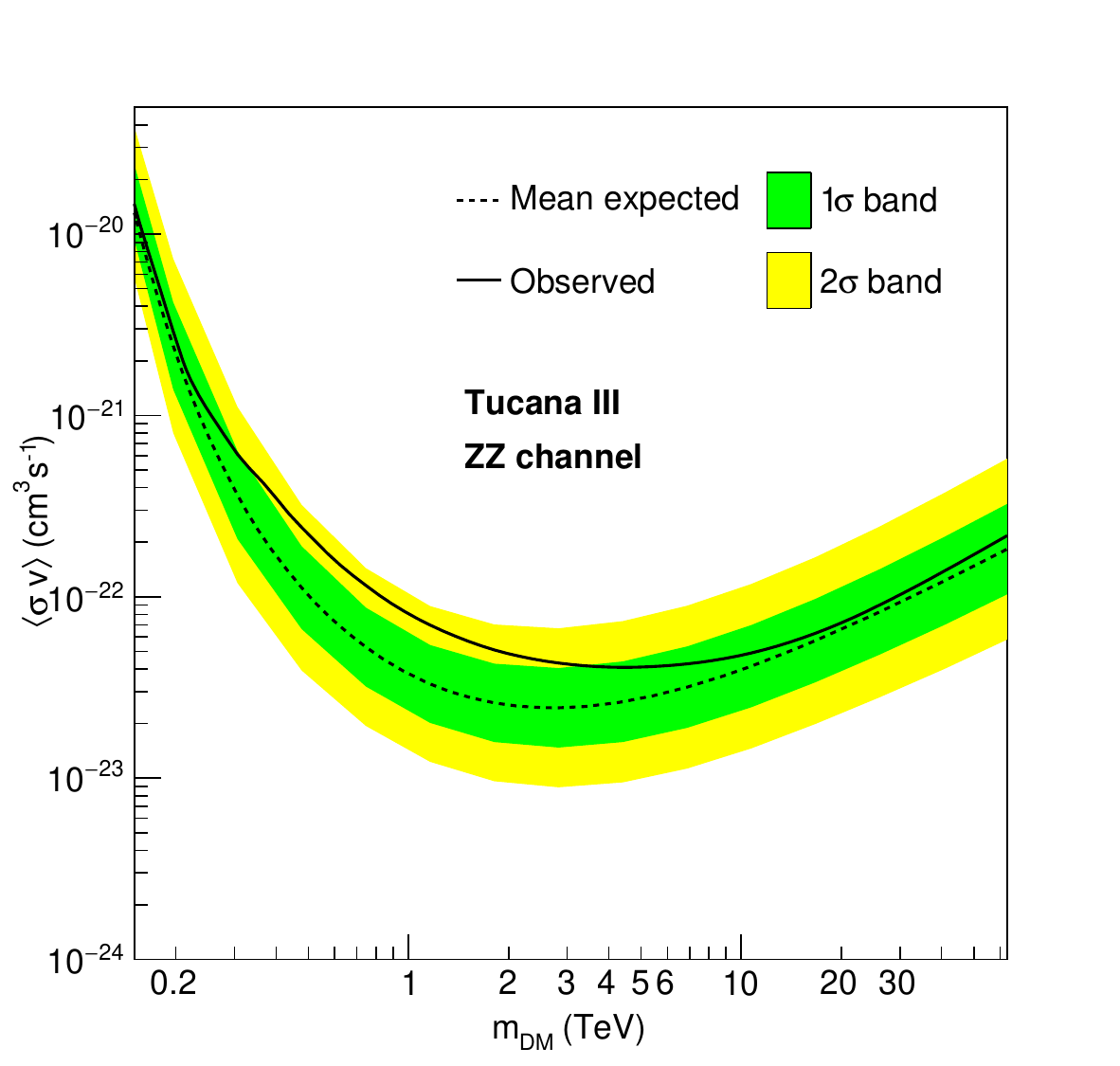}
\includegraphics[width=0.45\textwidth]{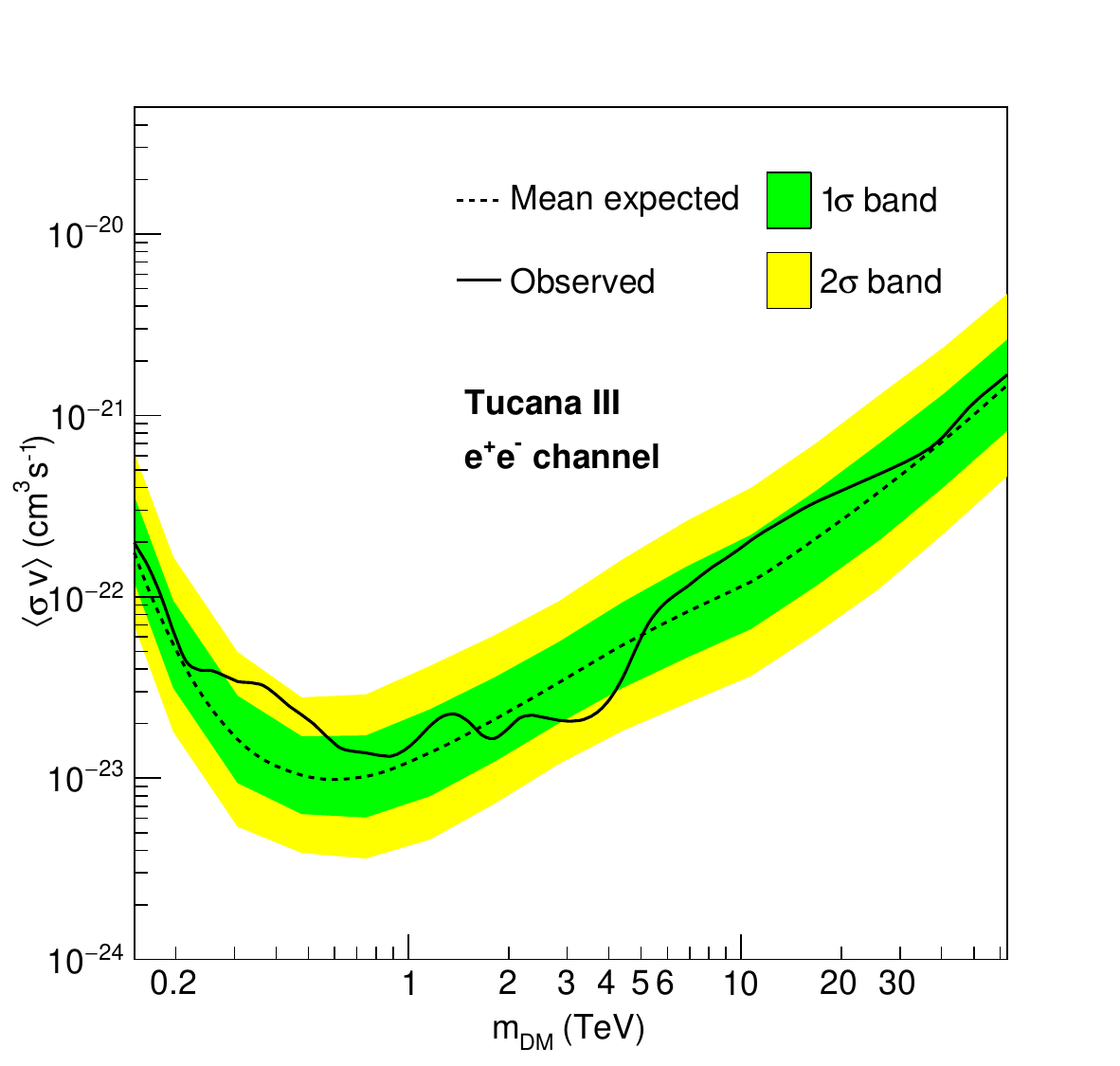}
\includegraphics[width=0.45\textwidth]{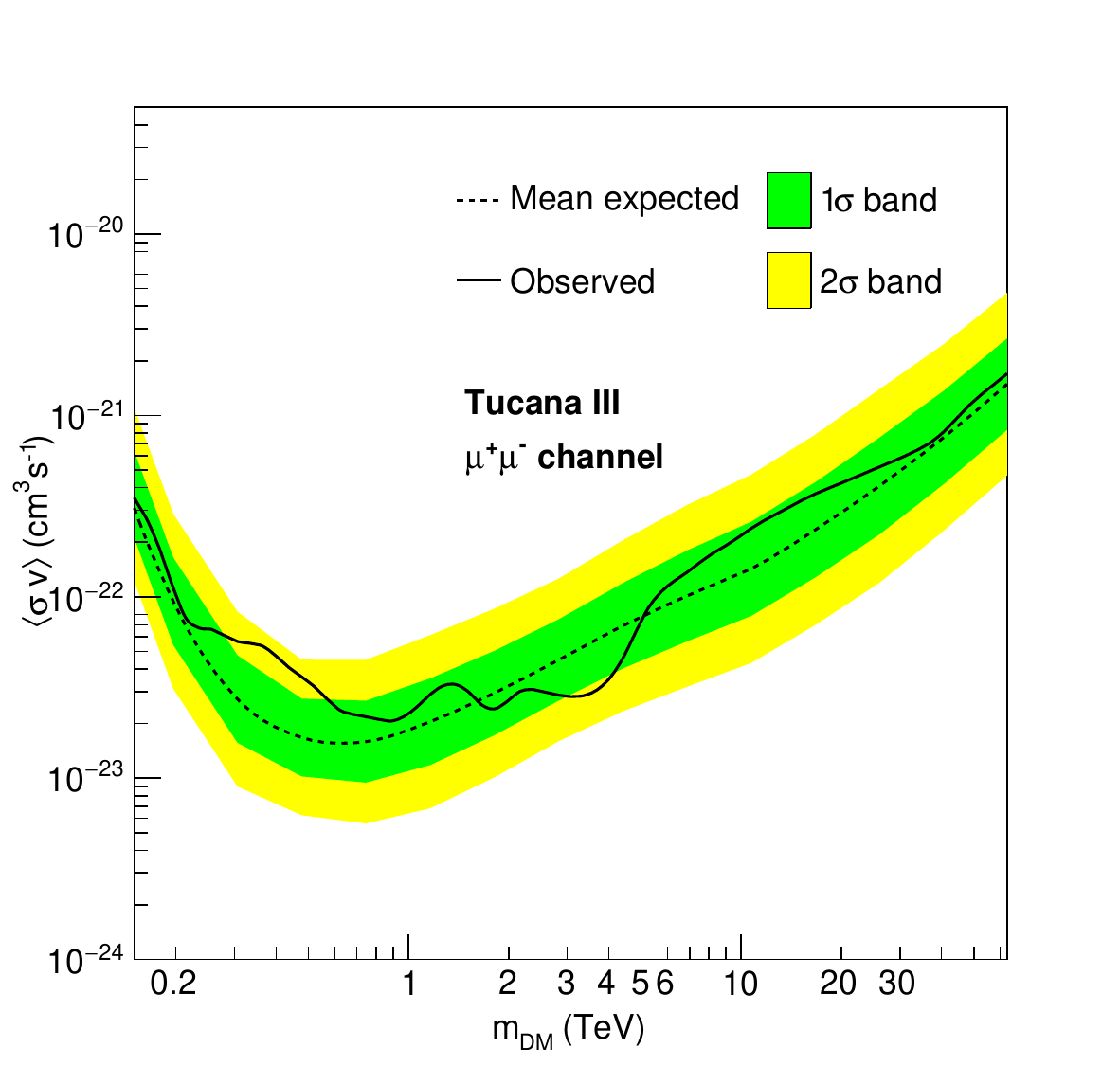}
\includegraphics[width=0.45\textwidth]{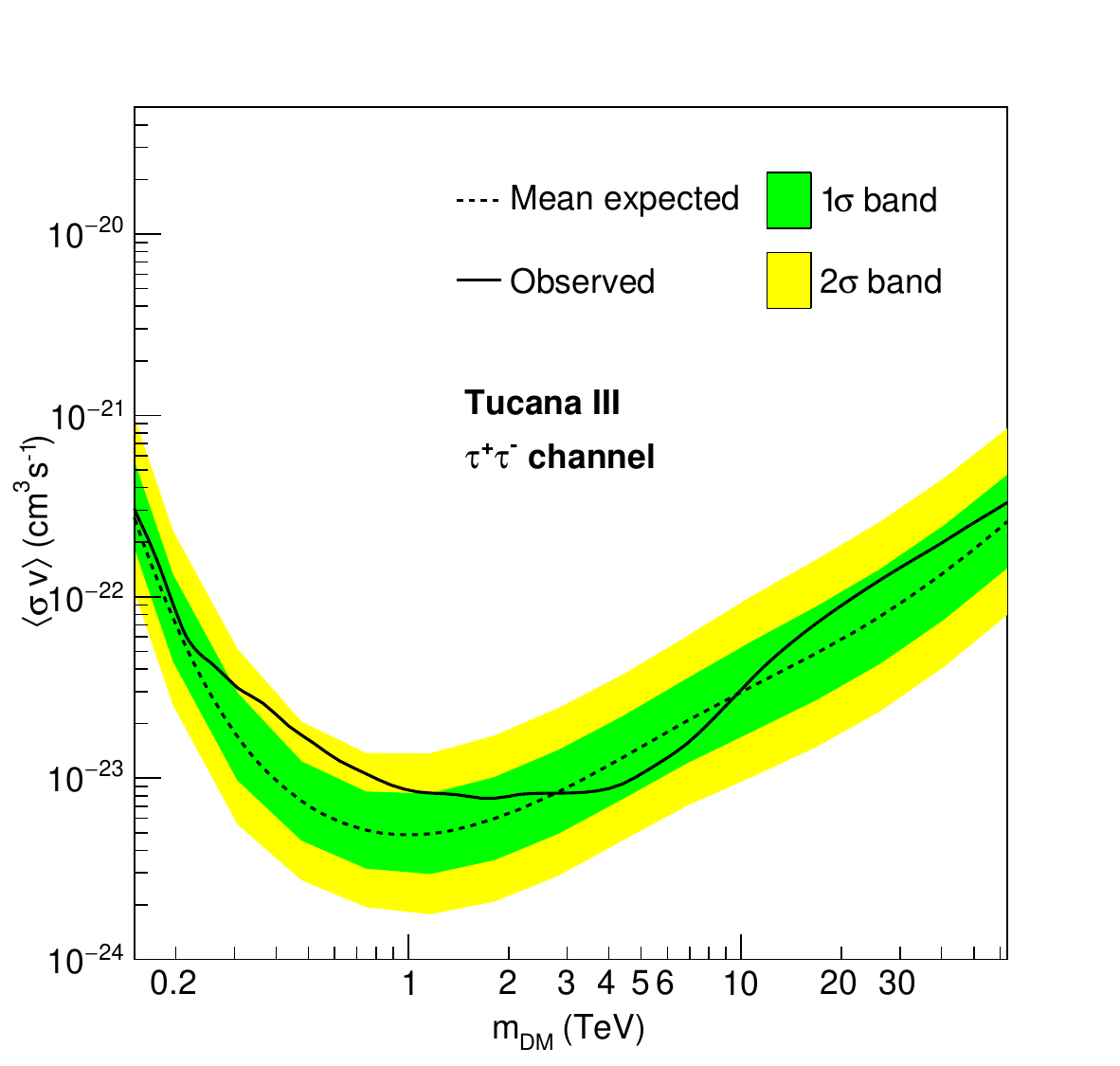}
\caption{95\% C.L. upper limits on the annihilation cross section $\langle\sigma v\rangle$ for Tuc~III in the $b\bar{b}$, $t\bar{t}$, $ZZ$, $e^+e^-$, $\mu^+\mu^-$, $\tau^+\tau^-$ annihilation channels, respectively, without the uncertainty on the $J$-factor. Observed limits (solid lines) together with mean expected limits (dashed line) and the 1$\sigma$ (green area) and 2$\sigma$ (yellow area) containment bands are shown, respectively.}
\label{fig:sigmavTucIII_all}
\end{center}
\end{figure*}

\begin{figure*}[htbp]
\begin{center}
\includegraphics[width=0.45\textwidth]{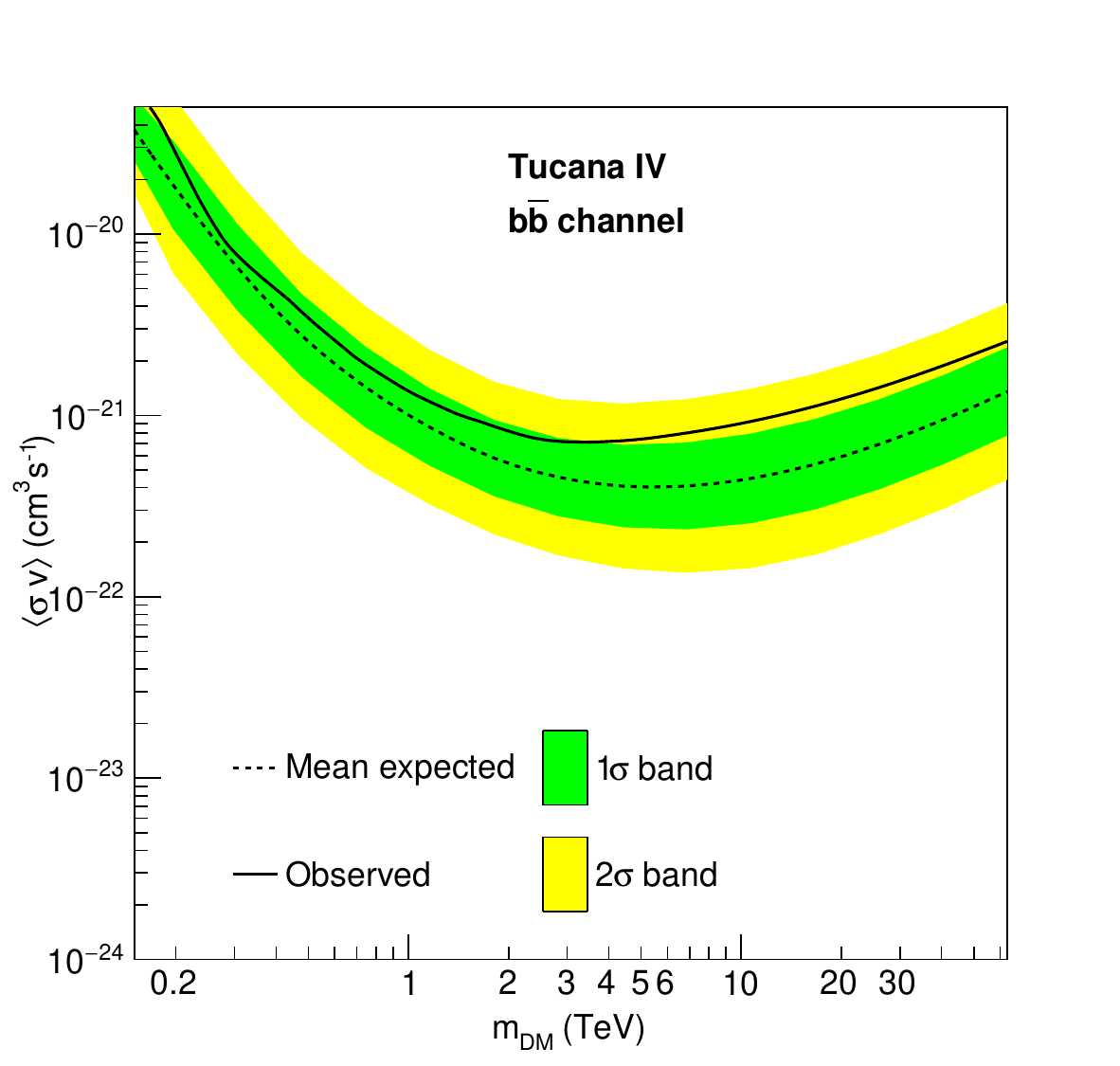}
\includegraphics[width=0.45\textwidth]{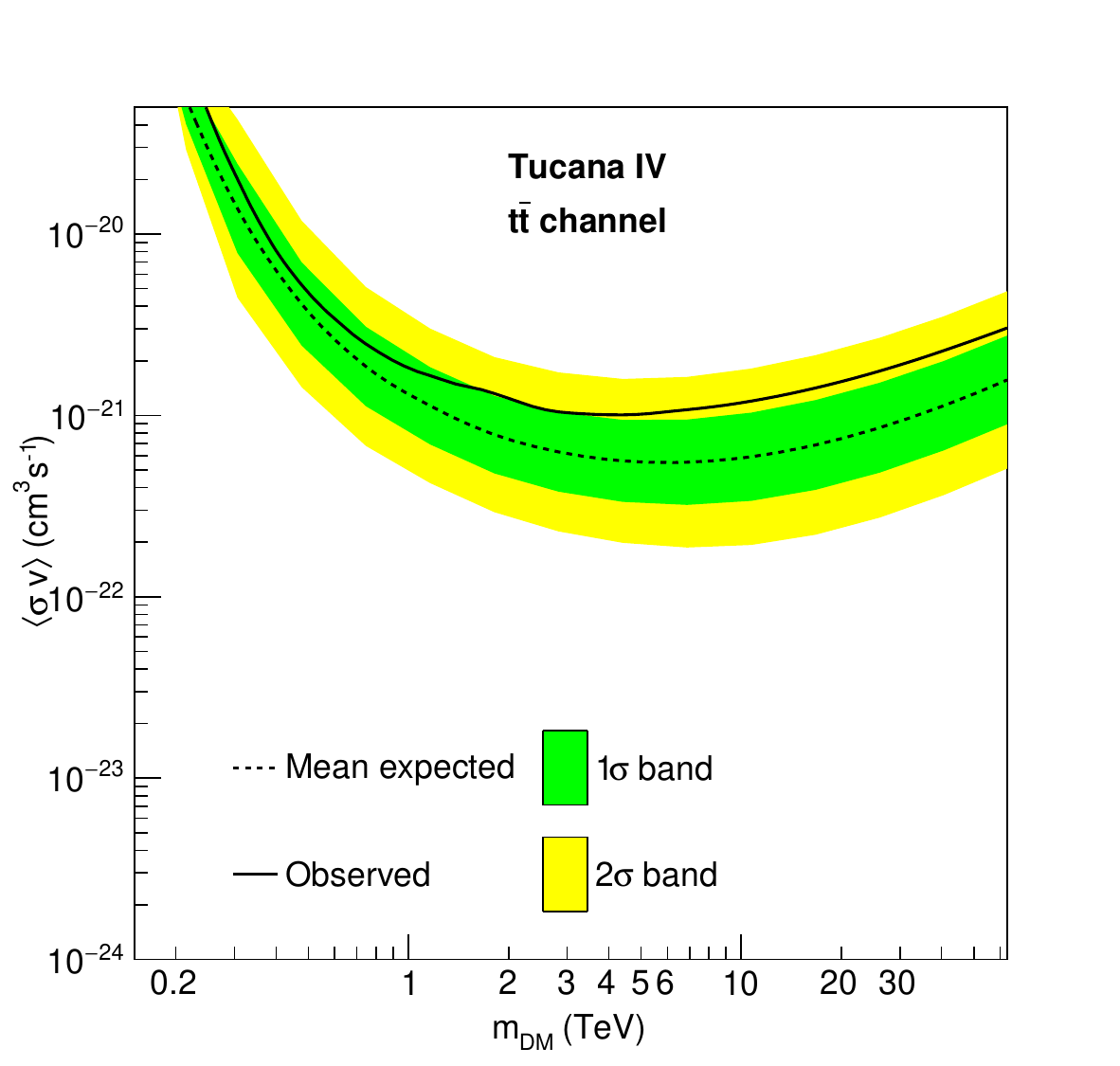}
\includegraphics[width=0.45\textwidth]{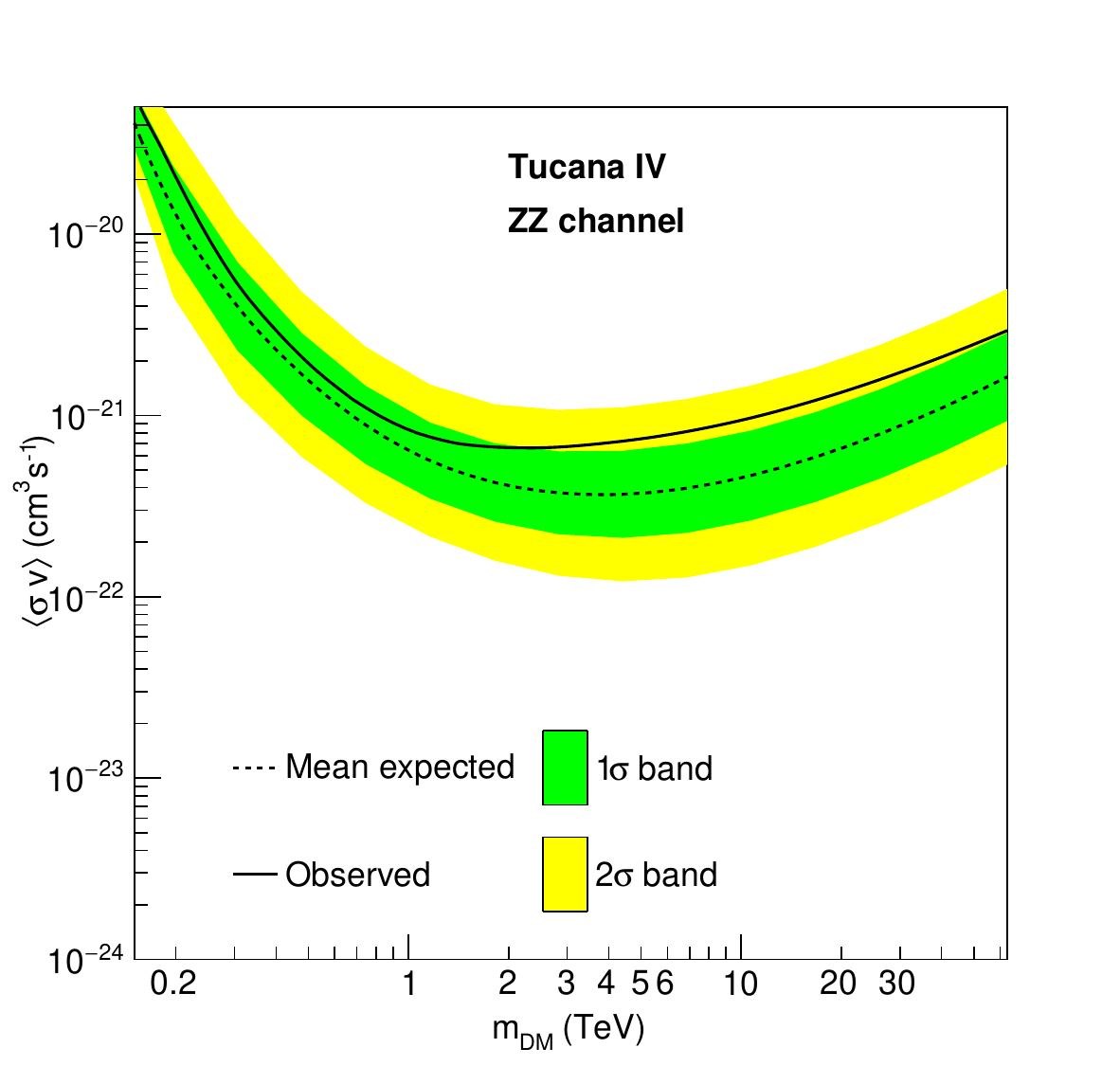}
\includegraphics[width=0.45\textwidth]{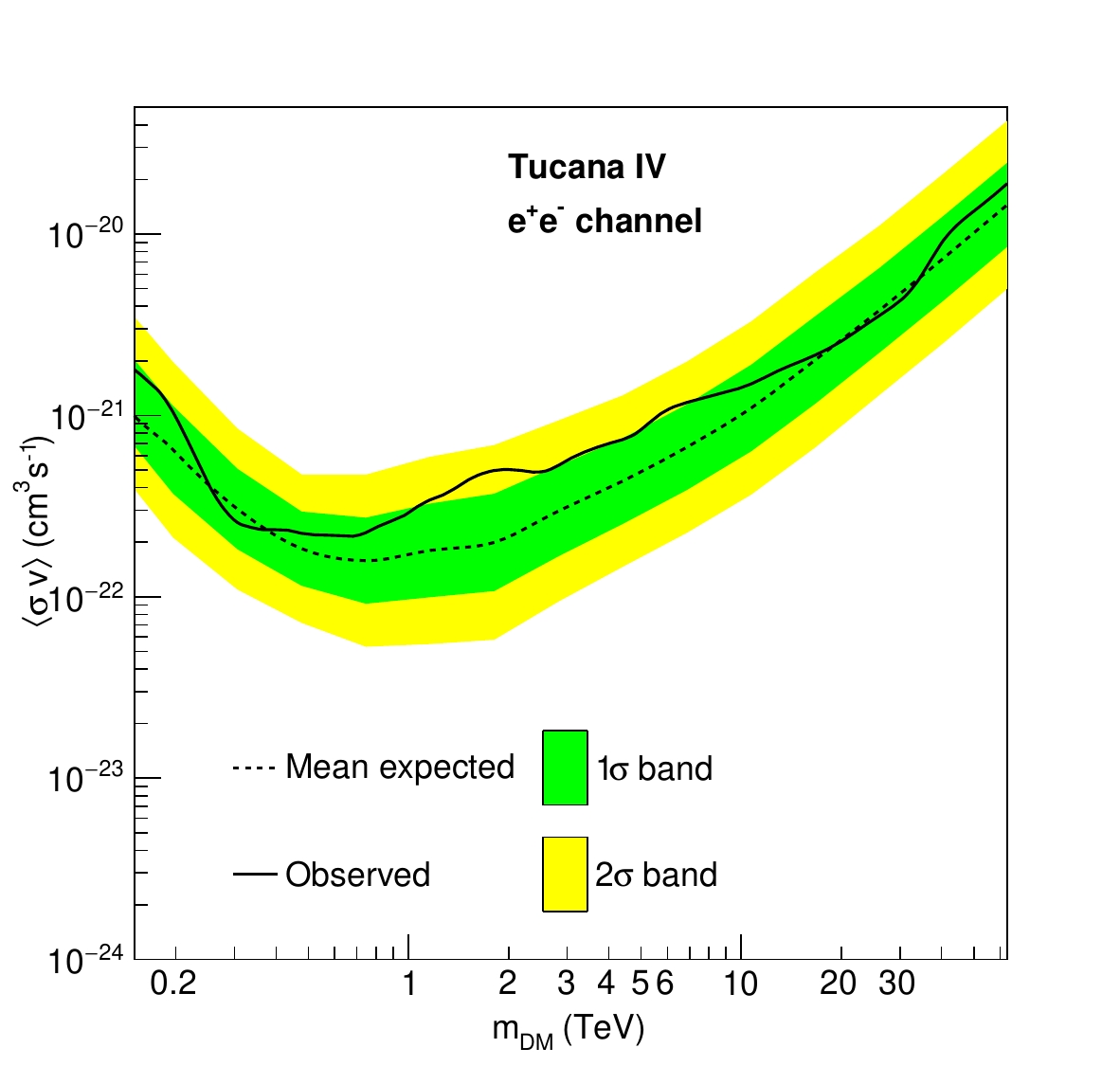}
\includegraphics[width=0.45\textwidth]{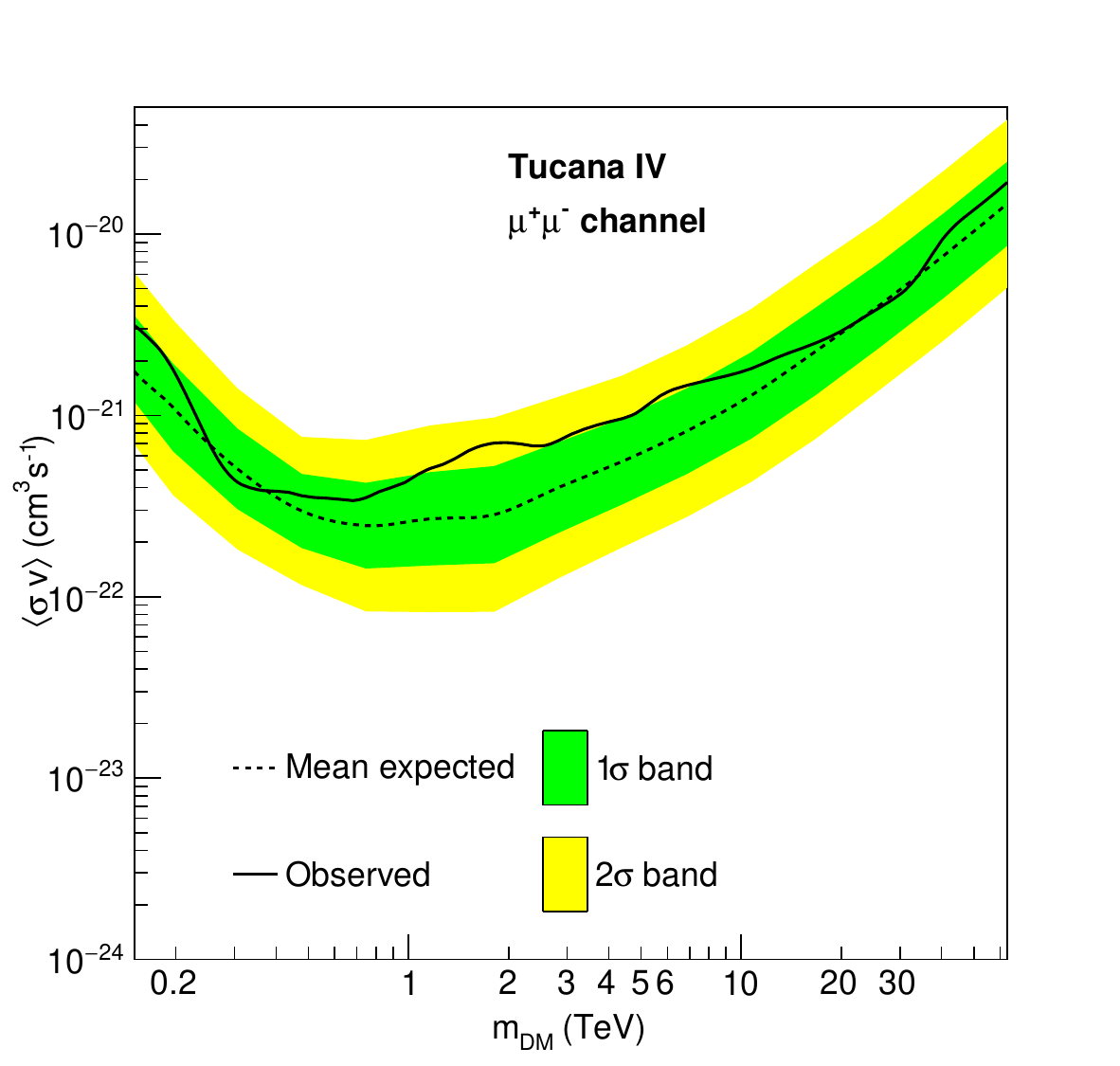}
\includegraphics[width=0.45\textwidth]{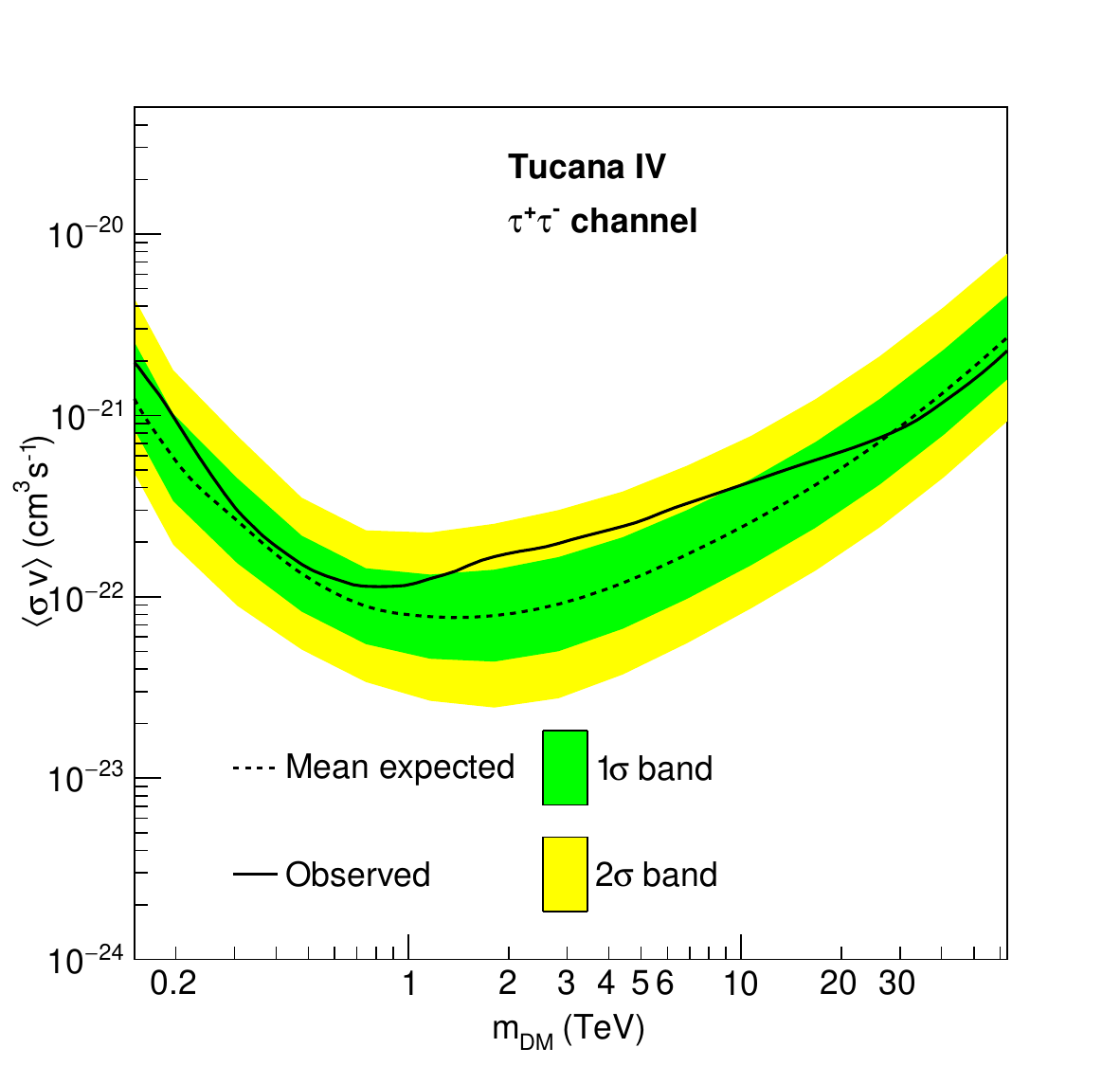}
\caption{95\% C.L. upper limits on the annihilation cross section $\langle\sigma v\rangle$ for Tuc~IV in the $b\bar{b}$, $t\bar{t}$, $ZZ$, $e^+e^-$, $\mu^+\mu^-$, $\tau^+\tau^-$ annihilation channels, respectively, without the uncertainty on the $J$-factor. Observed limits (solid lines) together with mean expected limits (dashed line) and the 1$\sigma$ (green area) and 2$\sigma$ (yellow area) containment bands are shown, respectively.}
\label{fig:sigmavTucIV_all}
\end{center}
\end{figure*}

\begin{figure*}[htbp]
\begin{center}
\includegraphics[width=0.45\textwidth]{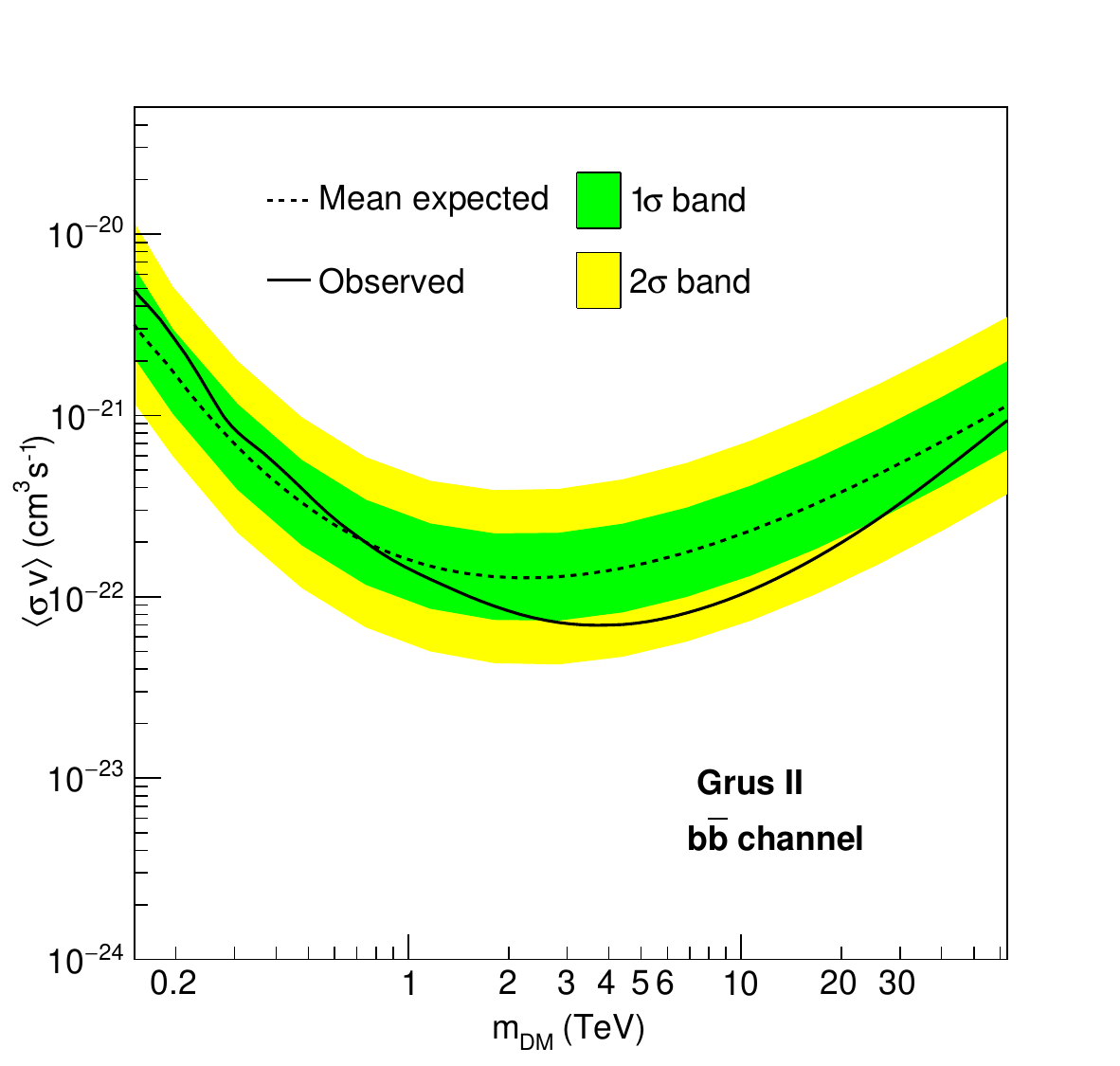}
\includegraphics[width=0.45\textwidth]{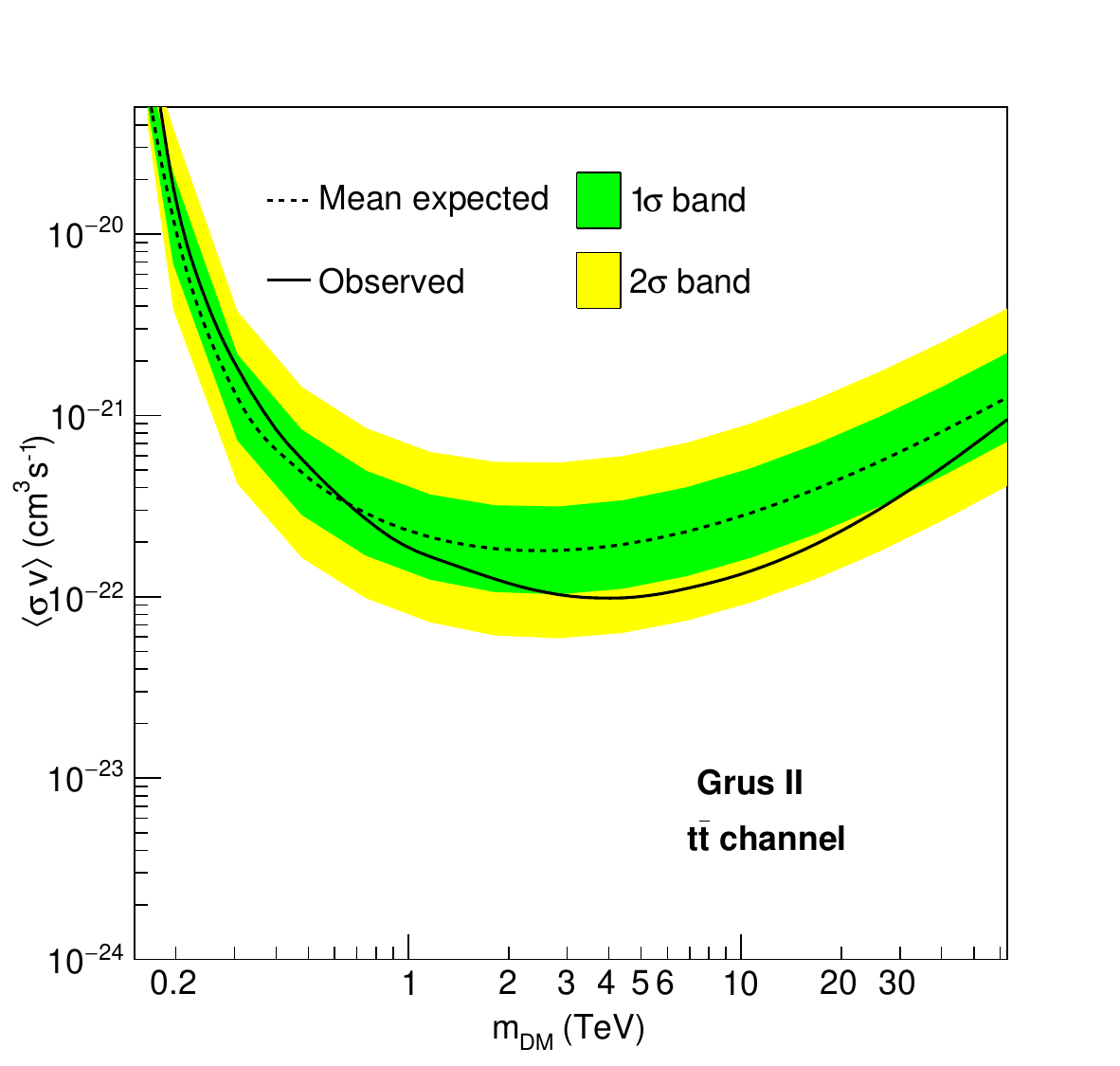}
\includegraphics[width=0.45\textwidth]{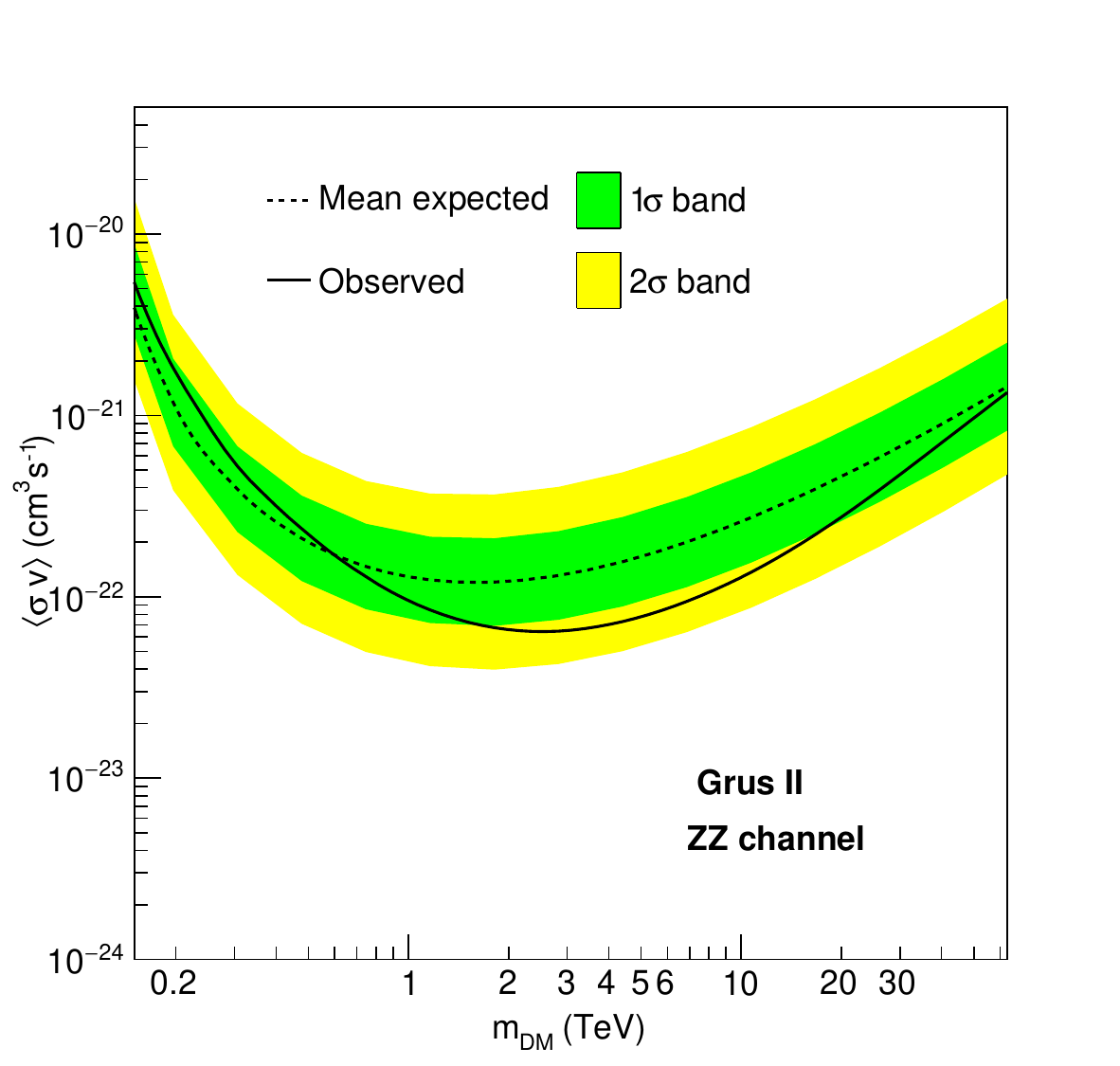}
\includegraphics[width=0.45\textwidth]{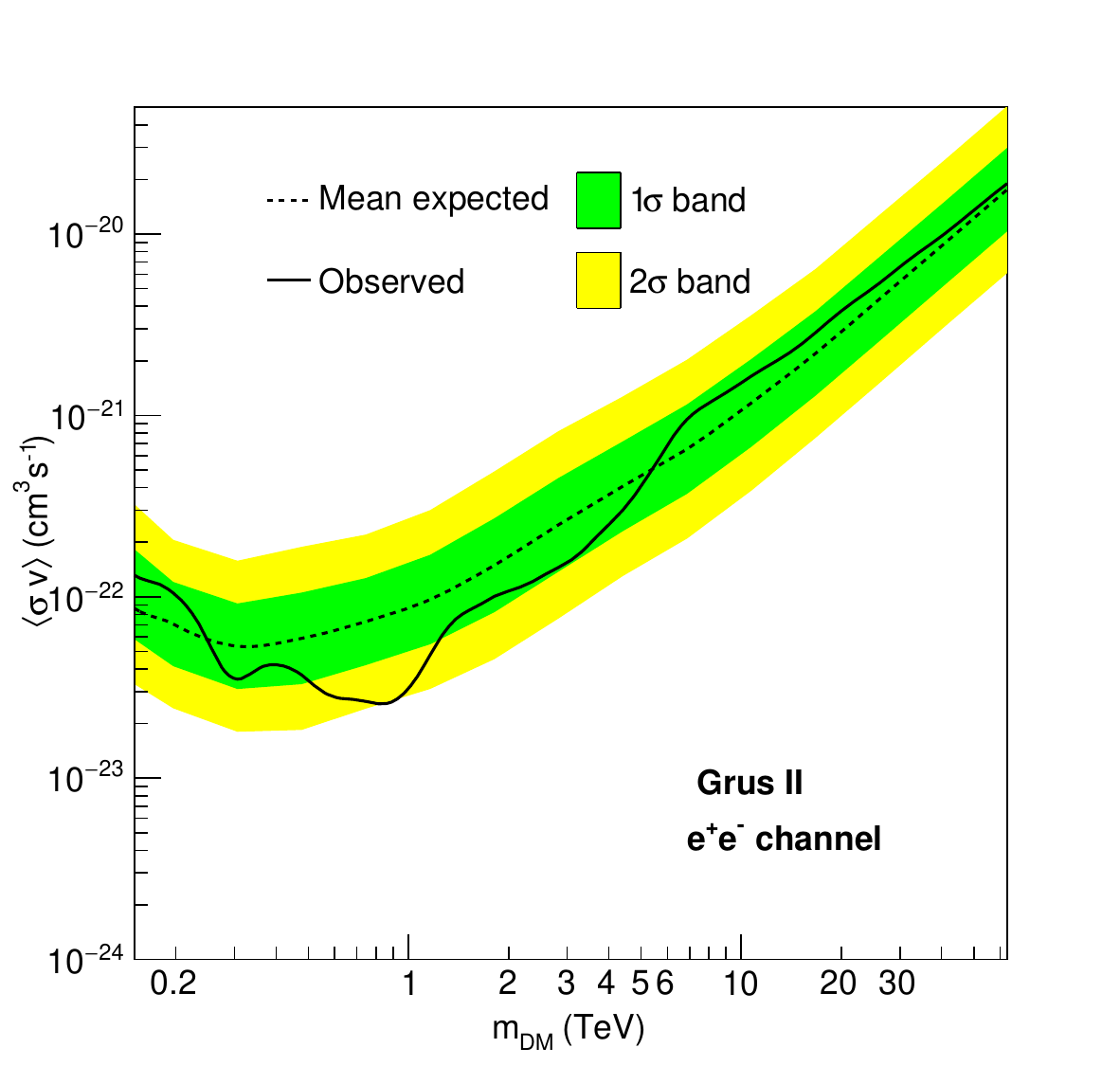}
\includegraphics[width=0.45\textwidth]{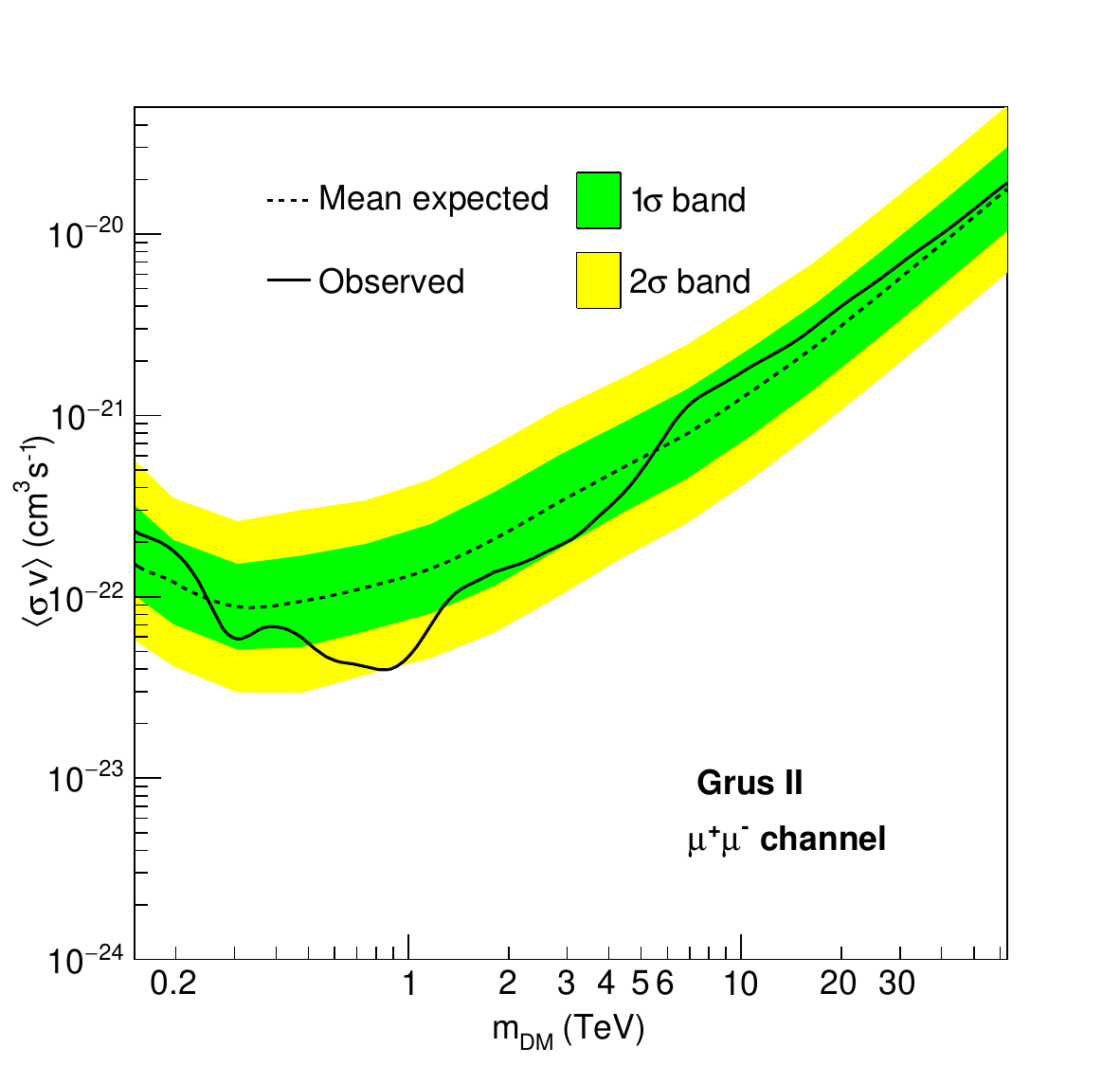}
\includegraphics[width=0.45\textwidth]{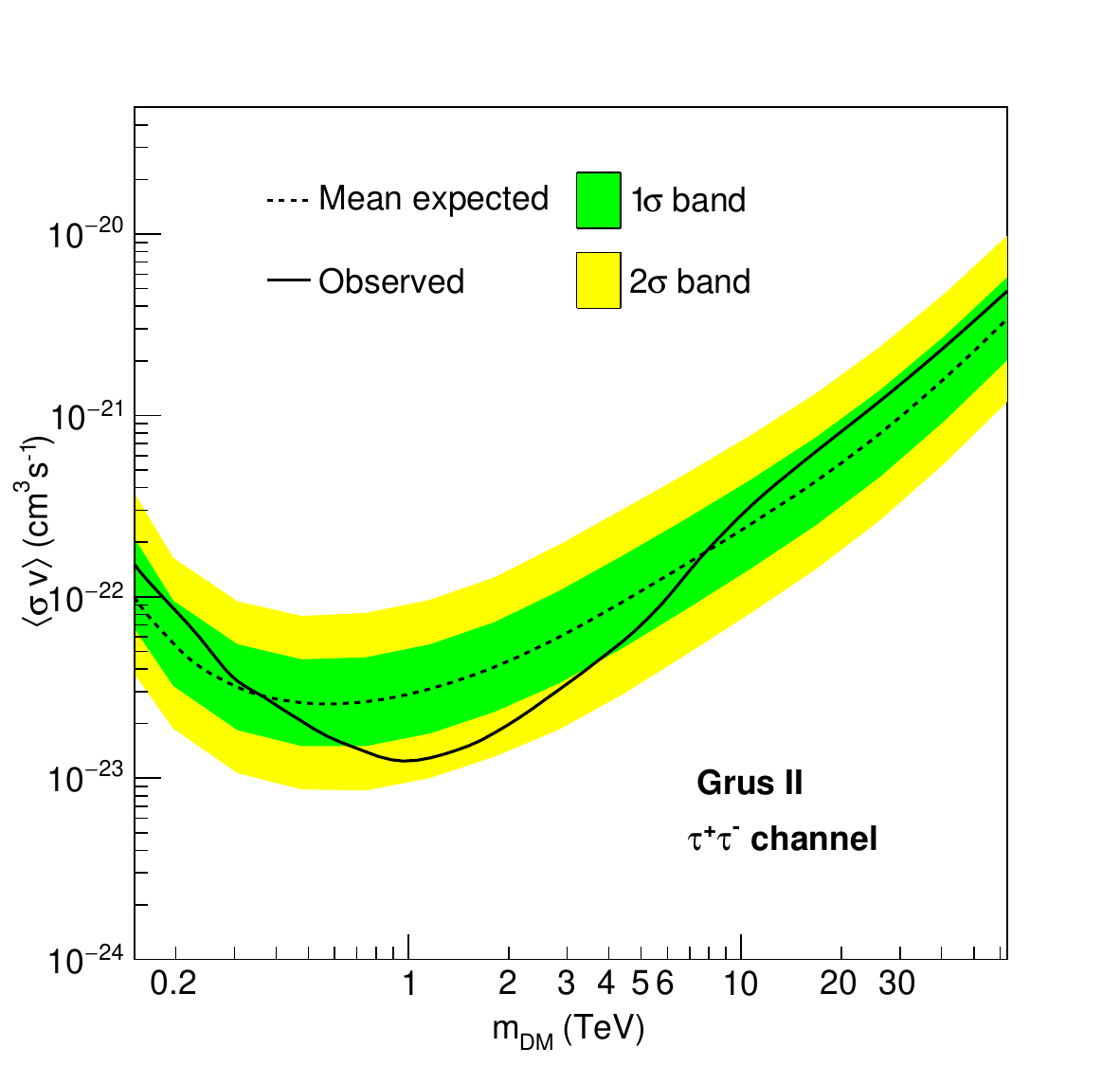}
\caption{95\% C.L. upper limits on the annihilation cross section $\langle\sigma v\rangle$ for Gru~II in the $b\bar{b}$, $t\bar{t}$, $ZZ$, $e^+e^-$, $\mu^+\mu^-$, $\tau^+\tau^-$ annihilation channels, respectively, without the uncertainty on the $J$-factor. Observed limits (solid lines) together with mean expected limits (dashed line) and the 1$\sigma$ (green area) and 2$\sigma$ (yellow area) containment bands are shown, respectively.}
\label{fig:sigmavGruII_all}
\end{center}
\end{figure*}

\end{document}